\documentclass{pasa}
\usepackage{graphicx}
\usepackage{amsmath,amssymb}
\usepackage{gensymb}
\usepackage{xfrac}
\usepackage[flushleft]{threeparttable}
\usepackage{aas-macros}
\usepackage{booktabs}
\usepackage{dblfloatfix}
\usepackage{rotating}
\usepackage[draft=false]{hyperref}
\hypersetup{
  colorlinks=true,
  linkcolor=blue,
  citecolor=blue,
  urlcolor=blue
}

\usepackage{acronym}

\acrodef{AGN}{active galactic nuclei}
\acrodef{VLSSr}[VLSSr]{Very Large Array Low-frequency Sky Survey Redux}
\acrodef{LOTSS}[LoTSS]{LOFAR Two-metre Sky Survey}
\acrodef{MWA}{Murchison Widefield Array}
\acrodef{LOFAR}{LOw-Frequency ARray}
\acrodef{NVSS}{NRAO VLA Sky Survey}
\acrodef{SDSS}{Sloan Digital Sky Survey}
\acrodef{GLEAM-X}{GaLactic and Extragalactic All-sky Murchison Widefield Array eXtended survey}
\acrodef{GLEAM}{GaLactic and Extragalactic All-sky Murchison Widefield Array survey}
\acrodef{CAMB}[CAMB]{Code for Anisotropies in the Microwave Background}
\acrodef{ISW}[ISW]{integrated Sachs-Wolfe effect}
\acrodef{BLLac}[BLLac]{BL Lacertae}
\acrodef{CMB}{cosmic microwave background}
\acrodef{RACS}[RACS]{Rapid ASKAP Continnum Survey}
\acrodef{FLASK}[FLASK]{Full-sky Lognormal Astro-fields Simulation Kit}
\acrodef{EoR}[EoR]{epoch of re-ionisation}
\acrodef{COBE}[COBE]{COsmic Background Explorer}
\acrodef{TGSS}[TGSS]{TIFR GMRT Sky Survey}
\acrodef{WMAP}[WMAP]{Wilkinson Microwave Anisotropy Probe }
\acrodef{LoTSS}[LoTSS]{LOFAR Two-metre Sky Survey}
\acrodef{UV}[UV]{ultraviolet}
\acrodef{ACF}[ACF]{angular correlation function}
\acrodef{GLASS}[GLASS]{Generator for large-scale Structure}
\acrodef{SKA}[SKA]{Square Kilometre Array}

\acrodef{SKAO}[SKAO]{SKA Observatory}%
\acrodef{SKA-Low}[SKA-Low]{the SKAO low-frequency radio telescope}%
\acrodef{SKA-Mid}[SKA-Mid]{the SKAO mid-frequency radio telescope}%

\acrodef{CAMB}[CAMB]{Code for Anisotropies in the Microwave Background}
\acrodef{SUMSS}[SUMSS]{Sydney University Molonglo Sky Survey}
\acrodef{APS}[APS]{angular power spectrum}
\acrodef{GXDS}[GXDS]{\ac{GLEAM-X} Data Subset}
\acrodef{LoFAR}[LOFAR]{Low Frequency Array}
\acrodef{GMRT}[GMRT]{Giant Meterwave Radio Telescope}
\acrodef{SKADS}[SKADS]{European SKA Design Study Simulated Skies}
\acrodef{DMO}[DMO]{dark matter only}
\acrodef{SPH}[SPH]{smoothed particle hydrodynamics}
\acrodef{SOAP}[SOAP]{Spherical Overdensity and Aperture Processor}
\acrodef{SFG}[SFG]{star forming galaxies}
\acrodef{FSRQ}[FSRQ]{flat-spectrum radio quasars}
\acrodef{SSAGN}[SSAGN]{steep-spectrum \ac{AGN}}
\acrodef{HERG}[HERG]{high excitation radio galaxies}
\acrodef{LERG}[LERG]{low excitation radio galaxies}
\acrodef{SFR}[SFR]{star formation rate}
\acrodef{PNG}[PNG]{primordial non-gaussianity}
\acrodef{CDF}[CDF]{cumulative density function}
\acrodef{T-RECS}[T-RECS]{Tiered Radio Extragalactic Continuum Simulatio}
\acrodef{RLF}[RLF]{radio luminosity function}
\acrodef{MCMC}[MCMC]{Markov-Chain Monte-Carlo}

\title[The GHOST simulation]{Generating all-sky radio continuum clustering simulations with GHOST}

\author[Venville et al.]
{Brandon Venville$^{1}$\thanks{email:brandon.venville@postgrad.curtin.edu.au},
Anna Bonaldi$^{2}$,
David Parkinson$^{3}$,
Natasha Hurley-Walker$^{1}$,
Tim Galvin$^{4}$,
Nick Seymour$^{1}$,

\affil{$^{1}$ICRAR-Curtin, Curtin University, Bentley, 6102, 
Western Australia, Australia}%
\affil{$^{2}$SKA Organization, Jodrell Bank, Lower Whitington, Macclesfield SK11 9DL, UK}%
\affil{$^{3}$Korea Astronomy and Space Science Institute,  Daejeon 34055, Republic of Korea}%
\affil{$^{4}$ATNF, CSIRO Space \& Astronomy, Bentley, WA, Australia}

}
\jid{PASA}
\doi{10.1017/pas.\the\year.xxx}
\jyear{\the\year}
\hypersetup{draft}

\begin{document}
\begin{frontmatter}
\maketitle
\begin{abstract}
    Techniques using multiple tracers of the large scale structure of the universe show great promise for examining the fundamentals of our Universe’s cosmology. Such techniques rely on the different relationship between the overdensity of tracers and the broader matter overdensity, enabling cosmic-variance–free tests of primordial non-Gaussianity in the initial curvature perturbations. There is a great opportunity for current and future all-sky extra-galactic radio  surveys to make use of this technique to test for non-Gaussianity at a precision greater than existing all-sky constraints from the cosmic microwave background. To realize this goal there is a need for accurate simulations. Previous radio galaxy simulations have either been realistic but covering only a small area (and so unhelpful for cosmological forecasts), or all-sky dark matter only cosmological simulations but having no connection to a real radio galaxy population. In this study, we use the FLAMINGO suite of cosmological surveys, as well as the matching of dark matter halos to radio galaxy population, to create an accurate sky simulation in order to examine the feasibility of multi-tracer techniques. We present an analysis of the  clustering (with a bias model for the simulation), as well as redshift distributions, source counts and radio luminosity functions, and discuss future work on non-Gaussianity detection. 
\end{abstract}
\end{frontmatter}
\section{Introduction}
Multi-tracer techniques show great promise in constraining \ac{PNG} when applied to next generation large-scale extra-galactic surveys. \ac{PNG} encodes the dynamics of the early Universe and offers a sharp test of inflationary physics. In the widely used `local' model \citep{Bartolo2004,Chen2010}, its amplitude is captured by $f_{\mathrm{NL}}$, the fractional non-linearity, which imprints a distinctive, scale-dependent modulation of galaxy bias $b$ on ultra-large scales $k$ and at redshift $z$, $\Delta b(k,z)\propto f_{\mathrm{NL}}\,k^{-2}$ \citep{Dalal2008, Gomes2020}, enhancing power on the largest scales. 

While attractor single-field slow-roll inflation predicts a nearly Gaussian field with 
$f_{\rm NL}\ll\!1$ \citep{Maldacena2003}, non-attractor, 
or other non-standard dynamics can naturally yield $f_{\rm NL}$ of order unity and distinctive 
bispectrum shapes \citep{Bartolo2004,Chen2010}. 
Current CMB bispectrum limits show no detection but still is consistent with local type non-Gaussianity $f_{\rm NL}^{\rm local}=-0.9\pm5.1$
\citep{Planck2018NG}, leaving room for further tests. Forecasts for upcoming surveys, for example with the \ac{SKAO} telescopes
- \ac{SKA-Low} and \ac{SKA-Mid} - and with {\it Euclid} \citep{2018LRR....21....2A},
indicate sub-unity precision on $f_{\rm NL}^{\rm local}$ under realistic assumptions,
motivating a multi-tracer approach \citep{Yamauchi2014,Gomes2020}\footnote{Here $f_{\rm NL}^{\rm local}$ denotes the amplitude of the “local” bispectrum shape; in general $f_{\rm NL}$ is shape-dependent (e.g. local, equilateral, orthogonal), but large-scale structure is maximally sensitive to the local case because it induces a scale-dependent galaxy bias $\Delta b(k)\propto k^{-2}$. For brevity, unless stated otherwise we use $f_{\rm NL}\equiv f_{\rm NL}^{\rm local}$ hereafter.
}.

Multi-tracer methods are uniquely suited to this regime: by combining galaxy populations with different linear biases and redshift kernels, one can cancel cosmic variance in shared modes and convert relative clustering into information on $f_{\mathrm{NL}}$. 
This typically involves identifying tracers of large scale structure which trace the underlying total matter distribution differently, and using the common distribution to negate the effects of cosmic variance. Large sky area surveys provide access to large angular scales, where the signals of non-Gaussianity are most likely to be found. Theoretical studies indicate we may be able to perform such techniques \citep{Gomes,Ferramacho}, where previously a lack of appropriately observed source populations, with identification of different tracers, has hampered efforts.

Practical sensitivity depends on population mixing, relative ratios in clustering between \ac{AGN} and \ac{SFG}, flux- and frequency-dependent selection, photometric-$z$ errors, survey masks and depth gradients, resolution/morphology effects, and calibration systematics that can mimic large-scale power \citep{AbramoLeonard2013, Ferramacho, Gomes}. A new, end-to-end forward simulation that ties halo populations, luminosity functions, and spectral assumptions to survey-specific observing effects is therefore essential to quantify biases, propagate uncertainties, and validate multi-tracer estimators in conditions that match real data. 

\medskip
\noindent
Prior studies such as \cite{Bonato2017} and the \ac{T-RECS} \citep{TRECS} underpin our approach. \cite{Bonato2017} matched redshift-dependent \ac{SFR} functions to 1.4-GHz  \acp{RLF} to calibrate the radio–\ac{SFR} mapping, explored linear vs. mildly non-linear relations (including scatter), updated the \cite{Massardi} evolutionary model for radio-loud \ac{AGN}, and used number counts to motivate a mild redshift evolution in the synchrotron/\ac{SFR} ratio. \ac{T-RECS} provides an end-to-end population synthesis for \ac{FSRQ}, BL Lac, and steep-spectrum \ac{AGN} and for \ac{SFG}s, including empirical/evolutionary \ac{RLF}s, spectral-index prescriptions, size and subclass proxies, and mock catalogues designed to reproduce source counts and clustering. In this work we adopt those physically motivated ingredients (\ac{RLF} parameterizations, \ac{SFR} to radio calibrations, and soft \ac{HERG}/\ac{LERG} mapping with mass weighting) but embed them in a halo-based framework tied to a cosmological lightcone \cite[\textsc{FLAMINGO};][]{2023MNRAS.526.4978S,Kugel2023} and a forward survey emulator. This lets us enforce consistency between \ac{RLF}s and the halo supply via abundance matching, and propagate realistic selection effects into counts, redshift distributions, and clustering—bridging population models and large-volume cosmology.

\medskip
\noindent
This study seeks to use the aformentioned N-body simulations and analytic halo population to verify the possibilities of using multi-tracer techniques in cosmology, with a realistic sky and experimental systematics. There will be two papers in this series: In this first paper we build \textsc{GHOST}, an all-sky, clustered, survey-ready radio mock catalogue tied to a cosmological lightcone, expressly to validate multi-tracer pipelines for $f_{\rm NL}$. We construct population- and survey-aware catalogues that reproduce wide-area measurements, quantify how selections reshape redshift kernels and deliver the calibrated inputs (galaxy positions, fluxes and halo mass mapping) required for end-to-end \ac{PNG} tests. By supplying tracer families with distinct linear biases and redshift kernels, the catalogue enables controlled tests of cosmic-variance cancellation as it concernes non-Gaussianity constraints, estimator choices in configuration and harmonic space, and the impact of realistic masks, photo–$z$ tails, and population misclassification \citep{Seljak2009, AbramoLeonard2013, Ferramacho, Yamauchi2014, Dalal2008, Gomes, Planck2018NG}.

\medskip
\noindent
Paper~II will use these products to perform the $f_{\rm NL}$ analysis on the simulated sky. The surveys of current instruments, such as the LOFAR Two-metre Sky Survey (\acs{LoTSS}; \citealt{LoTSS}) using the LOw-Frequency ARray (\acs{LOFAR}; \citealt{2013A&A...556A...2V}), as well as the planned surveys of future instruments like \textit{Euclid} and the \ac{SKAO} telescopes shall be used to inform source selection, identification and cosmological analysis, to predict the constraints that may be obtained on cosmological parameters.

\medskip
\noindent
\textsc{GHOST} also will provide a simulation in the vein of \ac{T-RECS} \citep{TRECS} and \ac{SKADS} \citep{2008MNRAS.388.1335W}, but with all-sky coverage as well as large-scale clustering. These improvements will support numerous cosmological applications, for example late-time CMB cross-correlation. Full-sky density fields with known selection allow pipeline development for the integrated Sachs–Wolfe effect and cross-correlation with CMB lensing convergence, including null tests under injected large-scale systematics (dipoles, depth gradients, bright-source masks) and validation of mitigation strategies using pseudo-$C_\ell$ machinery matched to survey footprints \citep[e.g.][]{DavidISW}.

\medskip
\noindent
Weak-lensing and magnification applications for \textsc{GHOST} are supported via add-on ray-tracing. Coupling the sky to a lensing engine \citep[e.g.][LIGER]{2017MNRAS.471.3899B} turns the catalogue into a forward model for magnification and shear of radio sources, enabling tests of number-count and flux-weighted magnification estimators and of their sensitivity to the population mix and selection \citep{2017MNRAS.471.3899B}.

\medskip
\noindent
The \textsc{GHOST} simulation is also suited to studies of the cosmic radio dipole and survey-induced dipoles. With contiguous sky coverage and explicit selection functions, users can inject a kinematic dipole and survey systematics to quantify amplitude/direction bias, or measure the density dipole of flux-limited samples under controlled population mixtures. This connects directly to the radio-dipole literature \citep[e.g.][]{2025arXiv250916732B,2025A&A...697A.112W}.
\noindent
For survey design, the large sky area of \textsc{GHOST} provides a sandbox for confusion, flux boosting, and spatially varying completeness. Because classes are labelled, the catalogue supports training and stress-testing of cross-identification and population-classification models under controllable class imbalance, and exploration of how selection affects downstream cosmology.

Finally, masking \textsc{GHOST} to medium/small fields enables direct forecasts of cosmic-variance floors and assessment of when multi-tracer or cross-correlation approaches deliver meaningful gains for deep tiers, while maintaining consistency with the wide-area selection and clustering that drive ultra-large-scale signals.

The subsequent Sections include a discussion of the \textsc{FLAMINGO} simulations \citep{2023MNRAS.526.4978S,Kugel2023} and how they were used here (Section~\ref{FLAMINGObasics}), the model descriptions (Section~\ref{sec:sfg-rlf}, \ref{sec:agn-pop} and \ref{sec:abmatch}) and results from the simulation in Section~\ref{sec:results}, including clustering (Section \ref{sec:clustering}), redshift distributions (Section~\ref{sec:nz}), source counts (Section~\ref{sec:counts}), and \ac{RLF} results (Section~\ref{sec:rlf}). The luminosity distribution with redshift is discussed in Section~\ref{sec:lumz}, and future improvements to \textsc{GHOST} in Section~\ref{sec:improvements}. We then finish with conclusions of the \textsc{GHOST} simulation in Section~\ref{sec:conclusion}.

\section{FLAMINGO Simulation Basics}\label{FLAMINGObasics}

The \textsc{FLAMINGO} simulations \citep{2023MNRAS.526.4978S, Kugel2023} comprise a suite of 28 runs — 16 hydrodynamical and 12 \ac{DMO} — spanning a range of particle numbers, particle masses, cosmological models, and box sizes. Developed by the Virgo Consortium, \textsc{FLAMINGO} aims to support upcoming large-scale structure surveys through high-fidelity predictions. The simulations use the \textsc{SWIFT} hydrodynamics code \citep{2024MNRAS.530.2378S}, solving the fluid equations with the SPHENIX flavour of \ac{SPH} \citep{2022MNRAS.511.2367B}. \textsc{FLAMINGO} features a novel approach to calibrating subgrid physics — including \ac{AGN} feedback — by applying machine learning techniques trained on observed local datasets. Halo identification, whether in \ac{DMO} or hydrodynamic simulations, was performed using the VELOCIraptor halo finder \citep{2019PASA...36...21E}. Subhalo properties were calculated using the \ac{SOAP} \citep{SOAP} code, developed specifically for \textsc{FLAMINGO}. Available data products include \ac{SOAP} catalogues for multiple redshift snapshots, lightcones for select runs, and \textsc{HEALPix} maps of derived lightcone properties. The full \textsc{FLAMINGO} simulation specifications are detailed by \cite{2023MNRAS.526.4978S}.\\

\medskip
\noindent
In this work, we use a \textsc{FLAMINGO} \ac{DMO} simulation based on the best-fit cosmology from the Dark Energy Survey Year 3 analysis, specifically the ``$3\times2\mathrm{pt} + \mathrm{All\ Ext}\ \Lambda\mathrm{CDM}$'' model. The corresponding cosmological parameters are listed in Table~\ref{tab:cosmo_params}.

\begin{table*}[t]
\centering
\caption{Cosmological parameters adopted for the fiducial flat $\Lambda$CDM model used in this work. Columns list the symbol, parameter name, and adopted value (all quoted at $z=0$). We use the dimensionless Hubble constant $h$; the total matter, dark-energy, and baryon density parameters $\Omega_{\mathrm{m}}$, $\Omega_{\Lambda}$, and $\Omega_{\mathrm{b}}$; the summed neutrino mass $\sum m_\nu$; the amplitude and spectral index of the primordial scalar power spectrum, $A_{\mathrm{s}}$ and $n_{\mathrm{s}}$; and the present-day r.m.s.\ density fluctuation in $8\,h^{-1}\,\mathrm{Mpc}$ spheres, $\sigma_8$. We also quote two derived parameters: $S_8 \equiv \sigma_8\sqrt{\Omega_{\mathrm{m}}/0.3}$ and the neutrino density parameter $\Omega_\nu$ defined via $\Omega_\nu h^2 = \sum m_\nu/(93.14~\mathrm{eV})$.}
\label{tab:cosmo_params}
\begin{tabular}{lr}
\hline
\textbf{Parameter} & \textbf{Value} \\
\hline
$h$                             & $0.681$ \\
$\Omega_{\mathrm{m}}$           & $0.306$ \\
$\Omega_{\Lambda}$              & $0.694$ \\
$\Omega_{\mathrm{b}}$           & $0.0486$ \\
$\sum m_{\nu}$                  & $0.06~\mathrm{eV}$ \\
$A_{\mathrm{s}}$                & $2.099\times10^{-9}$ \\
$n_{\mathrm{s}}$                & $0.967$ \\
$\sigma_{8}$                    & $0.807$ \\
$S_{8}$                         & $0.815$ \\
$\Omega_{\nu}$                  & $1.39\times10^{-3}$ \\
\hline
\end{tabular}
\end{table*}

\medskip
\noindent
We specifically use the L1000N3600 run, with a 1\,Gpc$^3$ box and $3600^3$ particles of mass $8.40\times10^8\ M_\odot$, with the intention of providing sufficient mass resolution to resolve the halos of less massive, low \ac{SFR} \ac{SFG}. $\sim 2.29\times 10^{9}$ halos were selected from a total of $\sim 1.14\times 10^{11}$ available in redshift $z=0$ to $z=10$ and assigned celestial coordinates (RA and Dec) based on their 3D positions in the simulation box at each redshift snapshot. The observer is located at $(0,0,0)$, the center of the coordinate system.

\medskip
\noindent
The \textsc{FLAMINGO} lightcone provides the two ingredients our $f_{\rm NL}$ tests are most sensitive to: ultra–large angular modes from a $\sim$Gpc volume so that the lowest multipoles and cross–population covariance are realized in a single, contiguous sky; and a well-sampled halo mass function down to the scales that host low–SFR \ac{SFG}. We intentionally anchor the abundance matching and host assignment to this specific halo field, because the \ac{PNG} signal enters large-scale clustering primarily through the tracers’ (mass–dependent) bias. Using the same halo supply for catalog construction and for clustering predictions ensures that $n(z)$, effective bias, and survey selections are mutually consistent when we calculate angular clustering statistics. While this paper employs a DMO run, baryonic effects are subdominant at the very large scales driving the scale dependent bias from non-Gaussianity; their residual impact is absorbed in the empirical validation against counts and $n(z)$, and in the bias calibration we carry forward to Paper~II.

\begin{figure*}[p]
    \centering
    \includegraphics[angle=90,width=0.7\textwidth,height=0.92\textheight,keepaspectratio]{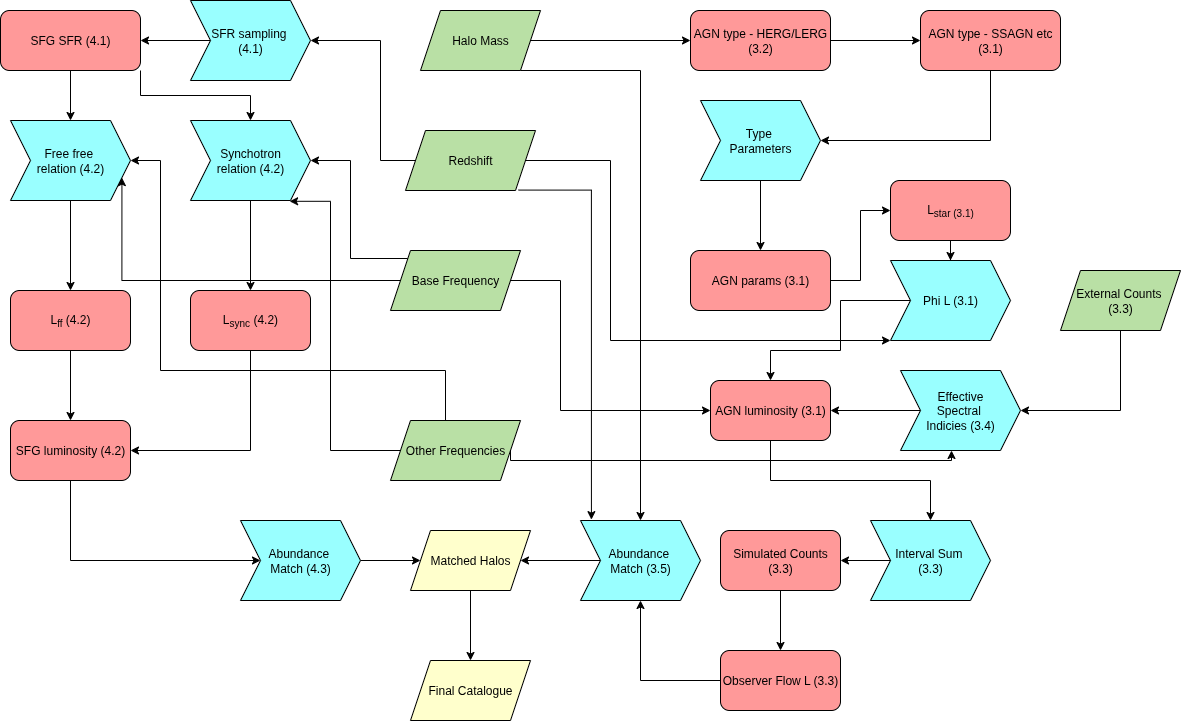}
    \caption{\textbf{Population workflow used in \textsc{ghost}.}
The diagram summarizes how radio sources are generated for the two branches, with relevant subsections of the paper indicated by the bracketed numbers: \textbf{\ac{SFG}} (left) and \textbf{AGNs} (right).
\emph{Green slanted boxes} are external inputs (halo mass, redshift, base/other frequencies, external counts);
\emph{cyan chevrons} are operations/relations (e.g.\ SFR sampling, synchrotron/free–free relations, evolving RLF $\Phi(L)$, spectral–index mapping, abundance matching);
\emph{red rounded boxes} are intermediate products or outputs (component luminosities $L_{\rm ff}$/$L_{\rm sync}$, \ac{SFG}/\ac{AGN} luminosities, simulated counts, observer-frame fluxes).
On the \ac{SFG} side, sampled SFRs are converted to radio luminosity via synchrotron and free–free relations, then propagated to other frequencies.
On the \ac{AGN} side, type parameters (HERG/\ac{LERG}, \ac{SSAGN}, etc.) and the evolving RLF (via $L_\star$ and $\Phi(L)$) set the \ac{AGN} luminosity, with spectral indices mapping to other bands.
Both branches feed an \emph{abundance-matching} step (using halo mass and redshift) to assign hosts and form the final matched-halo catalogue; the counts/interval-sum nodes provide predicted number counts used for verification.
Arrows indicate data flow from inputs to catalogue outputs.}

\end{figure*}

\section{Relation to \ac{T-RECS} and rationale for a new implementation}
\label{sec:relation-trecs}

Our goal is an all-sky, clustered, survey-ready radio mock catalogue built to validate multi-tracer pipelines on ultra–large scales and, ultimately, to constrain $f_{\rm NL}$.
This demands contiguous full-sky light-cones so that the lowest multipoles and cross-population covariances are realized, a halo assignment re-derived on our $N$-body field so that large-scale bias reflects the actual halo supply, and end-to-end, instrument-aware selection functions.
\ac{T-RECS} provides excellent population prescriptions and public clustered mocks, but its clustered realizations are limited in area and tied to a different halo pipeline, which is sub-optimal for the variance-cancellation regime targeted here \citep{TRECS}.
For $f_{\rm NL}$, where the signal appears as a $k^{-2}$ modulation of bias at the largest angular scales, these constraints are decisive.

\medskip
\noindent
We adopt the same high-level taxonomy and broad-band philosophy as \ac{T-RECS}, and inherit some of the modeling: radio \ac{AGN} split into \ac{FSRQ}, \ac{BLLac}, and \ac{SSAGN}, and \ac{SFG} with synchrotron and free–free contributions to their radio spectral energy distributions (SEDs).
We retain smoothly evolving \ac{RLF}s, class-dependent spectral-index distributions, and a minimalist Fanaroff–Riley (FR) proxy for internal labeling.
\textsc{GHOST} differs from \ac{T-RECS} in three ways most relevant to $f_{\rm NL}$.
First, we re-derive the luminosity–halo mapping against the \textsc{FLAMINGO} light-cone and use a rank-preserving, no-replacement match within each redshift slice; this enforces monotonicity, respects the available halo counts, and stabilizes large-scale bias.
Second, we generate contiguous, full-sky clustered realizations so the ultra–large angular modes are present.
Third, our validation focuses on the observables that feed \ac{PNG} analyses—Euclidean-normalized counts, redshift kernels $n(z)$, and angular clustering $w(\theta)\!/C_\ell$—for total and per-population samples under survey-like cuts.

\medskip
\noindent
For \ac{AGN}, \ac{T-RECS} draws host halo masses from \ac{HERG}/\ac{LERG}-based probabilities (inferred via stellar- to halo-mass relations) and associates each source to the nearest-mass halo in the same redshift slice, excluding already-assigned halos.
For \ac{SFG}, it fits an $L(M_h)$ relation by abundance matching and inverts it to map radio luminosities to halo masses, assigning those to halos per slice, with very faint sources below the halo-mass floor placed randomly on the sky.
In \textsc{GHOST} we instead perform a global, rank-order match within each slice (monotonic $L\!\leftrightarrow\!M_h$, no replacement), which prevents Eddington-style leakage across the steep end of the mass function, guarantees that the integral bias of each subsample matches the simulated halo supply, and reduces slice-to-slice bias fluctuations that would otherwise propagate into the $k^{-2}$ \ac{PNG} signature.

\subsection{Types of galaxies in \textsc{GHOST}}
\label{sec:types}
\noindent
\textsc{ghost} includes four observational families that dominate the extragalactic radio sky at metre–centimetre wavelengths. Radio \ac{AGN} appear in three classes—\ac{FSRQ}, \ac{BLLac}, and \ac{SSAGN} — which differ in beaming/orientation and spectral behaviour but are often unified as the same central engine viewed at different angles \citep{Urry1995,HeckmanBest}. Star–forming galaxies (\ac{SFG}) form a separate branch whose radio output tracks recent star formation. Below we summarize the properties and the labels used in this work.

\begin{itemize}
    \item \ac{FSRQ}:
      Flat–spectrum radio quasars show core–dominated, compact radio emission with flat spectra (assumed mean $\alpha=-0.1$\footnote{We define the radio spectral index via $S_{\nu}\propto(\nu/\nu_{0})^{\alpha}$; unless stated otherwise we take $\nu_{0}=1.4\,\mathrm{GHz}$. When quoted as $\alpha_{0}\pm\sigma_{\alpha}$ this denotes the population mean and its $1\sigma$ scatter; by this convention synchrotron sources have $\alpha<0$.}), bright optical continua, and broad emission lines \citep{Urry1995}. In orientation-based unification they are relativistic jets viewed close to the line of sight, hence they have Doppler-boosted luminosities and are compact.
    
      \item \ac{BLLac}:
      Blazars with weak/absent emission lines and flat radio spectra, likewise interpreted as beamed (Doppler-boosted) jets at small inclination; compared to FSRQ they typically have lower line equivalent widths and similarly compact morphologies \citep{Urry1995}.
    
      \item \ac{SSAGN}:
      Steep–spectrum radio AGN, representing the bulk of radio–loud AGN, with mean spectral index $\zeta=-0.8$. This class includes spectroscopic and morphological sub–taxonomies:
      \begin{itemize}
          \item \ac{LERG}: Jet–dominated systems with radiatively inefficient, low–Eddington accretion and weak/absent optical lines.
          \item \ac{HERG}: Radiatively efficient (“radiative–mode”) accretors with strong optical emission lines \citep{HeckmanBest,TRECS}.
          \item FR\,I: Edge–darkened morphology (brightness peaks nearer the core; jets fade outward), more common at lower radio power \citep{FR}.
          \item FR\,II: Edge–brightened lobes with terminal hotspots, more common at higher radio power \citep{FR}.
      \end{itemize}
      These schemes probe different physics and are not one–to–one (e.g.\ FR\,II \ac{LERG} exist). In our catalogue the FR tag is assigned via the size–ratio parameter $R_s$ (Section~\ref{sec:agn-pop}), defined as the separation of peak–brightness regions divided by the total source extent; $R_s>0.5$ is labelled FR\,II and $R_s<0.5$ FR\,I.
    
      \item \ac{SFG}:
      Radio emission powered predominantly by recent star formation: non–thermal synchrotron from cosmic–ray electrons accelerated in supernova remnants plus thermal free–free from H\,\textsc{ii} regions \citep{Condon1992,2011ApJ...732..126M}. At GHz frequencies the ensemble spectral index is steep (typ.\ $\zeta\!\approx\!-0.7$ to $-0.8$) with a mild flattening from free–free toward higher frequencies. In \textsc{ghost} the \ac{SFG} branch is modelled with a calibrated radio luminosity function at 1.4\,GHz and class–appropriate $K$–corrections; it is treated separately from \ac{AGN} because its radio output traces star–formation physics rather than black–hole accretion.
\end{itemize}

\noindent Radio \ac{AGN} typically inhabit more massive halos than \ac{SFG} and therefore have a higher large-scale bias; within \ac{AGN}, radiatively inefficient/jet-dominated systems tend to occupy the most massive environments, while radiatively efficient (quasar-like) systems are somewhat less clustered. Conversely, radio-selected \ac{SFG} are associated with lower-mass halos and exhibit a lower bias. These trends are well-established in wide/deep radio clustering studies (e.g.\ \citealt{2014MNRAS.440.1527L,2017MNRAS.464.3271M,Hale2018}) and are reproduced in our Section~\ref{sec:clustering}. The bias contrast between \ac{SFG} and \ac{AGN} is precisely the lever arm exploited by multi-tracer estimators; as such the presence of the trend in our simulation is of the upmost importance.

\subsection{Star–forming galaxy radio luminosity function (\ac{SFG} RLF)}
\label{sec:sfg-rlf}

We require an \ac{SFG} description that is dust–insensitive and straightforward to propagate through survey selections across metre–centimetre bands. Rather than constructing an SFR function from UV/IR tracers and mapping ${\rm SFR}\!\to\!L_\nu$, we calibrate the \emph{radio} luminosity function of \ac{SFG} directly at 1.4\,GHz and model its smooth redshift evolution. This keeps the baseline independent of assumptions about obscured star formation and the FIR–radio correlation, letting the radio data set the space density as a function of $L_\nu$ and $z$.

\medskip
\noindent
Relative to \ac{T-RECS}, we omit the IR/SFRF intermediate (no $L_{\rm IR}$ term, no explicit FIR–radio coupling); \textsc{GHOST} treats \ac{SFG} emission radio-first. This choice reduces dependence on dust corrections and the redshift evolution of the FIR–radio correlation. The trade-off is that we do not separately track an explicitly IR-luminous subpopulation; any such behaviour is absorbed into the smooth evolution of the radio RLF. As discussed in Section~\ref{sec:nz}, this tends to produce a smoother $n(z)$ at fixed flux (less pronounced features around $z\!\sim\!1$–2) while preserving number counts and large-scale clustering.

For the functional form we adopt a smoothly broken (double-power-law) RLF with redshift-dependent normalization and knee luminosity. All parameters are obtained by fitting to 1.4\,GHz \ac{SFG} RLF measurements in redshift slices, then validated against differential counts at 150\,MHz \citep{Franzen,Mandal2021}, 1.4\,GHz \citep{Bondi2008,Bridle1972,Ciardi2000,Fomalont2006,Gruppioni1999,Hopkins2003,Ibar2009,Kellermann2008,MitchellCondon1985,OwenMorrison2008,Richards2000,Seymour2008,White1997,deZotti2010} and 3\,GHz \citep{2017A&A...602A...2S} using the \ac{SFG} spectral prescription applied elsewhere in this work. Fluxes at other frequencies are generated via standard $K$-corrections with that spectral model; no infrared (IR) quantities enter the catalogue construction. The explicit evolution law and fitted coefficients are listed in Appendix~\ref{app:trecs-eqs}.

In the multi-tracer context the \ac{SFG} branch provides the lower-bias tracer with a broadly lower-redshift kernel set by the radio RLF and survey flux limit. Together with the higher-bias \ac{AGN} branch this delivers the bias contrast and distinct $P(z)$ kernels needed for cosmic-variance cancellation on ultra-large scales. Calibrating the \ac{SFG} RLF directly in radio reduces sensitivity to dust-related systematics that could otherwise imprint spurious large-scale structure in $n(z)$ and $w(\theta)$, improving robustness for $f_{\rm NL}$ forecasts.

\subsection{\ac{AGN} population}
\label{sec:agn-pop}
For each class we adopt the \ac{T-RECS} double–power-law radio luminosity function with smooth redshift evolution in normalisation and knee luminosity; the functional form is listed in Appendix~\ref{app:trecs-eqs}. Coefficients are refitted where required to remain consistent with the host assignment on our simulation and with the class mix used in this work; the resulting values and uncertainties are reported in Table~\ref{tab:trecs_params}. Observed flux densities at 150\,MHz, 1.4\,GHz, and 3\,GHz are obtained from $L_\nu$ via standard $K$–corrections using class-dependent spectral-index distributions (flat for \ac{FSRQ}/\ac{BLLac}; steep for \ac{SSAGN}), as in \ac{T-RECS}. We do not model variability.

Sizes and a soft edge-brightening proxy are assigned to \ac{SSAGN} to provide a morphology label used only for host weighting (Section~\ref{app:sizes}); this proxy is converted to a probabilistic \ac{HERG}/\ac{LERG} tag to allow overlap between morphology and excitation classes. Photometry is unaffected by these labels.

\medskip
\noindent
Hosts are assigned using the abundance-matching scheme defined in Section~\ref{sec:abmatch}. In brief, \ac{AGN} are matched to halos within redshift shells using a monotonic, no-replacement map, with class-dependent host probabilities and stellar-mass–weighted excitation fractions inspired by \cite{Janssen2012}. Tying the placement to our halo supply stabilises the large-scale bias of each class and ensures that the predicted $w(\theta)$ reflects the simulation’s actual halo population.

\medskip
\noindent
Beamed subclasses are handled by drawing jet orientations isotropically and enforcing a maximum viewing angle per class and redshift to reproduce the observed beamed fractions. This yields flat spectra and compact sizes for the beamed subset while leaving the large-scale clustering unchanged.

In summary, we inherit the \ac{T-RECS} class taxonomy, RLF functional form, and spectral prescriptions, while refitting the RLF coefficients where needed and replacing the original halo mapping with our abundance-matching scheme on the \textsc{FLAMINGO} light-cone.

\section{Host assignment and abundance matching}
\label{sec:abmatch}

All clustering results that follow use a host–assigned catalogue: every mock source is attached to a specific halo in the \textsc{FLAMINGO} DMO light-cone. We place this section before the clustering analysis to make explicit what catalogue underlies all $w(\theta)$ and $C_\ell$ measurements.

\medskip
\noindent
Given a survey selection, the requested number of sources in a $(\log L, z)$ bin is
\begin{equation}
\Delta N_{i,j} \;=\; \Omega \,\Phi(L\,|\,z_i)\,
\left.\frac{dV}{dz}\right|_{z_i}\,
\Delta\log L_{i,j}\,\Delta z_i ,
\label{eq:main-deltaN}
\end{equation}
with $\Phi$ the class radio luminosity function (double–power–law with smooth evolution; see Appendix~\ref{app:trecs-eqs}). For \ac{SFG} we realise these targets by fitting, per redshift slice $i$ and flux density interval $j$, a monotone halo–to–luminosity map $L_\nu(M_h; z)$ (Equation~\ref{eq:LofM}) with optional log–normal scatter and then rank–ordering halos to luminosities without replacement. For \ac{AGN} we first draw hosts with mass–weighted \ac{HERG}/\ac{LERG} properties (Appendix~\ref{app:trecs-eqs}), then assign $L_\nu$ monotonically within the selected subset to reproduce the class RLFs. When halos are scarce, we apportion the available hosts across populations in proportion to their requested counts (largest–remainder rule; Section~\ref{sec:abmatch-apportion}) so the hosted mixture tracks the selection–driven requested mixture. This alignment of mixture, bias hierarchy, and selection is the key ingredient for robust multi-tracer $f_{\rm NL}$ forecasts.

\subsection{Assumptions and limitations}
\label{sec:abmatch-limits}
We assign a single central radio source per halo (no explicit satellites). The scatter $\sigma_{\log L}$ is taken to be constant within a shell. \ac{HERG}/\ac{LERG} labels are proxies for host–mass weighting rather than strict spectroscopic classes. At the highest redshifts, finite halo supply can saturate the brightest \ac{SFG} targets; we quantify hosted versus requested counts in Section~\ref{sec:rlf}.

\subsection{Inputs and redshift shells}
\label{sec:abmatch-shells}
We partition redshift into disjoint shells: 
$
\mathcal{Z}_i \equiv [z_i, z_{i+1}) \quad (i=0,\dots,N-1),
$
so that $\bigcup_{i=0}^{N-1}\mathcal{Z}_i = [z_0,z_N)$ and $\mathcal{Z}_i\cap\mathcal{Z}_j=\varnothing$ for $i\neq j$.
 (the same grid used for $n(z)$ and RLFs). In each shell we compute the comoving shell volume per steradian and the target numbers per luminosity bin and population from the relevant RLF under the survey selection (cf. Equation~\ref{eq:main-deltaN}). Candidate hosts are all halos in the shell, each with mass $M_h$ and light–cone position. We assign a stellar–mass proxy via the redshift–dependent $M_\star(M_h, z)$ relation of \citet{2015ApJ...810...74A}.

\subsection{\ac{SFG} branch: monotonic $L$–$M_h$ abundance matching}
\label{sec:abmatch-sfg}
For \ac{SFG} we fit a smooth, monotonic mapping so that the host–assigned \ac{SFG} luminosity histogram reproduces the radio \ac{SFG} RLF $\Phi_{\rm \ac{SFG}}(L_\nu, z)$ (Section~\ref{sec:sfg-rlf}). We adopt
\begin{equation}
L_\nu(M_h; z)\;=\;N(z)\,\Bigg[\Big(\frac{M_h}{M_\star(z)}\Big)^{\gamma}+\Big(\frac{M_h}{M_\star(z)}\Big)^{\beta}\Bigg]^{-1},
\label{eq:LofM}
\end{equation},
where $M_h$ is the halo mass, and $M_*(z)$, $\gamma$,$\beta$ and $N(z)$ are free parameters.
with log–normal scatter $\epsilon\sim \mathcal{N}(0,\sigma_{\log L}^2)$. The parameters $\{N, M_\star, \gamma, \beta, \sigma_{\log L}\}$ are inferred independently in each shell by maximising a Poisson likelihood comparing the target counts (from the RLF) to the counts produced by rank–ordering halos in $M_h$ and mapping to $L_\nu$ via Equation~\ref{eq:LofM}. We explore the posterior with an ensemble \ac{MCMC} \citep[emcee,][]{emcee} and adopt median parameters for assignment.

\subsection{\ac{AGN} branch: mass–weighted host selection and monotone assignment}
\label{sec:abmatch-agn}
For radio \ac{AGN} we draw hosts with a stellar–mass–dependent probability that encodes \ac{HERG}/\ac{LERG} tendencies and then assign luminosities monotonically within the selected subset so that the class RLFs are reproduced.

\medskip
\noindent
Each \ac{SSAGN} is given a stochastic size from \cite{Dipompeo} and edge–brightening proxy $R_s$ from \cite{Lin2010}; we label $R_s>0.5$ as FR\,II and $R_s<0.5$ as FR\,I and use this as a soft proxy for excitation class. Following \cite{Janssen2012,TRECS}, we weight hosts by stellar mass, with a high–mass cap (“saturation”) for the LERG branch:
\begin{align}
P_{\rm LERG}(M_\star|z)\;\propto\;
\begin{cases}
\left(M_\star/M_0\right)^{2.5}, & M_\star \le M_{\rm sat},\\[3pt]
\left(M_{\rm sat}/M_0\right)^{2.5}, & M_\star > M_{\rm sat},
\end{cases}
\qquad\\
P_{\rm HERG}(M_\star|z)\;\propto\;\left(\dfrac{M_\star}{M_0}\right)^{1.5},
\end{align}
with $M_{\rm sat}\simeq 10^{11.6}\,M_\odot$. By “saturating” we mean the LERG triggering weight stops increasing above $M_{\rm sat}$. We renormalise $P_t(M_\star|z)$ over candidate hosts in each redshift shell and draw hosts so that \ac{HERG}/\ac{LERG} totals match the RLF targets for that shell (\ac{FSRQ} $\rightarrow$ HERG-like; \ac{BLLac} $\rightarrow$ LERG-like; \ac{SSAGN} split by $R_s$).
Hosts are sampled without replacement. Within each class and shell we sort hosts by $M_h$, sort luminosities by $L_\nu$, and map by rank (with optional log–normal $L$ scatter) to reproduce the class RLF while anchoring large–scale bias to the actual halo field. FSRQ/BL\,Lac orientations are drawn isotropically (uniform in $\cos\theta$). 

\subsection{Population–weighted host apportionment}
\label{sec:abmatch-apportion}
Let $\mathcal{L}_t(s)$ be the set of luminosities requested in slice $s$ for population $t\in\{\mathrm{FSRQ},\mathrm{BL\,Lac},\mathrm{SSAGN},\mathrm{SFG}\}$, with size $R_t(s)=|\mathcal{L}_t(s)|$. Let $K(s)$ be the number of available hosts (unassigned halos) in the slice. If $K(s)\ge R_{\rm tot}(s)\equiv\sum_t R_t(s)$, all requests are hosted.

When $K(s)<R_{\rm tot}(s)$, we apportion the $K(s)$ hosts across populations in proportion to requests using a largest–remainder rule:
\begin{equation}
q_t \;=\; \frac{R_t(s)}{R_{\rm tot}(s)}\,K(s), \qquad
K_t \;=\; \min\!\bigl(R_t(s),\,\lfloor q_t\rfloor \bigr),
\end{equation}
then distribute the remaining $K(s)-\sum_t K_t$ one by one to populations with the largest fractional remainders $q_t-\lfloor q_t\rfloor$ (never exceeding $R_t$). This guarantees $\sum_t K_t=\min\!\bigl(K(s),R_{\rm tot}(s)\bigr)$ and $K_t\le R_t$ for all $t$.

Within each population, sort luminosities descending, $L_{t,(1)}\ge\cdots\ge L_{t,(R_t)}$, and map rank to a host–mass index inside the quota via
\begin{equation}
i(r) \;=\; \Bigl\lceil r\,\frac{K_t}{R_t}\Bigr\rceil \in \{1,\dots,K_t\}.
\end{equation}
Across all populations we collect these (predicted–mass, population) pairs, sort by predicted mass, and assign the top $K(s)$ to the actual top $K(s)$ halos (strictly preserving mass rank). Leftover requests $R_t-K_t$ are written as “no–halo’’ rows (\texttt{*NH}) at the slice mid–redshift, preserving per–population totals and making host scarcity explicit.

\subsection{Algorithmic summary}
\label{sec:abmatch-steps}
In each shell we compute target counts per class and luminosity bin from the relevant RLF (Equation~\ref{eq:main-deltaN}) under the survey selection, gather all halos and assign $M_\star(M_h, z)$ from \citet{2015ApJ...810...74A}, and then perform assignment by branch. For \ac{SFG} we fit $L_\nu(M_h)$ via Equation~\ref{eq:LofM} using an ensemble MCMC on a Poisson likelihood and assign by rank without replacement, adding log–normal scatter. For \ac{AGN} we draw \ac{HERG}/\ac{LERG} hosts using $P_{\rm \ac{HERG}/\ac{LERG}}(M_\star)$ and, within each class, assign $L_\nu$ monotonically to match the class RLF. We then attach sky positions from hosts, apply spectral prescriptions and $K$–corrections, and finally apply the survey selection (flux/SNR survey selection functions and completeness surrogates).

\section{Initial Results and Catalogue Validation}\label{sec:results}
In this section, we present representative results from the simulated sky catalogue described above. We perform consistency checks against the input luminosity and redshift distributions, assess population statistics, and provide example sky visualizations. We will discuss applications to multi-tracer techniques. The simulated catalogue contains a total of $\sim2.29\times10^{9}$ galaxies, comprising both \ac{AGN} and \ac{SFG} across $4\pi$\,steradians to a sensitivity of $1\,\mu\mathrm{Jy}$ at $1.4~\mathrm{GHz}$. The \ac{AGN} population includes $\sim2.32\times10^{7}$ \ac{FSRQ}, $\sim2.08\times10^{7}$ \ac{BLLac}s, and $\sim6.60\times10^{7}$ \ac{SSAGN}, while the \ac{SFG} population contributes $\sim2.18\times10^{9}$ sources.
 The redshift distribution spans the range $z \in [0, \sim6]$, with the majority of sources concentrated below $z \sim 4$.

\medskip
\noindent
We use the same survey selections for both the redshift–distribution $n(z)$ and clustering analyses to enable like-for-like comparisons with the literature. We re-classify for comparison with \cite{Magliocchetti2004} because the rule is a single, explicit luminosity threshold that can be mirrored exactly. We do not re-classify to compare with \cite{Hale2018} because those clustering measurements treat the sample as a mixed population without a uniform \ac{AGN}/\ac{SFG} split. For a different reason, we do not re-classify for comparison with \cite{Hale2018} because their labels are derived from multi-band diagnostics that are not available uniformly in our mock; we therefore emulate only the detection flux density threshold and separate population-mix effects (our classes) from selection effects (their $\mathrm{S/N}$ cut).

\subsection{Selections used for validation}
\label{sec:validation-selections}
 This Section defines the cuts and clarifies how each survey separated \ac{AGN} from \ac{SFG} (where they did), and how we mirrored those choices in \textsc{GHOST}. The three emulations are: the VLA–COSMOS 1.4\,GHz luminosity threshold used by \citet{2017MNRAS.464.3271M}; the LoTSS-DR2 \citep{2024MNRAS.527.6540H,2022A&A...659A...1S} wide-area 150\,MHz flux and peak–S/N survey selection functions; and the VLA–COSMOS 3\,GHz peak–S/N threshold and multi-wavelength classes of \citet{Hale2018}. Table~\ref{tab:class-schemes} summarises the survey selection functions and our emulation strategy. As emphasised in Section~\ref{sec:clustering}, these selections compress or broaden the redshift kernel $n(z)$ and thus reshape the projected clustering amplitude/slope without altering the intrinsic bias ordering, so we present raw and selected catalogues side by side in what follows.

\medskip
\noindent
\subsubsection{LoTSS–DR2 wide 150\,MHz \citep{LoTSSDR2,2024MNRAS.527.6540H}.}\label{sec:lotssselection}
We emulate the LoTSS selection by combining an integrated–flux cut $S_{150}\ge 1.5~\mathrm{mJy}$ with a peak–S/N threshold ${\rm S/N}_{\rm peak}\ge 7.5$. The image rms varies across the mosaic due to multiplicative effects, such as primary beam sensitivity, mosaicking overlap,  and calibration/dynamic range, so we model the local rms $\sigma$ as log–normal in an effort to replicate this, $\ln\sigma\sim\mathcal{N}(\mu_{\ln},\sigma_{\ln}^2)$. Anchoring the distribution to the median $\tilde{\sigma}=83~\mu\mathrm{Jy\,beam^{-1}}$ and 95th percentile $171~\mu\mathrm{Jy\,beam^{-1}}$ gives
\begin{equation}
\mu_{\ln}=\ln\tilde{\sigma},\qquad
\sigma_{\ln}=\frac{\ln(171~\mu\mathrm{Jy})-\mu_{\ln}}{1.645},
\end{equation}
which reproduces the observed quantiles by construction. For a source of integrated flux $S$ (we take $S_{\rm peak}\!\approx\!S$ at 1.5\,mJy; for resolved sources this is conservative), the S/N cut is equivalent to $\sigma\le\sigma_{\rm thr}=S/7.5$, yielding a detection probability
\begin{equation}
p_{\rm keep}(S)=\Phi\!\left(\frac{\ln\sigma_{\rm thr}-\mu_{\ln}}{\sigma_{\ln}}\right),
\end{equation}
with $\Phi$ the standard normal CDF. Using the class counts $dN_X/dS$, we define the area–weighted acceptances
\begin{equation}
{\rm acc}_X=
\frac{\displaystyle \int_{S_{\min}}^\infty \frac{dN_X}{dS}(S)\,p_{\rm keep}(S)\,dS}
     {\displaystyle \int_{S_{\min}}^\infty \frac{dN_X}{dS}(S)\,dS},
\qquad X\in\{\mathrm{SFG},\mathrm{AGN}\},
\end{equation}
and form the effective mixture entering the total $n(z)$,
\begin{equation}
w_{\rm \ac{SFG}}=\frac{{\rm acc}_{\rm \ac{SFG}}\,N_{\rm \ac{SFG}}}
{{\rm acc}_{\rm \ac{SFG}}\,N_{\rm \ac{SFG}}+{\rm acc}_{\rm AGN}\,N_{\rm AGN}},
\end{equation}
where $S_{\min}=1.5~\mathrm{mJy}$ and $N_X$ are class totals from the un–gated $n(z)$. At $S_{150}=1.5~\mathrm{mJy}$ the flux threshold removes only a few percent of sources in both classes, so the resulting $n(z)$ is dominated by the intrinsic \ac{SFG}/\ac{AGN} mix (Figure~\ref{fig:nz_lotss15}). \cite{2024MNRAS.527.6540H} does not supply a homogeneous \ac{AGN}/\ac{SFG} split for this clustering sample, so we emulate the selection only and compare the \emph{total} $n(z)$ and clustering; by–class predictions shown elsewhere use our internal labels to isolate population–mix effects from detection selection.

\medskip
\noindent
\subsubsection{VLA–COSMOS 1.4\,GHz \citep{2017MNRAS.464.3271M}.}\label{sec:maglioccettiselection} We mirror the flux density limit $S_\mathrm{1.4GHz}\ge 0.15$ mJy and re-classify sources using their redshift-dependent radio-luminosity boundary. Radio powers are computed as $P_{1.4}=F_{1.4}\,D_A^2\,(1+z)^{3+\alpha}$ (W\,Hz$^{-1}$ sr$^{-1}$) with $\alpha=0.7$ and $D_A$ the angular diameter distance, and the AGN/\ac{SFG} division is
\begin{equation}
\log_{10} P_{\rm cross}(z)=
\begin{cases}
21.7 + z, & z\le 1.8\\
23.5, & z>1.8
\end{cases}
\end{equation}
(we convert to isotropic luminosity via $L_{\rm cross}=4\pi P_{\rm cross}$). Objects with $L_{1.4}\ge L_{\rm cross}(z)$ are labelled “AGN-like”, the rest “SFG-like”. This reproduces their threshold classifier exactly and makes the comparison sensitive to population mix and redshift tails rather than to differences in detection survey selection functions; it also emphasises the overlapping mJy regime where both classes contribute (see Figure~\ref{fig:nz_mag_split}).

\medskip
\noindent
\subsubsection{VLA–COSMOS 3\,GHz \citep{Hale2018}.}\label{sec:haleselection} We emulate the catalogue’s peak–flux threshold ${\rm S/N}\ge 5.5$ with a field median rms $\tilde{\sigma}\simeq 2.3~\mu{\rm Jy\,beam^{-1}}$, modelling field-to-field rms as lognormal with
\begin{equation}
\mu_{\ln}=\ln\tilde{\sigma},\qquad \sigma_{\ln}=0.30.
\end{equation}
For a source of flux density $S$, the rms required to pass is $\sigma_{\rm thr}=S/5.5$; the area fraction that admits detection is
\begin{equation}
p_{\rm keep}(S)=\Phi\!\left(\frac{\ln(S/5.5)-\mu_{\ln}}{\sigma_{\ln}}\right).
\end{equation}
We then compute a class-wise, area-weighted acceptance by averaging $p_{\rm keep}$ over the per-population 3\,GHz counts histogram $H_X(S_i)$ (from the simulation),
\begin{equation}
{\rm acc}_X=\frac{\sum_i H_X(S_i)\,p_{\rm keep}(S_i)}{\sum_i H_X(S_i)},\qquad X\in\{\mathrm{SFG},\mathrm{AGN}\},
\end{equation}
and form the effective mixture
\begin{equation}
w_{\rm \ac{SFG}}=\frac{{\rm acc}_{\rm \ac{SFG}}\,N_{\rm \ac{SFG}}}
{{\rm acc}_{\rm \ac{SFG}}\,N_{\rm \ac{SFG}}+{\rm acc}_{\rm AGN}\,N_{\rm AGN}}.
\end{equation}
At the Hale depth the selection function is weak; in our run the acceptances are close to unity for both classes (numerically ${\rm acc}_{\rm \ac{SFG}}\!\approx\!0.89$, ${\rm acc}_{\rm AGN}\!\approx\!0.94$), so the predicted $n(z)$ is governed mainly by the intrinsic \ac{SFG}/\ac{AGN} mix. Because the Hale classification relies on deep multi-wavelength diagnostics (X-ray, MIR colour, UV–IR SED, rest-frame NUV colour, and radio excess relative to SFR), we do not attempt to reproduce their labels inside \textsc{GHOST}; we emulate only the $\mathrm{S/N}$ selection with a final flux cut of $12.5\,\mu$Jy at $3\,$GHz and compare the \emph{total} $n(z)$, using our internal classes where a by-class illustration is helpful.

\begin{table*}[t]
\centering
\caption{Classification schemes of comparison surveys and the emulation used in this work.}
\label{tab:class-schemes}
\begin{tabular}{p{2.5cm} p{3.0cm} p{4.0cm} p{4.0cm}}
\toprule
Dataset & Selection & Their classification & Our emulation \\
\midrule
VLA-COSMOS, 1.4\,GHz \citep{2017MNRAS.464.3271M} &
$F_{1.4}\!\ge\!0.15$\,mJy &
Luminosity split: $L_{1.4}\gtrless L_{\rm cross}(z)$ with $\log L_{\rm cross}(z)=21.7+z$ (to $z\le1.8$), else $23.5$ &
We apply the same $L_{1.4}$ vs $L_{\rm cross}(z)$ split to GHOST\\
\addlinespace
LoTSS DR2, 150\,MHz \citep{LoTSSDR2} &
$S_{150}\!\ge\!1.5$\,mJy + peak S/N threshold &
LoTSS DR2 per-catalogue classes (data-driven mix) &
We emulate the survey selection function \\
\addlinespace
VLA--COSMOS 3\,GHz (Hale+18) &
$S/N_{\rm peak}\!\ge\!5.5$ &
Multi-wavelength (SED/line) \ac{AGN} and \ac{SFG} classes &
We emulate the S/N threshold and compare total and by-class using our internal classes \\
\bottomrule
\end{tabular}
\end{table*}

\subsection{Abundance–matching outcomes}
\label{sec:abmatcheffects}

All results that follow use the host–assigned catalogue from Section~\ref{sec:abmatch}: every mock source is either attached to a specific \textsc{FLAMINGO} halo or, if the slice’s host supply is exhausted, written as a "no–halo" (\texttt{*NH}) row. When halos are scarce we apportion the available hosts across populations in proportion to their requested counts (largest–remainder rule; Section~\ref{sec:abmatch-apportion}) and then assign luminosities monotonically within each population so that the class \ac{RLF}s are reproduced.

At fixed flux limit the accessible radio luminosity rises with redshift,
\begin{equation}
L_{\rm lim}(z,\zeta)=4\pi\,D_L^2(z)\,S_{\rm lim}\,(1+z)^{\alpha-1},
\label{eq:Llim}
\end{equation}
so the observed $(L_\nu,z)$ domain is bounded jointly by Equation~\ref{eq:Llim} and the evolving halo mass function. With the corrected slice projection, both the abundance-matching targets and the host–assigned catalogue populate this domain smoothly across all slices.

\begin{figure*}[p]
  \centering
  \includegraphics[width=\textwidth,height=0.8\textheight,keepaspectratio]{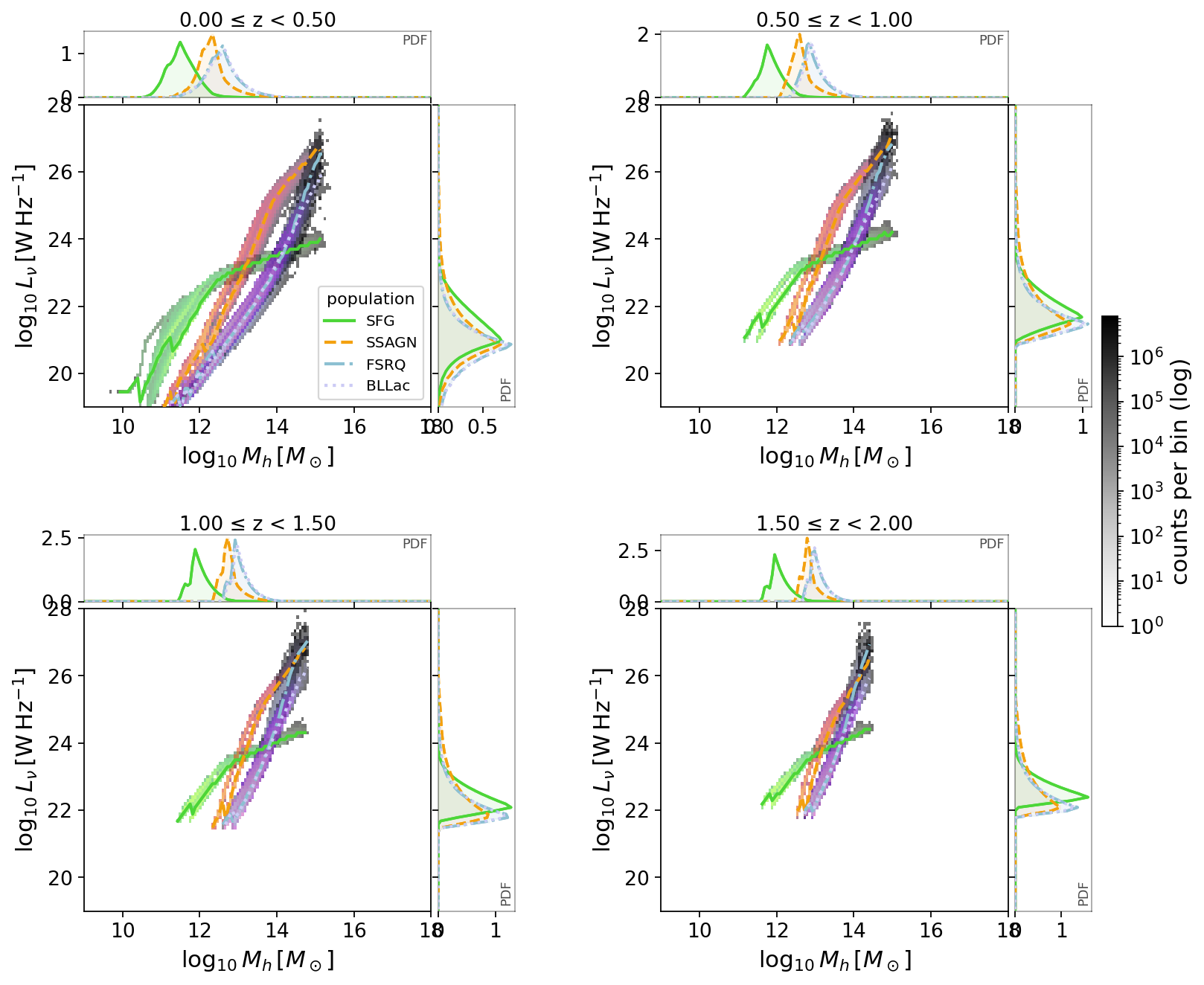}
    \caption{\textbf{All populations — radio luminosity vs.\ halo mass with per–population marginals (shared density scale).}
Each panel shows 2D binned abundances in a fixed redshift window; colored semi–transparent maps overlay populations (\ac{SFG} \emph{solid}, \ac{SSAGN} \emph{dashed}, FSRQ \emph{dash–dot}, BL\,Lac \emph{dotted}). All density layers share the same logarithmic colorbar, so intensity is directly comparable across populations and panels. Solid lines trace the median (“ridgeline”) of $\log_{10}L_\nu$ at fixed $\log_{10}M_h$.
The small top (right) axes show, for each population, the \emph{normalized} marginal PDFs $p(\log_{10}M_h)$ and $p(\log_{10}L_\nu)$—with $\int p\,\mathrm{d}\log M_h=1$ and $\int p\,\mathrm{d}\log L=1$—using the same color/linestyle as the ridgelines.
\emph{Reading the figure:} \ac{SFG} peak at lower host masses and luminosities; \ac{SSAGN} extend to higher $M_h$ with a steeper $L_\nu$–$M_h$ trend; FSRQ and BL\,Lac preferentially inhabit the most massive halos and dominate the bright radio tail. With increasing redshift, ridgelines shift to higher $M_h$ and the PDFs narrow, reflecting both selection effects and the evolving halo–occupation mix.}
  \label{fig:LMh_all_overlay}
\end{figure*}

Figure~\ref{fig:LMh_all_overlay} summarizes the delivered mapping. Star–forming galaxies occupy lower–mass halos and lower $L_\nu$; steep–spectrum \ac{AGN} extend to higher $M_h$ with a steeper $L$–$M$ trend; beamed subclasses (\ac{FSRQ},\ac{BLLac}) preferentially inhabit the most massive halos and dominate the bright tail. With increasing redshift, median ridge-lines shift as expected from $K$–corrections and the evolving halo supply, and the per–population marginals narrow modestly.

We track requested versus hosted counts per slice and population (Section~\ref{sec:abmatch-apportion}). The hosted fraction is high across all panels; occasional \texttt{*NH} rows appear only at very low~$z$ and $L_\nu$ and are negligible for the selections used. In the RLF grids we show the model projected on the same $(L,z)$ grid alongside the host–assigned estimate; the two agree within uncertainties over the full dynamic range.

The host–assigned catalogue embeds a clear bias hierarchy—\ac{SFG} $<$ \ac{SSAGN} $<$ FSRQ/BL\,Lac—set by the actual halo field under the survey selection. Because our population–weighted apportionment keeps the hosted mixture close to the requested mixture for any flux/SNR cut, changes in $w(\theta)$ across selections reflect projection kernels rather than a reversal of intrinsic bias ordering. This alignment of mixture, bias hierarchy, and selection is exactly what multi-tracer $f_{\rm NL}$ estimators require.

\subsection{The angular correlation function}
\label{sec:clustering}
\noindent
Our goal in this section is to verify that the catalogue places each population in environments consistent with their expected large-scale bias, and to produce clustering summaries that feed the \ac{PNG} analysis in Paper~2. We therefore measure the two-point angular correlation of the simulated sky, both in configuration space, $w(\theta)$, and in harmonic space, $C_\ell$. Given a redshift kernel $\phi(z)$ (from the $n(z)$ of each population), these observables probe the projection of the 3D matter power spectrum modulated by the tracers’ bias, so their amplitudes and slopes are sensitive to the typical host halo masses of \ac{SFG}s versus \ac{AGN} subclasses and how survey selections reshape $\phi(z)$.

Concretely, we compute $w(\theta)$ for \ac{AGN} and \ac{SFG} samples (and for specific survey-like cuts), fit a compact power-law model $w(\theta)=A(\theta/\theta_0)^{1-\gamma}$ over $0.02^\circ\text{--}1^\circ$ to quote comparable amplitudes and slopes, and cross-check in harmonic space by fitting $C_\ell$ with fixed $\phi(z)$ templates to infer effective linear bias (both constant and gently evolving with redshift).

\subsubsection{Definitions and modelling}
We quantify angular clustering with the two–point correlation function $w(\theta)$, related to the angular power spectrum $C_\ell$ by
\begin{equation}
w(\theta)\;=\;\sum_{\ell=0}^{\infty}\frac{2\ell+1}{4\pi}\,C_\ell\,P_\ell(\cos\theta)
\;\simeq\;\int\frac{\ell\,\mathrm{d}\ell}{2\pi}\,J_0(\ell\theta)\,C_\ell,
\label{eq:wtheta_def}
\end{equation}
where the Bessel form is accurate for small angles. In the Limber regime, the auto–spectrum for a population $X\in\{\mathrm{AGN},\mathrm{SFG}\}$ is

\begin{equation}\label{eq:Cl_limber}
\begin{split}  
C_\ell^{XX}
=\int \mathrm{d}z\;
\underbrace{\frac{H(z)}{\chi^2(z)}\,\phi_X^2(z)}_{\text{projection}}
\;\underbrace{b_X^2(k,z)\,D^2(z)}_{\text{bias/growth}}\;\\
\times P_m\!\left(k=\frac{\ell+1/2}{\chi(z)},z\right),
\end{split}
\end{equation}

\noindent
with comoving distance $\chi(z)$, Hubble rate $H(z)$, linear growth $D(z)$, matter spectrum $P_m$, and a normalized redshift kernel $\phi_X(z)$ (unit area).

Across a finite angular decade the observed slope of $w(\theta)$ reflects a weighted average of the 3D logarithmic slope $ \mathrm{d}\ln[P_m\,b_X^2]/\mathrm{d}\ln k $ evaluated at $k\simeq \ell/\chi(z)$, convolved with $\phi_X^2$.
For direct comparison to the literature and to quote compact summary numbers we fit a power law
\begin{equation}
w(\theta)=A\left(\frac{\theta}{\theta_0}\right)^{1-\gamma},
\label{eq:powerlaw_fit}
\end{equation}
using full covariances where available and a pivot $\theta_0$ taken to be the geometric mean of the fitted $\theta$ to minimise the $A$–$\gamma$ covariance. Unless stated otherwise we fit over $\theta\in[0.02^\circ,\,1.0^\circ]$ ($\ell\simeq 50$–$400$).

To interpret slope changes induced by selections, we also quote \emph{effective} redshifts and distances computed with Limber–like weights
\begin{equation}
W_X(z)\;\propto\;\frac{H(z)}{\chi^2(z)}\,\phi_X^2(z)\,D^2(z),
\qquad
\end{equation}
\begin{equation}
z_{\rm eff,X}=\frac{\int \! W_X(z)\,z\,\mathrm{d}z}{\int \! W_X(z)\,\mathrm{d}z},
\qquad
\end{equation}
\begin{equation}
\chi_{\rm eff,X}=\frac{\int \! W_X(z)\,\chi(z)\,\mathrm{d}z}{\int \! W_X(z)\,\mathrm{d}z}.
\label{eq:zeff_chieff}
\end{equation}
A fixed angular separation then corresponds to a characteristic projected comoving scale
\begin{equation}
r_\perp(\theta)=\theta\,\chi_{\rm eff,X}.
\label{eq:rperp}
\end{equation}

\begin{figure}[t]
  \centering
  \includegraphics[width=\linewidth]{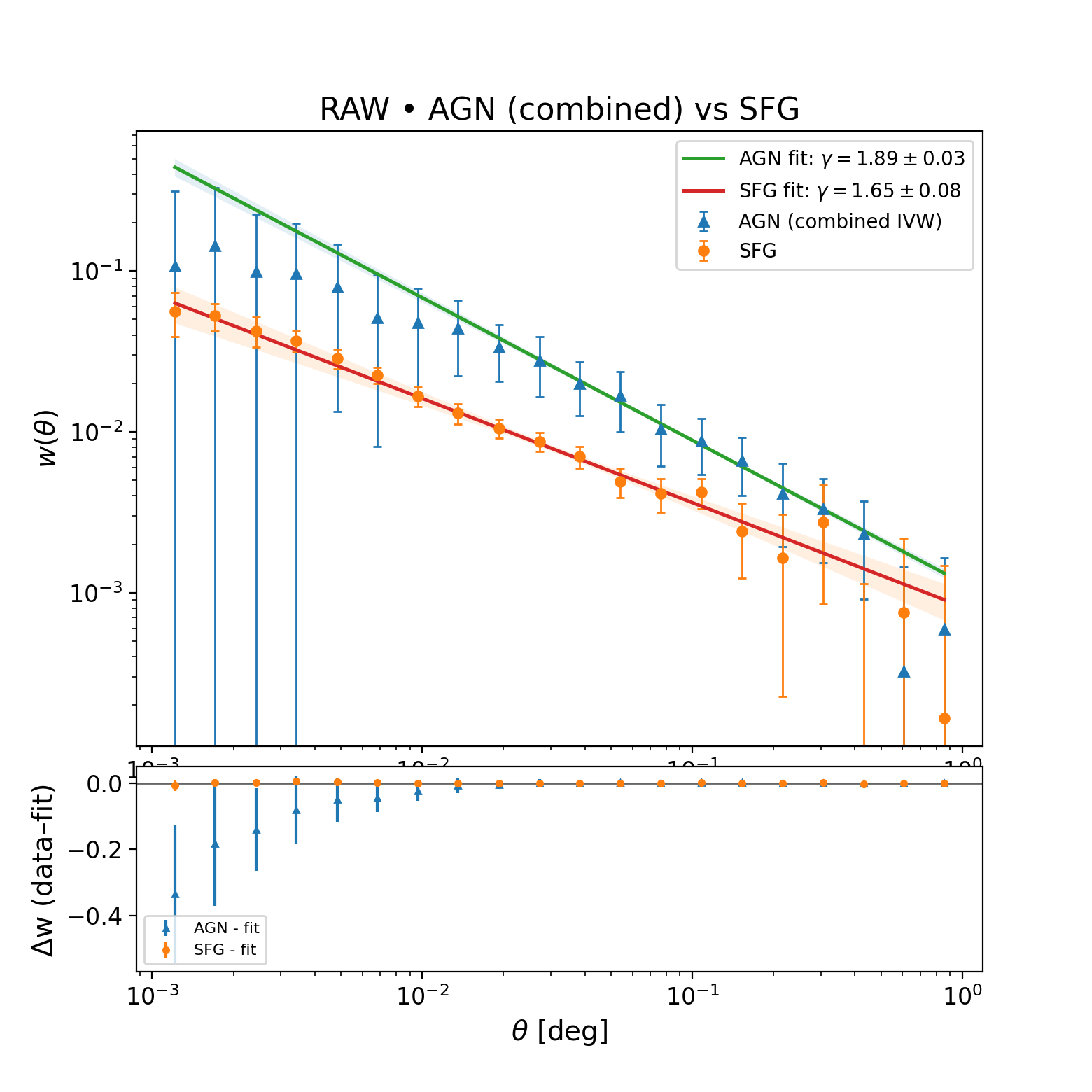}
  \caption{raw \textsc{GHOST} sample: Angular correlation function $w(\theta)$ for \ac{AGN} (blue triangles) and \ac{SFG} (orange circles). Solid lines show inverse-variance weighted power-law fits $w(\theta)=A(\theta/\theta_0)^{1-\gamma}$ over $\theta\in[0.02^\circ,\,1.0^\circ]$; shaded bands are the propagated $1\sigma$ model uncertainties from the $(A,\gamma)$ covariance. Best-fit slopes are $\gamma_{\rm AGN}=1.886\pm0.028$ and $\gamma_{\rm \ac{SFG}}=1.647\pm0.076$ . The lower panel shows data$-$fit residuals. The amplitude ordering AGN$>$\ac{SFG} reflects the intrinsic bias hierarchy, while the slightly shallower \ac{SFG} slope is consistent with a broader, higher-$z$ kernel $\phi(z)$ that mixes more $k=\ell/\chi(z)$ in projection (Equations~\ref{eq:wtheta_def}--\ref{eq:Cl_limber}).}
  \label{fig:acf_raw}
\end{figure}

\subsubsection{Raw \textsc{GHOST} sample.}
In Figure~\ref{fig:acf_raw}, we present the angular correlation function of the full \textsc{GHOST} sample showing \ac{AGN} lying above \ac{SFG} in amplitude across all $\theta$, consistent with their higher bias and host–halo masses.
Power–law fits give
$\gamma_{\rm AGN}=1.886\pm0.028$ and
$\gamma_{\rm \ac{SFG}}=1.647\pm0.076$.
The effective depths are
$(z_{\rm eff},\chi_{\rm eff})=(0.318,\,1.22~\mathrm{Gpc})$ for \ac{AGN} and $(0.602,\,2.12~\mathrm{Gpc})$ for \ac{SFG}, so $r_\perp(0.05^\circ)=(1.06,\,1.85)~\mathrm{Mpc}$, respectively.
The shallower \ac{SFG} slope in the raw \textsc{GHOST} sample is naturally explained by their broader, higher–$z$ kernel $\phi(z)$, which mixes a wider range of $k=\ell/\chi$ in projection (Equations~\ref{eq:Cl_limber}–\ref{eq:zeff_chieff}).

\begin{figure}[h]
  \centering
  \includegraphics[width=\linewidth]{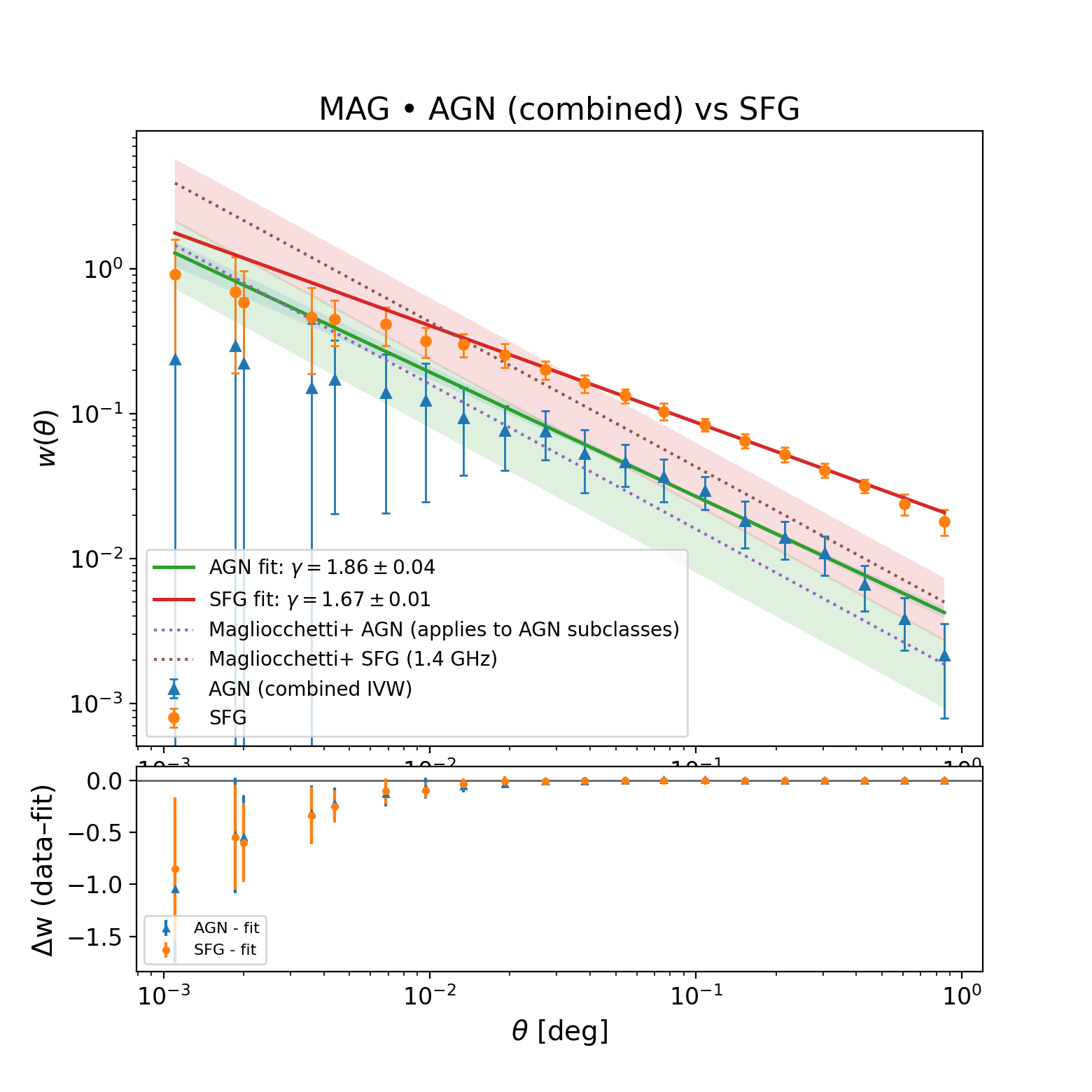}
  \caption{VLA-COSMOS selection (1.4\,GHz): $w(\theta)$ for \ac{AGN} and \ac{SFG} after the $L_{1.4}\gtrless L_{\rm cross}(z)$ selection function detailed in Section~\ref{sec:maglioccettiselection}. Solid curves are our power-law fits over $\theta\in[0.02^\circ,\,1.0^\circ]$ (shaded $1\sigma$ bands); dotted lines and translucent bands show the literature reference for \ac{AGN} and \ac{SFG}, respectively. We obtain $\gamma_{\rm AGN}=1.858\pm0.041$ and $\gamma_{\rm \ac{SFG}}=1.667\pm0.010$. The \ac{SFG} amplitude lies above the reference at large scales and the slope is slightly flatter. From the effective-depth summary (Table~\ref{tab:proj_numbers}) the \ac{SFG} kernel compresses to $z_{\rm eff}\simeq0.094$ and $\chi_{\rm eff}\simeq0.41$\,Gpc, so a fixed $\theta$ probes small $r_\perp$ (e.g.\ $r_\perp(0.05^\circ)\approx0.35$\,Mpc), flattening the angular scaling (Equations~\ref{eq:zeff_chieff}--\ref{eq:rperp}).}
  \label{fig:acf_mag}
\end{figure}

\subsubsection{VLA–COSMOS 1.4\,GHz }
Applying the $L_{1.4}\gtrless L_{\rm cross}(z)$ selection, as detailed in Section \ref{sec:maglioccettiselection}, compresses the \ac{SFG} kernel to very low $z$.
We find $\gamma_{\rm AGN}=1.858\pm0.041$ and $\gamma_{\rm \ac{SFG}}=1.667\pm0.010$, with $(z_{\rm eff},\chi_{\rm eff})=(0.339,\,1.30~\mathrm{Gpc})$ for \ac{AGN} and $(0.094,\,0.41~\mathrm{Gpc})$ for \ac{SFG}, implying $r_\perp(0.05^\circ)=(1.14,\,0.35)~\mathrm{Mpc}$. In Fig.~\ref{fig:acf_mag}, the \ac{SFG} correlation lies above the AGN over the fitted range.
This does not imply a larger intrinsic \ac{SFG} bias; it follows from the projection weights in the Limber integral, which scale as $\phi^2(z)/\chi^2(z)$. Our selection compresses the \ac{SFG} kernel to very low redshift ($z_{\rm eff}\simeq0.094$, $\chi_{\rm eff}\simeq0.41~\mathrm{Gpc}$) while the AGN kernel sits higher ($z_{\rm eff}\simeq0.339$, $\chi_{\rm eff}\simeq1.30~\mathrm{Gpc}$).
The resulting geometric factor $(\chi_{\rm AGN}/\chi_{\rm \ac{SFG}})^2\simeq(1.30/0.41)^2\!\approx\!10$ boosts the projected \ac{SFG} amplitude and can outweigh the larger AGN bias, thereby producing the observed inversion in $w(\theta)$.
The residual shape mismatch — most evident in the flatter \ac{SFG} slope—likely reflects blending of one- and two–halo terms when projecting such a low-$z$ kernel over $0.02^\circ\!-\!1^\circ$ ($r_\perp\!\sim\!0.14\!-\!7~\mathrm{Mpc}$), and modelling choices that suppress small-scale power (e.g. our current radio-SFG occupation lacks an explicit satellite component).
These effects account for the difference in gradient while preserving the relative bias hierarchy that underpins the multi-tracer analysis.

\begin{figure}[h]
  \centering
  \includegraphics[width=\linewidth]{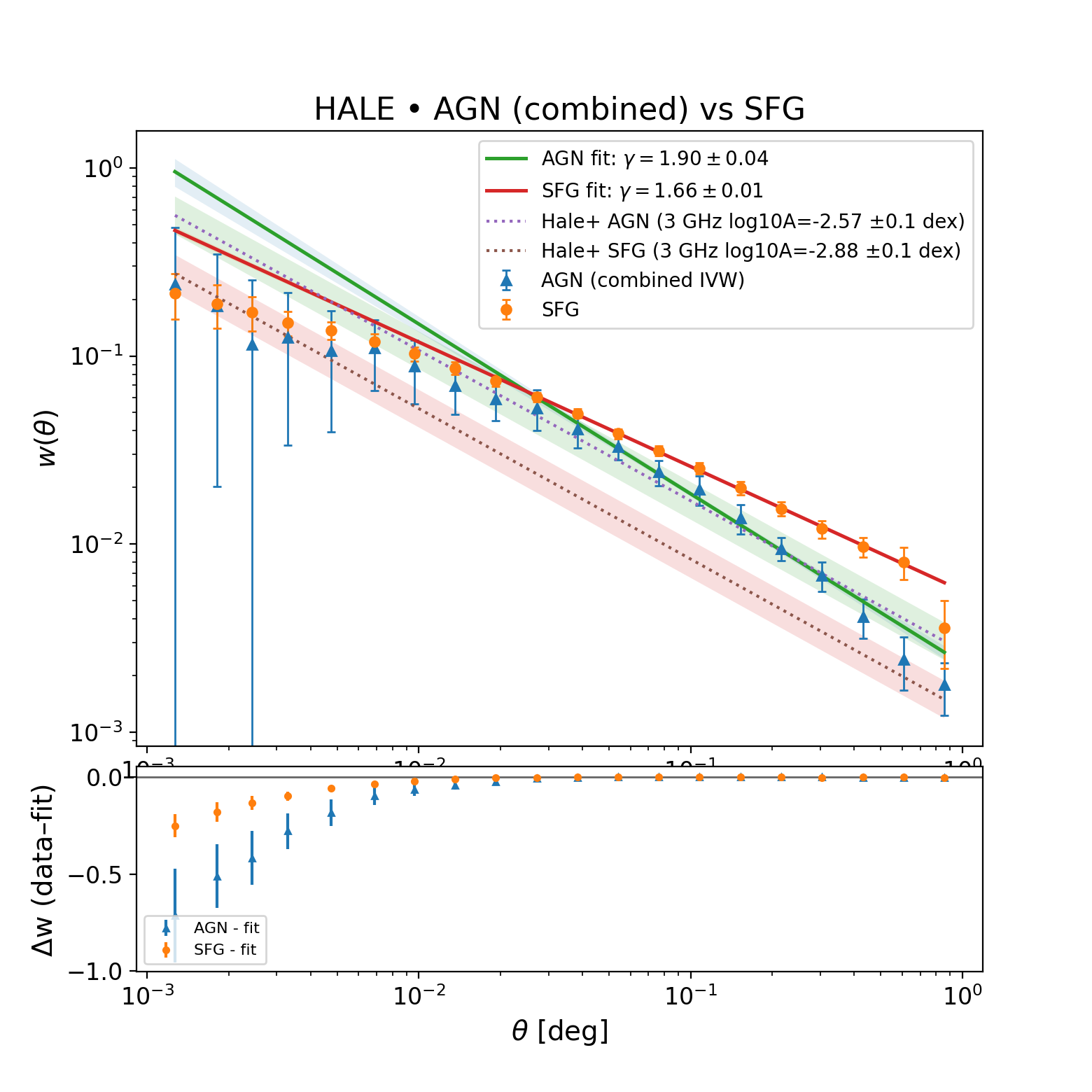}
  \caption{VLA-COSMOS selection (3\,GHz): Same as Fig.~\ref{fig:acf_mag} but for the 3\,GHz cut (Section~\ref{sec:haleselection}. Our fits give $\gamma_{\rm AGN}=1.903\pm0.037$ and $\gamma_{\rm \ac{SFG}}=1.662\pm0.010$. \ac{AGN} closely follow the band from \cite{Hale2018}, while \ac{SFG} sit high and slightly flatter. The projection summary (Table~\ref{tab:proj_numbers}) shows $z_{\rm eff,\,SFG}\simeq0.161$ and $\chi_{\rm eff,\,SFG}\simeq0.67$\,Gpc (so $r_\perp(0.05^\circ)\approx0.59$\,Mpc), again explaining the slope change as a selection-driven projection effect rather than frequency-dependent systematics. Residuals (lower panel) are shown relative to the fitted power laws.}
  \label{fig:acf_hale}
\end{figure}

\subsubsection{VLA–COSMOS 3\,GHz}
The same pattern repeats at 3\,GHz, with the selection described in Section~\ref{sec:haleselection}:
$\gamma_{\rm AGN}=1.903\pm0.037$ and $\gamma_{\rm \ac{SFG}}=1.662\pm0.010$, with
$(z_{\rm eff},\chi_{\rm eff})=(0.313,\,1.21~\mathrm{Gpc})$ for AGN and
$(0.161,\,0.67~\mathrm{Gpc})$ for \ac{SFG}, hence $r_\perp(0.05^\circ)=(1.05,\,0.59)\,\mathrm{Mpc}$.
Relative to the Hale et al.\ 3\,GHz bands our \ac{SFG} points lie higher and show a slightly
shallower slope, eventually overtaking the AGN at large angles. This behavior is expected from the different redshift kernels: the \ac{SFG} selection sits at lower $\chi_{\rm eff}$,
so the projection integral schematically
$w(\theta)\!\propto\!\int dz\,\phi^2(z)\,b^2(z)\,D^2(z)/\chi^2(z)$
weights it more strongly, and the same angular range probes smaller $r_\perp$,
mixing some one–halo power into the fitted scales and flattening the apparent slope.
By contrast, the AGN kernel projects to larger $r_\perp$ where the two–halo term
dominates and the slope is steeper. Residual differences with the \cite{Hale2018} bands are therefore plausibly explained by modest kernel mismatches (e.g. a slight \ac{SFG}
overweight at $z\!\lesssim\!0.3$ in the mocks), survey–specific treatments of integral-constraint and close pairs/multi-component sources that suppress small-angle power, and minor AGN/SFG cross-contamination near the luminosity split.
The cross–frequency consistency (compare 1.4\,GHz) supports the interpretation
that these are selection–driven projection effects rather than frequency–dependent
systematics.

\subsubsection{Picture across frequencies}
The differences we see across the raw, Magliocchetti and Hale–like samples are driven primarily by how each selection reshapes the redshift kernel $\phi(z)$, not by any change in the intrinsic bias ordering. In Equation~\ref{eq:Cl_limber} the amplitude scales as $b^2\,\phi^2/\chi^2$ and the slope reflects \emph{which} 3D wavenumbers are projected ($k\simeq \ell/\chi$) given the kernel. Cuts that compress the \ac{SFG} kernel to lower redshift (e.g.\ luminosity splits or high S/N survey selection functions) increase $\phi^2/\chi^2$ and shift the map toward larger $k$, yielding the observed “lift and flattening’’ of \ac{SFG} $w(\theta)$. \ac{AGN} kernels move much less under the same cuts, so the intrinsic hierarchy $b_{\rm AGN}>b_{\rm \ac{SFG}}$ remains intact. 

\medskip
\noindent
This explains the apparent discrepancy between the Magliocchetti and Hale measurements: once their selection functions are emulated, both behaviours arise naturally from projection effects with a fixed bias ordering. The cross–frequency consistency (1.4 vs 3\,GHz) further shows the effect is selection–driven rather than band–dependent. For multi-tracer $f_{\rm NL}$ applications, the lesson is operational: clustering inferences must be made with tracer–specific $\phi(z)$ that match the actual survey survey selection functions; otherwise slope and amplitude shifts can be misread as bias changes. Here we carry forward bias parameters together with the matched kernels to ensure the large–scale modes relevant for $f_{\rm NL}$ are interpreted consistently. The statistics of the \ac{ACF} are summarized in Table~\ref{tab:proj_numbers}.

\begin{table*}[t]
\centering
\small
\caption{Projection summary for \ac{AGN} and \ac{SFG} by selection. Errors are $1\sigma$ from power–law fits over $\theta\in[0.02^\circ,0.5^\circ]$. Effective depths use the Limber–like weights of Equation~(\ref{eq:zeff_chieff}). We quote the characteristic projected comoving separation $r_\perp(\theta)=\theta\,\chi_{\rm eff}$ at three angles.}
\label{tab:proj_numbers}
\begin{tabular}{l l c c c c c c}
\toprule
Cut & Group & $\gamma$ & $z_{\rm eff}$ & $\chi_{\rm eff}$ [Gpc] & $r_\perp(0.02^\circ)$ [Mpc] & $r_\perp(0.05^\circ)$ [Mpc] & $r_\perp(0.10^\circ)$ [Mpc] \\
\midrule
raw \textsc{GHOST} sample                      & \ac{AGN} & $1.886\pm0.028$ & 0.318 & 1.220 & 0.426 & 1.065 & 2.130 \\
                         & \ac{SFG} & $1.647\pm0.076$ & 0.602 & 2.117 & 0.739 & 1.847 & 3.695 \\
\addlinespace
Magliocchetti (1.4\,GHz) & \ac{AGN} & $1.858\pm0.041$ & 0.339 & 1.303 & 0.455 & 1.137 & 2.275 \\
                         & \ac{SFG} & $1.667\pm0.010$ & 0.094 & 0.406 & 0.142 & 0.354 & 0.708 \\
\addlinespace
Hale--like (3\,GHz)      & \ac{AGN} & $1.903\pm0.037$ & 0.313 & 1.208 & 0.422 & 1.054 & 2.108 \\
                         & \ac{SFG} & $1.662\pm0.010$ & 0.161 & 0.674 & 0.235 & 0.588 & 1.176 \\
\bottomrule
\end{tabular}
\end{table*}

\subsection{Angular power spectra and bias fits}
\label{sec:cl_bias}

\begin{figure*}[t]
  \centering
  \includegraphics[width=\textwidth]{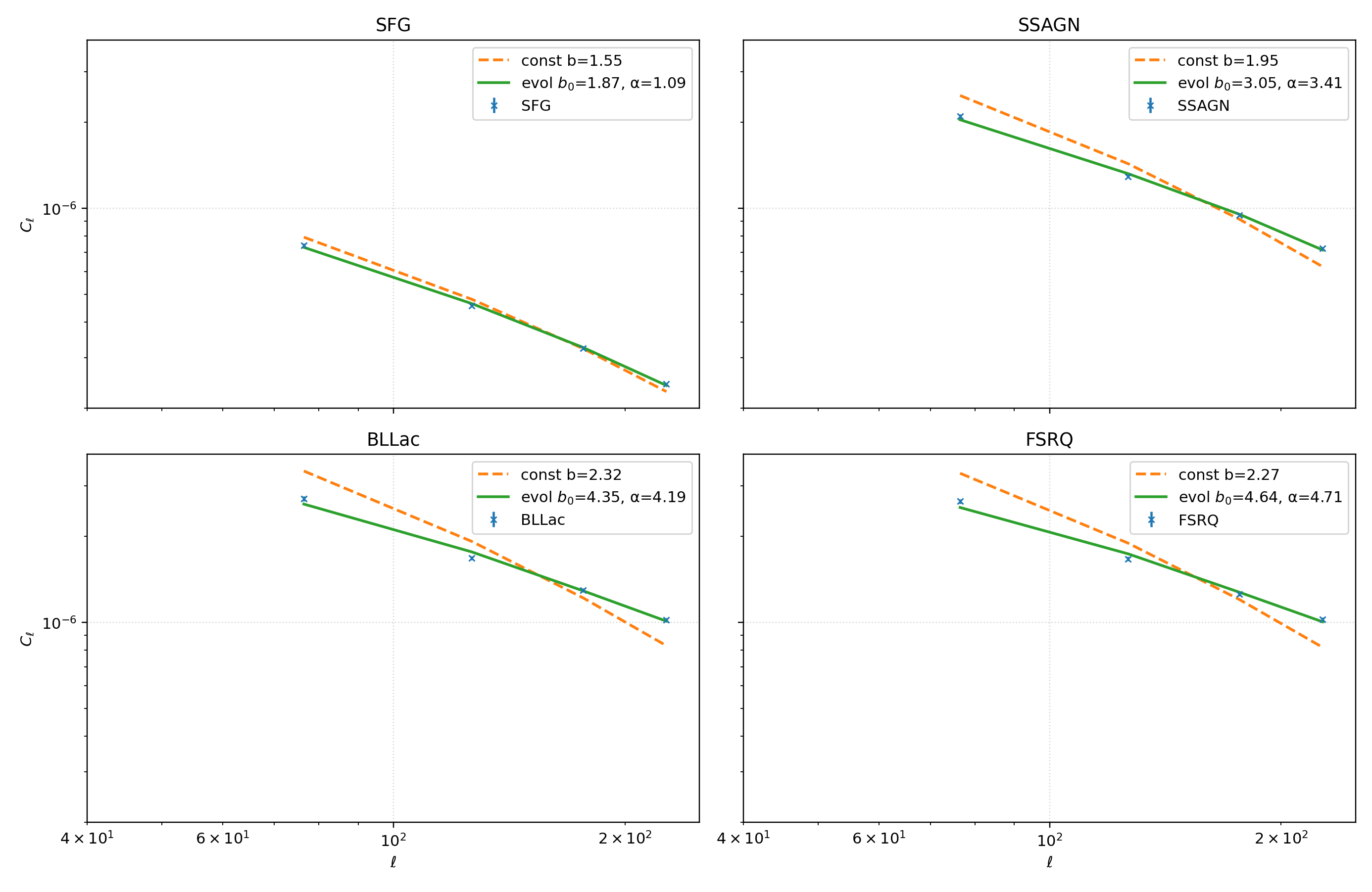}
  \caption{\textbf{$C_\ell$ bias fits by population.} Points show measured bandpowers; dashed lines are best–fit constant–bias models ($b=$ const) and solid lines are evolving models $b(z)=b_0(1+z)^\zeta$, both fit over $\ell\in[40,250]$. Numerical values are listed in Table~\ref{tab:cl_bias_fits}. The amplitude ordering \ac{SFG} $<$ \ac{SSAGN} $\lesssim$ FSRQ $\simeq$ \ac{BLLac} and the positive $\zeta$ for \ac{AGN} are consistent with the halo–mass distributions and the configuration–space results (Section~\ref{sec:clustering}).}
  \label{fig:cl_bias_grid}
\end{figure*}

We model the auto–spectra of each radio population $X\in\{\mathrm{SFG},\mathrm{SSAGN},\mathrm{BLLac},\mathrm{FSRQ}\}$ in the Limber regime using Equation~\ref{eq:Cl_limber}. We construct, for each tracer, a \emph{template} $T_\ell^X$ by evaluating equation~\ref{eq:Cl_limber} with a unit, redshift–independent bias (i.e.\ $b(z)\!\equiv\!1$). The measured spectrum in a given $\ell$–band then determines an overall amplitude related to the bias.

We fit two bias models over $\ell\in[\ell_{\min},\ell_{\max}]$ (default $40$–$250$), after sanitizing $n(z)$ (Section~\ref{sec:nz}) and optionally subtracting a shot component from the auto-correlations when available.

\subsubsection{Constant bias}
We assume $b_X(z)\!=\!b_X$ and fit an amplitude $a_X\!=\!b_X^2$ against the fixed template $T_\ell^X$.
Using (optionally) Knox weights $w_\ell=[2/((2\ell+1)f_\mathrm{sky})]/\sigma^2_\ell$ or per–band errors we solve the weighted least–squares problem
\begin{equation}
\widehat{a}_X \;=\; \frac{\sum_\ell w_\ell\,T_\ell^X\,C_\ell^{\mathrm{obs}}}{\sum_\ell w_\ell\,(T_\ell^X)^2}
,\qquad
\widehat{b}_X\;=\;\sqrt{\widehat{a}_X}\,,
\label{eq:const_bias_fit}
\end{equation}
with error propagation $\sigma^2_{a}\simeq 1/\!\sum_\ell w_\ell(T_\ell^X)^2$ and $\sigma_b\simeq \sigma_a/(2\,\widehat{b})$. Goodness of fit is summarized by $\chi^2=\sum_\ell w_\ell(C_\ell^{\rm obs}-\widehat{a}\,T_\ell)^2$.

\subsubsection{Evolving bias}
We test a minimal two–parameter form
\begin{equation}
b_X(z)=b_{0,X}\,(1+z)^{\zeta_X}.
\label{eq:evolving_bias}
\end{equation}
Within our pipeline this redshift dependence is folded into the \emph{shape} of the projection kernel by replacing $\phi_X(z)\!\to\!\phi_X(z)\,(1+z)^{\zeta_X}$ (and re–normalizing to unit area), while keeping the overall amplitude as $b_0^2$.
For each tracer we form the profile likelihood in $\zeta$ by
optimizing $b_0$ at fixed $\zeta$ using Equation~\ref{eq:const_bias_fit},
i.e. $\widehat b_0(\zeta)=\arg\min_{b_0}\chi^2(b_0,\zeta)$.
We then select $(\widehat b_0(\widehat\zeta),\widehat\zeta)$ with
$\widehat\zeta=\arg\min_\zeta \chi^2(\widehat b_0(\zeta),\zeta)$.

\subsubsection{Results}
Table~\ref{tab:cl_bias_fits} summarises the fits. In the constant model we find the expected hierarchy $b_\mathrm{SFG}<b_\mathrm{SSAGN}\lesssim b_\mathrm{FSRQ}\simeq b_\mathrm{BLLac}$.
The evolving model always yields a slightly lower $\chi^2$ statistic, indicating better fit, with $(\zeta,\ b_0)$ increasing from \ac{SFG} to radio–loud AGN, consistent with rarer, more strongly biased hosts at higher redshift.
Figure~\ref{fig:cl_bias_grid} shows the measured bandpowers with the best–fit constant (dashed) and evolving (solid) curves.

\subsubsection{Implications}
The bias ordering mirrors the halo–mass distributions inferred from the catalogue: \ac{SFG} peak at lower masses ($\sim 10^{12}M_\odot$) while \ac{AGN} subclasses peak at higher masses (few $\times 10^{12.5-13}M_\odot$), naturally producing larger large–scale clustering amplitudes. The mild preference for positive $\zeta$ in \ac{AGN} ($\zeta\sim3$–5) indicates bias growth with redshift, aligning with the picture of \ac{AGN} inhabiting rarer halos at earlier times. Overall, the $C_\ell$ fits and the configuration–space results (Section~\ref{sec:clustering}) are mutually consistent: amplitudes trace $b^2$ while the $w(\theta)$ slope variations are explained by projection effects rather than changes in the underlying bias hierarchy.
\begin{table*}[t]
\centering
\small
\caption{Bias fits from $C_\ell$ over $\ell\in[40\,,\,250]$. We fit a constant-bias model ($b=$ const) and an evolving form $b(z)=b_0(1+z)^\zeta$. Quoted $\chi^2$ values are for the best fit of each model. The last column gives the constant-bias ratio $b/b_{\rm \ac{SFG}}$.}
\label{tab:cl_bias_fits}
\begin{tabular}{l c c c c c c}
\toprule
Tracer & $b$ (const) & $\chi^2_{\rm const}$ & $b_0$ (evol) & $\zeta$ & $\chi^2_{\rm evol}$ & $b/b_{\rm \ac{SFG}}$ \\
\midrule
\ac{SFG}    & $ 1.55 \pm 0.04 $ & 0.7 & $ 1.87 \pm 0.05 $ & 1.09 & 0.0 & 1.00 \\
\ac{SSAGN}  & $ 1.95 \pm 0.06 $ & 4.2 & $ 3.05 \pm 0.09 $ & 3.41 & 0.1 & 1.26 \\
BLLac  & $ 2.32 \pm 0.07 $ & 8.1 & $ 4.35 \pm 0.12 $ & 4.19 & 0.2 & 1.50 \\
FSRQ   & $ 2.27 \pm 0.07 $ & 8.5 & $ 4.64 \pm 0.13 $ & 4.71 & 0.3 & 1.46 \\
\bottomrule
\end{tabular}
\end{table*}

\subsection{Host halo mass distributions}
\label{sec:mass_pdfs}

\begin{figure}[t]
  \centering
  \includegraphics[width=\columnwidth]{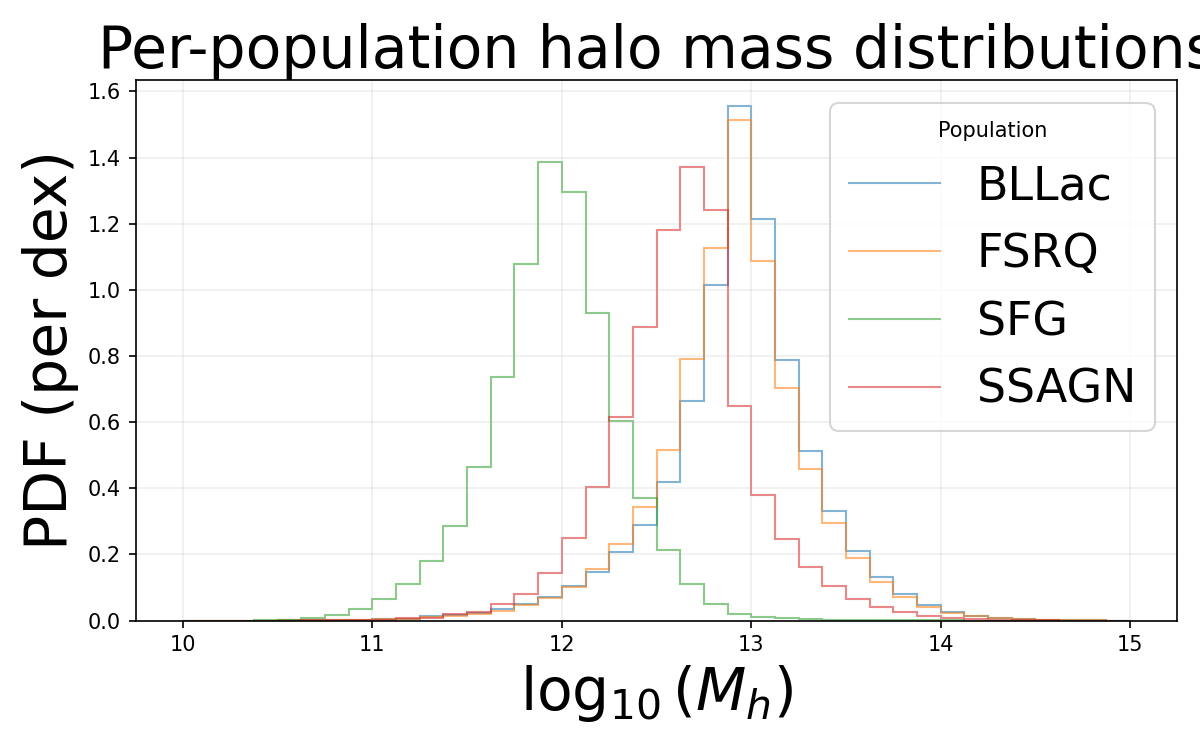}
  \caption{Host halo mass PDFs by population.
  Normalized distributions of $\log_{10}(M_h/M_\odot)$ (per dex) for \ac{SFG}, \ac{SSAGN}, BLLac, and \ac{FSRQ}.
  Distributions for the \ac{AGN} subclasses peak near $10^{12.8-13}\,M_\odot$, while \ac{SFG} peak near $10^{12}\,M_\odot$.
  The mass ordering maps onto the bias hierarchy measured in Section~\ref{sec:cl_bias}.}
  \label{fig:mass_hists}
\end{figure}

To connect clustering amplitudes with halo occupation, we examine the distribution of host halo masses $M_h$ for each population. Figure~\ref{fig:mass_hists} shows per–population PDFs of $\log_{10}(M_h/M_\odot)$ (normalized per dex). Table~\ref{tab:mass_summary} lists the median, the central 68\% interval, and the sample size.

The ordering is clear: \ac{SFG} inhabit lower–mass halos (median $\log_{10}M_h \simeq 11.97$), while all \ac{AGN} subclasses peak near $\sim\!10^{12.6-13.0}\,M_\odot$ (\ac{SSAGN} $\simeq12.65$, BLLac $\simeq12.95$, FSRQ $\simeq12.92$). The interquartile widths (IQR) are similar ( $\sim$0.63–0.68 dex ), indicating that the dominant difference is a shift in typical mass rather than a change in spread.

Because the large–scale bias $b(M,z)$ is a monotonic function of halo mass (e.g. in peak-background split models),
\begin{equation}
\begin{split}
b(M,z)\;=\;1+\frac{a\nu^2-1}{\delta_c}+\frac{2p}{\delta_c\,[1+(a\nu^2)^p]}\,,
\qquad\\
\nu\equiv \frac{\delta_c}{\sigma(M,z)} ,
\label{eq:bMz}
\end{split}
\end{equation}
(with $\delta_c\simeq1.686$ and $(a,p)\approx(0.707,0.3)$ as a common choice),
the $\sim$0.7–1.0 dex upward shift from \ac{SFG} to radio–loud \ac{AGN} naturally implies larger $b$ at the redshifts that dominate our projection. This is precisely what we measure from $C_\ell$ in Section~\ref{sec:cl_bias}: the constant-bias fits obey
$b_{\rm \ac{SFG}} < b_{\rm \ac{SSAGN}} \lesssim b_{\rm FSRQ} \simeq b_{\rm BLLac}$,
and the evolving model prefers positive $\zeta$ for AGN, consistent with rarer halos at earlier times.

Overall, the mass PDFs corroborate the bias/clustering hierarchy found in both configuration and harmonic space: \ac{AGN} populate more massive halos than \ac{SFG}, hence exhibit higher large-scale clustering amplitudes, while the small changes in $w(\theta)$ slope across selections are driven by projection (Section~\ref{sec:clustering}) rather than a reversal of the intrinsic bias–mass ordering.

\begin{table*}[t]
\centering
\small
\caption{Summary of host halo mass distributions per population. We report the sample size $N$, the median $\mathrm{med}[\log_{10}M_h/M_\odot]$, the central 68\% interval (16th--84th percentiles), the IQR width, and the median offset relative to \ac{SFG}.}
\label{tab:mass_summary}
\begin{tabular}{l r c c c c}
\toprule
Population & $N$ & $\mathrm{med}[\log_{10}M_h]$ & $[16,84]$ perc. & IQR width & $\Delta$ med vs \ac{SFG} \\
\midrule
\ac{SFG}   & 2184221793 & 11.97 & 11.64\,--\,12.27 & 0.63 & 0.00 \\
\ac{SSAGN} & 65986633 & 12.65 & 12.31\,--\,12.96 & 0.65 & 0.69 \\
BLLac & 20794032 & 12.95 & 12.59\,--\,13.27 & 0.68 & 0.8 \\
FSRQ  & 23240345 & 12.92 & 12.56\,--\,13.24 & 0.68 & 0.95 \\
\bottomrule
\end{tabular}
\end{table*}

\begin{figure}
    \centering
    \includegraphics[width=0.8\linewidth]{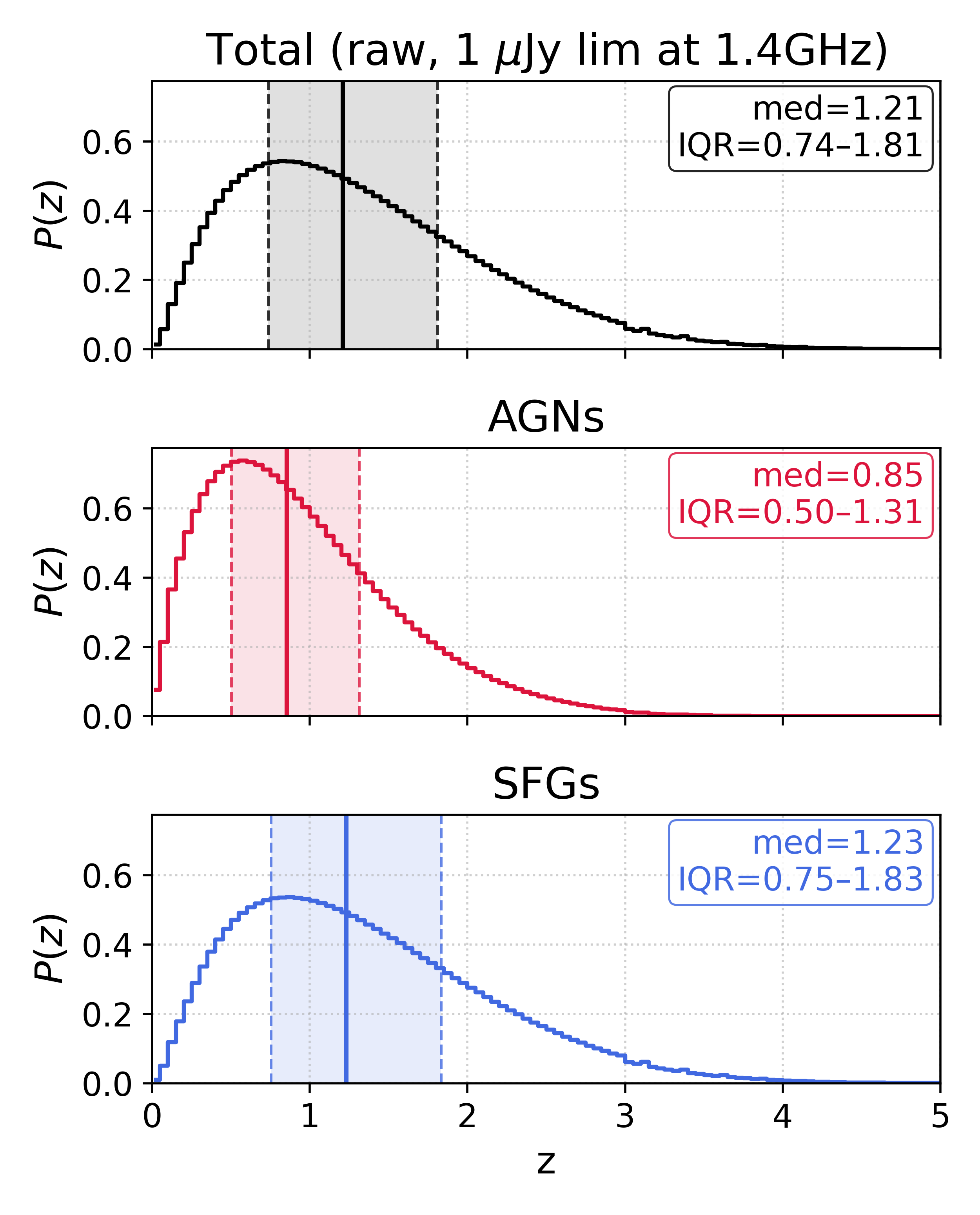}
    \caption{
    Normalized PDFs $P(z)$ (unit area) for the full \textsc{GHOST} catalogue (top) and by class
    (AGN, \ac{SFG}; middle/bottom). Vertical lines mark medians (solid) and IQRs (dashed).
    The total peaks near $z\!\sim\!0.8$ with a long high-$z$ tail, establishing the
    baseline cosmic distribution before survey selection.}
    \label{fig:nz}
\end{figure}

\subsection{Redshift distributions $n(z)$}\label{sec:nz}
Figure~\ref{fig:nz} shows normalized redshift distributions for the full
\textsc{ghost} catalogue and for \ac{AGN} and \ac{SFG} sub-samples (area under each curve
equals unity). Before any survey selection, the total population peaks near
$z\!\sim\!0.8$ with a long tail to $z>3$ (median $z=1.21$, IQR $0.74$--$1.81$).
The \ac{AGN} distribution is narrower (median $z=0.85$, IQR $0.50$--$1.31$), while \ac{SFG}
peak at slightly higher redshift (median $z=1.23$, IQR $0.75$--$1.83$) and
maintain the broader high-$z$ tail. This is consistent with radio-loud AGN
dominating the bright counts at low $z$ and \ac{SFG} contributing increasingly
toward the sub-mJy regime.

At our pre-selection stage with the raw \textsc{GHOST} catalogue, the $n(z)$ for \ac{SFG} and \ac{AGN} look consistent with the behavior from deep $\mu$Jy radio surveys: \ac{SFG} dominate at faint fluxes and exhibit a broad distribution that extends to $z\approx4$, while \ac{AGN} show a more concentrated, lower z peak. Similar behavior is seen in the \ac{LoTSS} Deep fields \citep{2023MNRAS.523.1729B,2013MNRAS.436.3759B,2017A&A...602A...2S}. Disagreements that emerge once survey
cuts are applied should therefore be read chiefly as mixture/evolution issues rather than a failure of the overall redshift density.

Our \ac{SFG} $p(z)$ shape---a broad peak around $z\!\sim\!1$ with a long, shallow high-$z$ tail---is qualitatively consistent with the component in empirical \ac{SKAO} telescope forecasting work (e.g. \cite{Mancuso}, their Fig.\,10). Two differences help explain why our curves are smoother and show no pronounced `bump' near $z\!\sim\!1$ at $\sim\!\mu$Jy depths:
\begin{enumerate}
  \item \emph{Population partitioning.} In those forecasts, star-forming
  galaxies are split into at least two \ac{SFG} families with distinct evolution: ``late-type'' discs (normal+starburst) that dominate at $z\!\lesssim\!1$--1.5, and high-$z$ dusty star-forming systems (often tagged as ``proto-spheroids''). The sum of two or more evolving families can imprint a mild feature near the transition redshift \citep{Mancuso}. In \textsc{ghost} we treat \ac{SFG} as a single, smoothly evolving family, which naturally washes out such structure.
    \item \emph{How dust enters.} Although radio SFR tracers are intrinsically insensitive to dust attenuation, many forecasting frameworks first construct an SFR function from UV, H$\alpha$, and IR data (the IR term capturing the obscured star formation) and then map SFR$\!\to\!L_{\nu}$ via radio–SFR calibrations \citep{2014ARA&A..52..415M, Bonato2017, Mancuso}. In such models, a distinctly evolving, dust–rich \ac{SFG} population that grows rapidly near the cosmic SFR peak ($z\!\sim\!1$--2) can imprint mild features in $n(z)$ at fixed flux (e.g. late–type versus proto–spheroidal contributions) \citep{Mancuso}. By contrast, our implementation fits the \ac{SFG} radio luminosity function directly at 1.4\,GHz with smooth redshift evolution, so a given flux limit projects to a correspondingly smooth redshift distribution.
\end{enumerate}


\subsubsection{Redshift distributions $n(z)$: rationale for the comparisons}
Our goal in this Section is verification of the \textsc{GHOST} mock catalogue. There is no all–sky radio survey to $\sim\!1~\mu$Jy, so we cannot validate the pre–selection $n(z)$ directly on the full sky. Instead, we forward–apply the selection functions of existing surveys (flux/SNR thresholds, peak–vs–integrated survey selection functions, band-passes) to the same underlying catalogue and compare the resulting $n(z)$ to the corresponding data products. This matters because $n(z)$ sets the projection kernels $\phi(z)$ that weight our angular statistics and biases (e.g.\ the Limber integrals for $C_\ell$ and the response to the $k^{-2}$ \ac{PNG} term); counts alone cannot diagnose class mixture or redshift tails. Agreement across multiple, independent selections implies that the global LF+SED calibration produces a plausible cosmic history for each class, and our survey emulation is realistic. Conversely, coherent offsets across selections isolate population–level issues (e.g.\ \ac{SFG}/\ac{AGN} mix, high–$z$ tails, spectral/K–correction choices) rather than artefacts of any one data set. The comparisons below are therefore tests designed to stress our population model and selection operators in regimes where data exist, not predictions for a hypothetical all–sky $\mu$Jy survey.

\subsubsection{LoTSS-like selection ($S_{150}\ge1.5$\,mJy)}
\label{sec:nz_lotss}

\begin{figure}[t]
  \centering
  \includegraphics[width=0.8\linewidth]{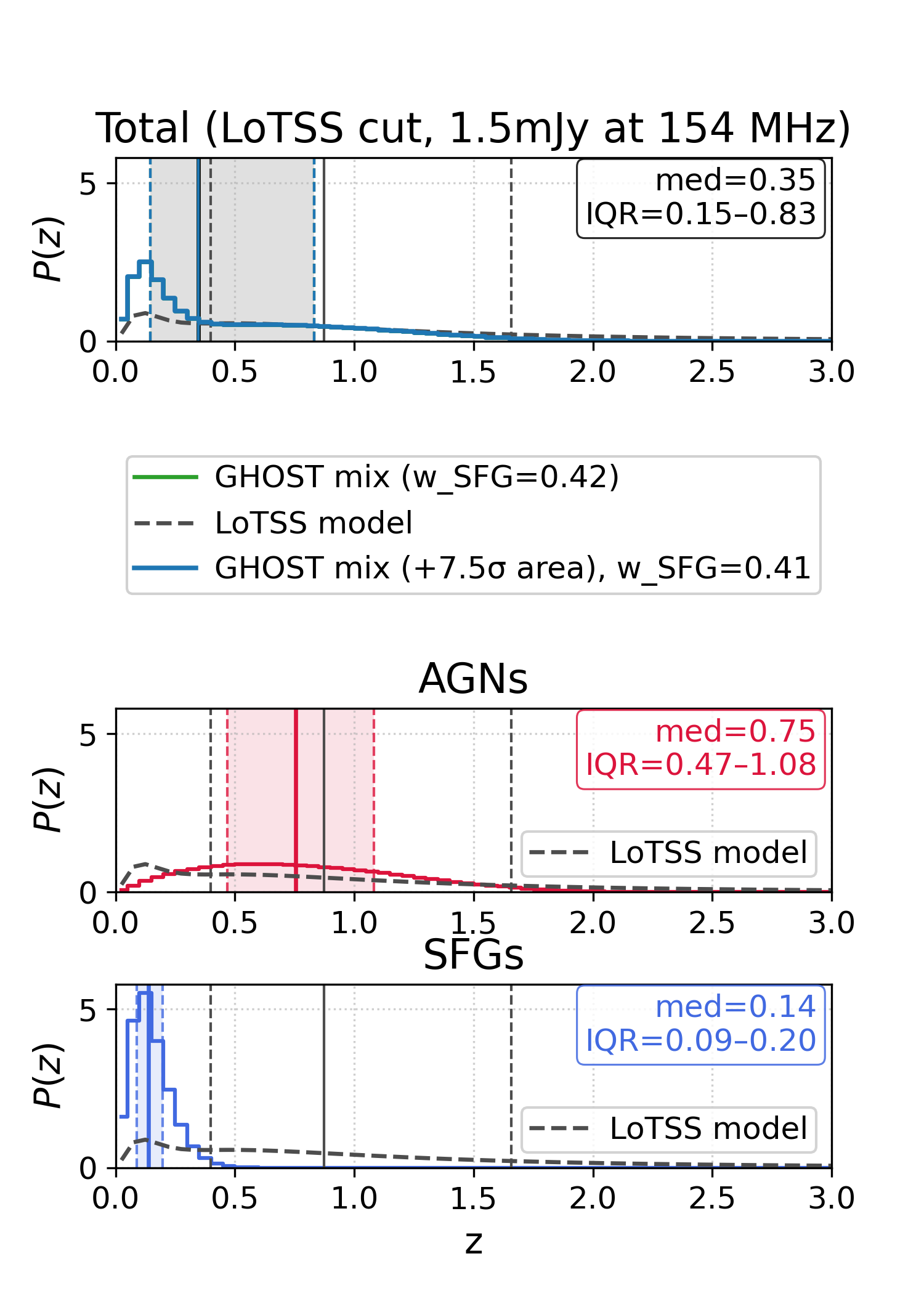}
\caption{Redshift distributions after applying the flux threshold $S_{150}\!\ge\!1.5\,\mathrm{mJy}$ (Section~\ref{sec:validation-selections}).
Top: total $P(z)$ from \textsc{ghost} (black). The green curve reconstructs the total as
$w_{\rm \ac{SFG}}\,P_{\rm \ac{SFG}}(z)+w_{\rm AGN}\,P_{\rm AGN}(z)$ using the intrinsic flux-limited class
fractions $(w_{\rm \ac{SFG}},w_{\rm AGN})$; the blue curve uses the same $P_{\rm \ac{SFG}}(z),P_{\rm AGN}(z)$ but
reweights the class fractions by the LoTSS peak–S/N$\ge 7.5$ area acceptance from a log-normal rms model
($\tilde{\sigma}=83~\mu\mathrm{Jy\,beam^{-1}}$, 95th percentile $171~\mu\mathrm{Jy\,beam^{-1}}$).
The dashed line shows the LoTSS-DR2 parametric model. The green and blue reconstructions are
essentially identical, indicating that the peak–S/N cut only weakly perturbs the class mix at 1.5\,mJy.
Middle/bottom: \ac{AGN} and \ac{SFG} PDFs (unit area) with medians and IQRs indicated. Relative to LoTSS,
the simulated total places excess weight at $z\!\lesssim\!0.4$ and underpowers the $z\!\gtrsim\!0.6$ shoulder,
suggesting a slightly \ac{SFG}-heavy mix.}

  \label{fig:nz_lotss15}
\end{figure}

We emulate the LoTSS\,DR2 \citep{2024MNRAS.527.6540H} wide-area selection with an integrated-flux threshold
$S_\mathrm{150MHz}\ge1.5$\,mJy and a peak-S/N threshold, as described in Section \ref{sec:lotssselection}. After replicating the $1.5$\,mJy sensitivity the selection function removes only a few percent of sources, so the resulting $n(z)$ is driven by the intrinsic class mixture, shown in Figure \ref{fig:nz_lotss15}. Where shown, the “LoTSS model” is the parametric $p(z)$ derived from Deep Fields. The total $p(z)$ is low-$z$ dominated (median $z=0.35$, IQR $0.15$--$0.83$), with a conspicuous \ac{SFG} contribution at $z\lesssim0.3$ (\ac{SFG} median $z=0.14$, IQR $0.09$--$0.20$). The \ac{AGN} component peaks broadly around $z\sim0.5$--$1$ (median $z=0.75$, IQR $0.47$--$1.08$). Using the measured 150\,MHz counts-by-population to set the mixture yields $w_{\rm \ac{SFG}}\approx0.42$. Figure~\ref{fig:nz_lotss15} compares the resulting PDFs to the LoTSS model. Relative to the LoTSS model, the total median is lower and the high-$z$ shoulder is underpowered, indicating a \ac{SFG}-heavy mix at the mJy level and an underrepresentation of moderate-$z$ \ac{AGN} in the total.

\medskip
\noindent
Increasing the flux of the cut shifts the prediction as expected (more AGN-dominated, higher median $z$), yet even at $10$\,mJy the median remains below the LoTSS curve (Table~\ref{tab:smin_scan}). The direction and size of the offset are consistent with the RLF diagnostics (Section Section \ref{sec:rlf}; possible population-level origins are deferred to sections \ref{sec:nz}.

\begin{table}[h]
\centering
\setlength{\tabcolsep}{6pt}
\begin{tabular}{rcc}
\hline
$S_{\min}$ (mJy) & $w_{\rm \ac{SFG}}$ & median $z$ \\
\hline
1.5  & 0.41 & 0.35 \\
2.0  & 0.37 & 0.41 \\
3.0  & 0.29 & 0.51 \\
5.0  & 0.22 & 0.59 \\
10.0 & 0.14 & 0.66 \\
\hline
\end{tabular}
\caption{Effect of increasing the 150\,MHz flux threshold on the mixture weight
($w_{\rm \ac{SFG}}$) and the total-sample median redshift.}
\label{tab:smin_scan}
\end{table}
\subsubsection{VLA-COSMOS split at 1.4\,GHz \citep{2017MNRAS.464.3271M}}
\label{sec:nz_mag17}
\begin{figure}[t]
  \centering
  \includegraphics[width=0.8\linewidth]{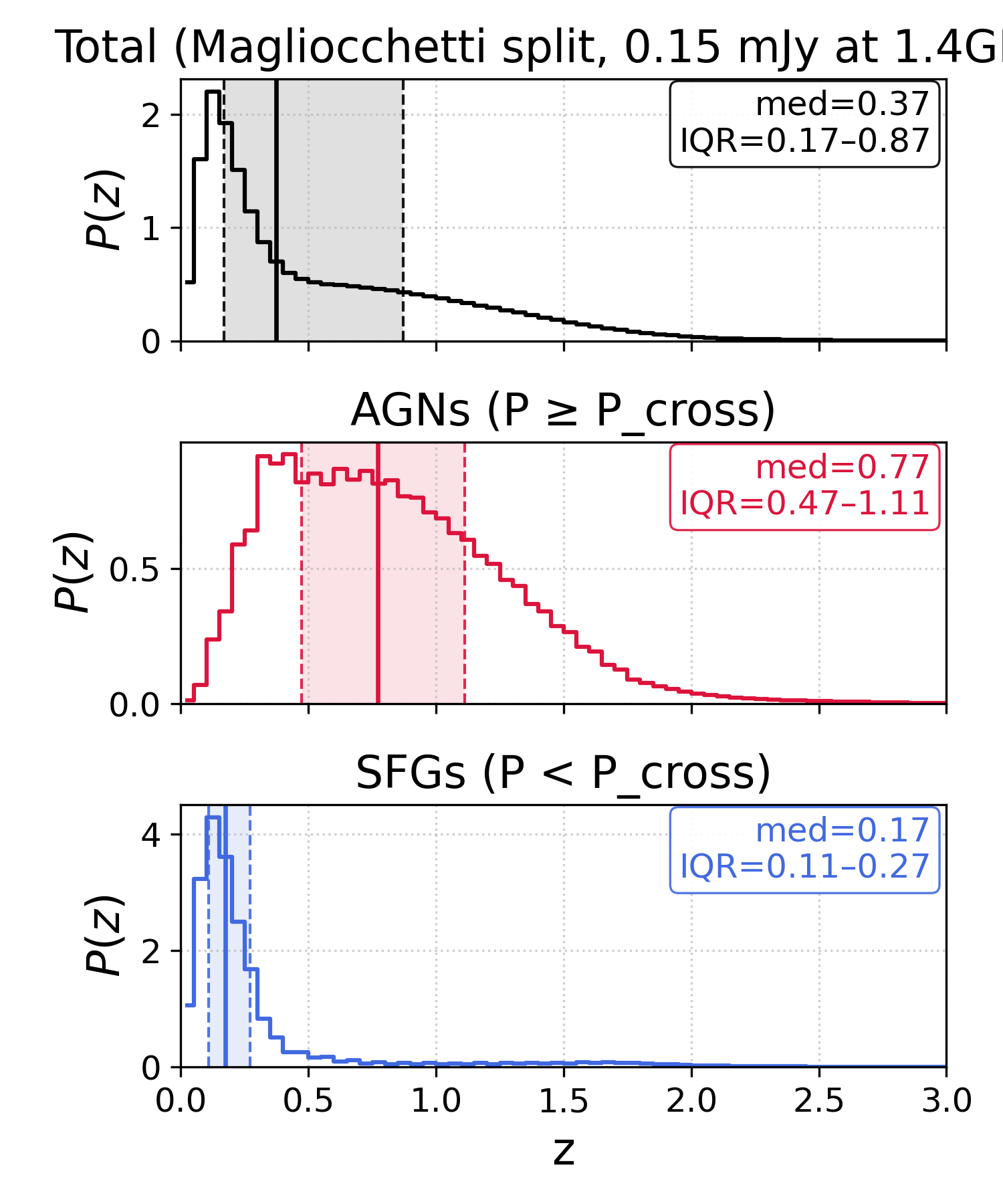}
    \caption{VLA-COSMOS-like $n(z)$ with 1.4~G\,Hz selection.
    We apply $F_{1.4}\!\ge\!0.15$\,mJy and classify sources via
    $L_{1.4}\gtrless L_{\rm cross}(z)$ with $L_{\rm cross}=4\pi P_{\rm cross}$ and
    $\log_{10}P_{\rm cross}(z)=21.7+z$ ($z\!\le\!1.8$) or $23.5$ otherwise (Section~\ref{sec:maglioccettiselection}).
    Top: total $P(z)$ with median and IQR; middle/bottom: AGN-like and \ac{SFG}-like PDFs.
    The expected qualitative  in redshift - with \ac{SFG} tightly clustered and \ac{AGN} more broad - is recovered, but the integrated \ac{SFG}-like
    fraction is high, again pointing to a slightly \ac{SFG}-heavy mix at the mJy level.}
  \label{fig:nz_mag_split}
\end{figure}
To mirror VLA--COSMOS \citep{2017MNRAS.464.3271M}, we apply a flux cut $F_{1.4}\ge0.15$\,mJy and partition
sources using a redshift-dependent luminosity boundary, as detailed in Section~\ref{sec:validation-selections}.

The total resulting distribution, as shown in Figure \ref{fig:nz_mag_split} remains low-$z$ dominated (median $z=0.37$, IQR
$0.17$--$0.87$). The AGN-like PDF peaks at $z\sim0.6$--$1$ with a tail to
$z\sim2$, while the \ac{SFG}-like PDF is tightly concentrated at $z\lesssim0.3$.
Integrating above/below $L_{\rm cross}(z)$ gives an \ac{SFG} fraction for this selection of
\begin{equation}
f_{\rm \ac{SFG}} \equiv \frac{N_{\rm \ac{SFG}}}{N_{\rm \ac{SFG}}+N_{\rm AGN}} \;=\; 0.49 .
\end{equation}

The selection function split recovers the expected qualitative separation in $n(z)$ prevalent in \cite{2017MNRAS.464.3271M}, with \ac{SFG} tightly clustered at low redshift and a broader \ac{AGN} distribution, but the integrated \ac{SFG} share is high for this selection, again pointing to a \ac{SFG}-heavy mix around the mJy level. Because this diagnostic does not depend on peak-vs-integrated flux (unlike LoTSS), it isolates the population modelling: too many low-$z$ \ac{SFG} and/or too few moderate-$z$ radio-loud AGN.

Relative to the COSMOS-based fractions of \citep{2017MNRAS.464.3271M}, our $f_{\rm \ac{SFG}}$ is larger in the same sense indicated by the LoTSS test. A hard $L$-threshold is an imperfect classifier (e.g.\ it can move luminous \ac{SFG} at $z\sim1$ into the AGN-like bin), so exact numerical agreement is not expected; the direction of the difference is robust and consistent across
selections.

\subsubsection{VLA-COSMOS-like selection at 3\,GHz \citep{Hale2018}}
\label{sec:nz_hale}
\begin{figure}[t]
  \centering
  \includegraphics[width=0.8\linewidth]{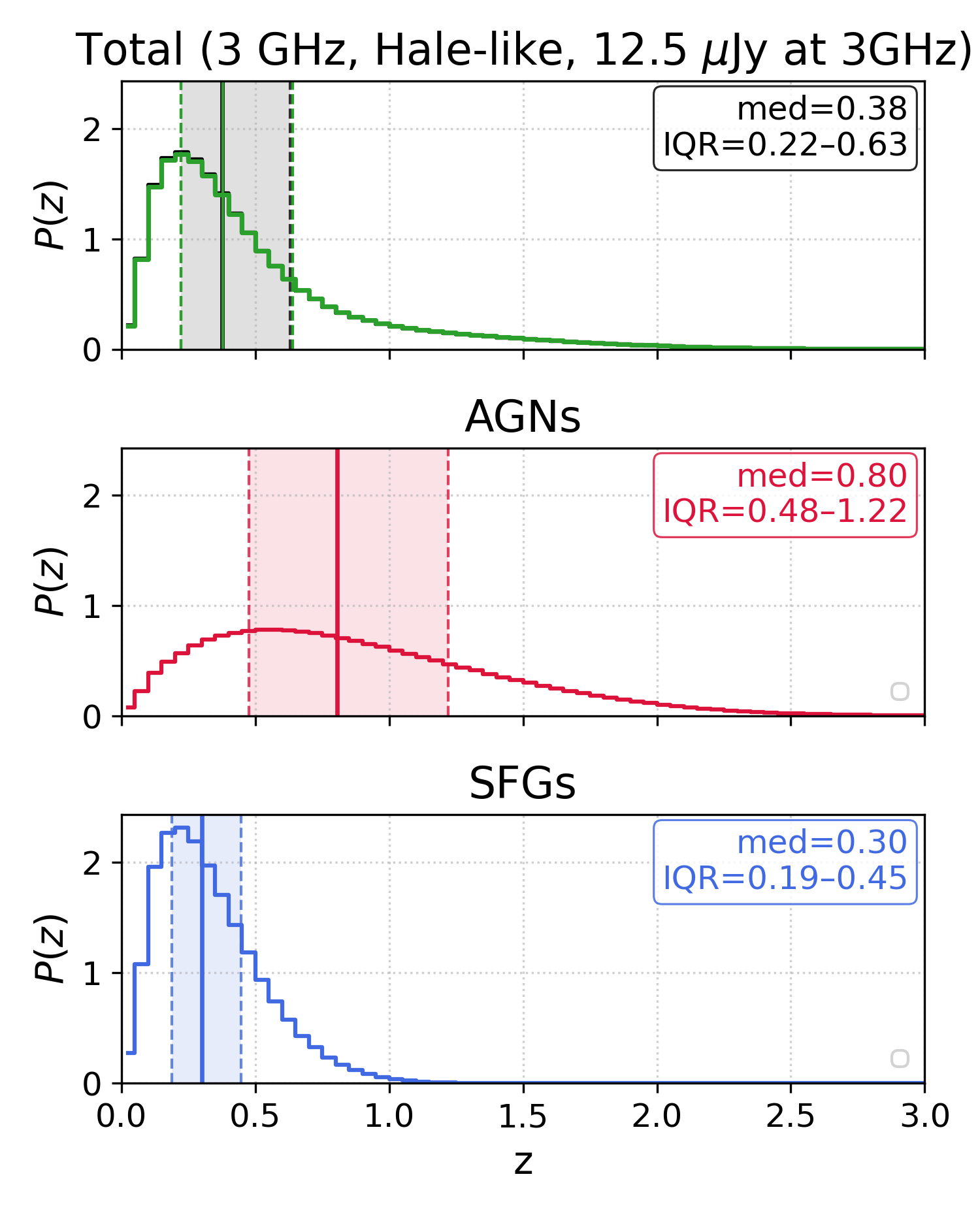}
    \caption{VLA–COSMOS-like $n(z)$ at 3\,GHz.
    GHOST total (top) and class PDFs (middle/bottom) for a
    $\mathrm{S/N}\!\ge\!5.5$ selection function (Section~\ref{sec:haleselection}); the top panel also shows the effective
    counts-weighted mixture (green). Medians and IQRs are indicated.
    The total median is low and the \ac{SFG} fraction is high compared to \cite{Hale2018},
    consistent with the trend that \textsc{ghost} overweights nearby \ac{SFG} relative
    to moderate-$z$ radio-loud AGN.}
  \label{fig:nz_hale3}
\end{figure}
We emulate the $\mathrm{S/N}\ge5.5$ 3\,GHz selection (median rms
$\simeq2.3\,\mu$Jy\,beam$^{-1}$) and compare to the multi-wavelength
SFG/\ac{AGN} classes of \cite{Hale2018}.
At the \cite{Hale2018} depth the applied flux threshold and selection function detailed in Section~\ref{sec:validation-selections} is weak: numerically the acceptances are
 close to unity for both classes (in our run ${\rm acc}_{\rm \ac{SFG}}\!\approx\!0.89$,
${\rm acc}_{\rm AGN}\!\approx\!0.94$), so the predicted $n(z)$ is governed mainly
by the intrinsic \ac{SFG}/\ac{AGN} mix. The total has median $z=0.38$. Class medians are $z_{\rm med}^{\rm \ac{SFG}}=0.30$ (IQR $0.19$--$0.45$) and $z_{\rm med}^{\rm AGN}=0.80$ (IQR $0.48$--$1.22$). The effective \ac{SFG} fraction is $w_{\rm \ac{SFG}}=0.69$.

At this depth and frequency the selection is intrinsically \ac{SFG}-rich, but the
simulated mixture is still somewhat \ac{SFG}-heavy relative to \cite{Hale2018}, again
implicating the population balance rather than the survey selection function. \cite{Hale2018} classify $\sim$79\% of sources via multi-wavelength diagnostics and
report a \ac{SFG} share of $\simeq3704/ 3704+2937)\approx0.56$ in the masked field, together with a higher total median $z$ (see their Fig.\,3). Our Hale-like selection therefore echoes the LoTSS and Magliocchetti results: \textsc{ghost} places relatively more weight in nearby \ac{SFG} and underweights moderate-$z$ radio-loud AGN. The sign and scale of the difference are consistent with the
\ac{RLF} and clustering comparisons.

\begin{figure*}[p]
    \centering
    \includegraphics[width=\textwidth,height=0.92\textheight,keepaspectratio]{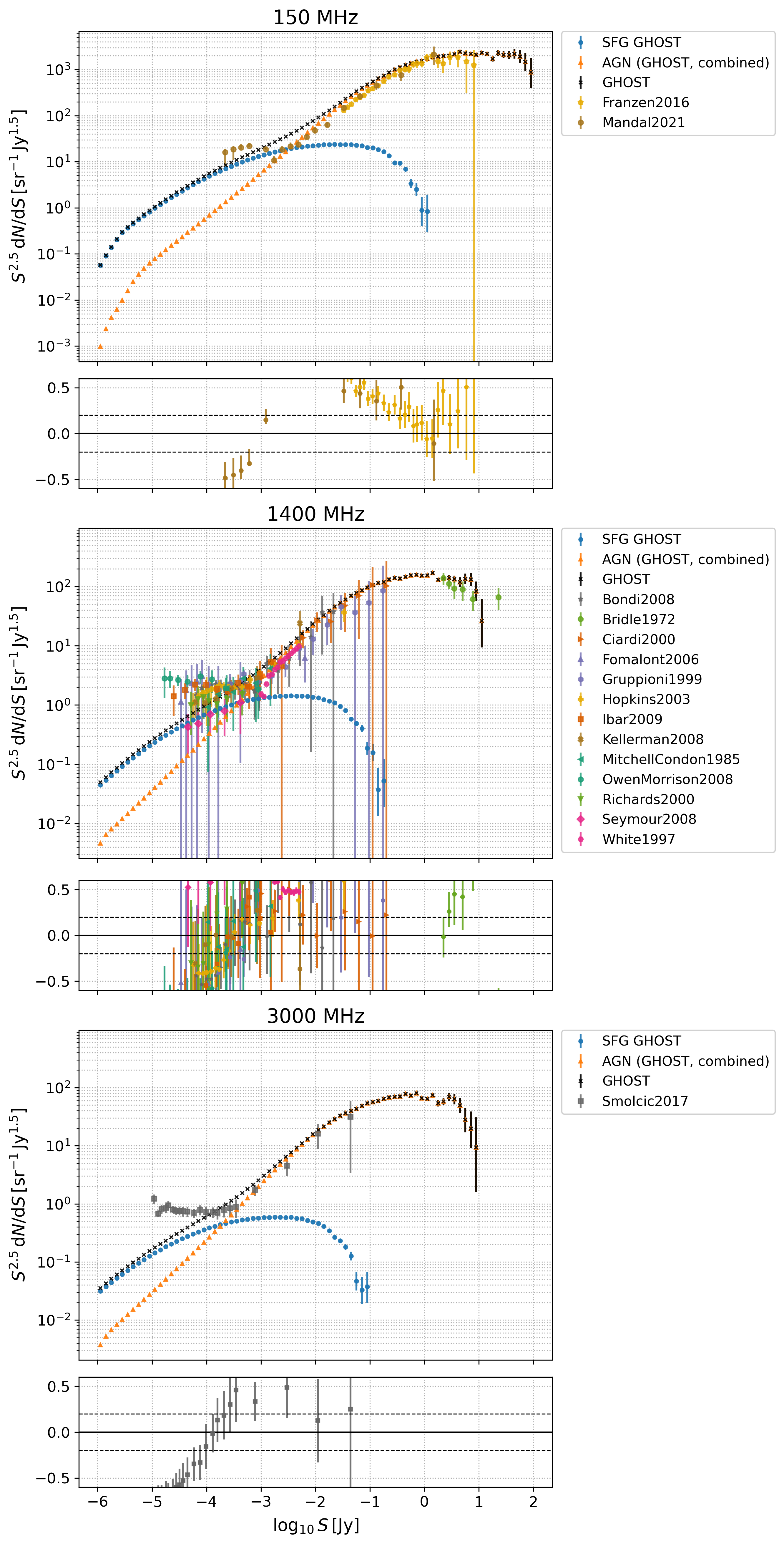}
    \caption{Euclidean-normalized differential counts, $S^{2.5}\,\mathrm{d}N/\mathrm{d}S$,
    for \textsc{ghost} (black markers) compared to measurements at
    150\,MHz \citep{Franzen,Mandal2021}, 1.4\,GHz \citep{Bondi2008,Bridle1972,Ciardi2000,Fomalont2006,Gruppioni1999,Hopkins2003,Ibar2009,Kellermann2008,MitchellCondon1985,OwenMorrison2008,Richards2000,Seymour2008,White1997,deZotti2010} and 3\ GHz \citep{2017A&A...602A...2S}. Error bars are $1\sigma$ measurement
    uncertainties, incorporating the estimations of \cite{Gehrels} if $N<20$ for any bin; shaded bands (if shown) indicate the field-to-field variance from
    \textsc{ghost} mock realizations with the same area and flux cuts as the data.
    Residual panels show $(\text{model}-\text{data})/\text{data}$.}
    \label{fig:counts}
\end{figure*}

\subsection{Differential and Euclidean-normalized number counts}
\label{sec:counts}

This section tests the total source density in \textsc{GHOST} against the literature via Euclidean-normalized differential counts, $S^{2.5}\,dN/dS$, at 150\,MHz, 1.4\,GHz, and 3\,GHz. Because number counts integrate over luminosity and redshift, they primarily probe the combination of luminosity-function normalisations and radio SEDs, largely independent of the host–halo assignment used for clustering.

Figure~\ref{fig:counts} shows that over nearly seven decades in flux density the simulation follows the observed trends at all three frequencies. At the bright end ($S\gtrsim10$ mJy), the counts are close to Euclidean ($S^{2.5}\,dN/dS\!\approx\!\mathrm{const}$), consistent with a population dominated by radio-loud AGN.

A tension appears around the classical sub-mJy upturn, where the data rise due to the emergence of \ac{SFG}. In our model this rise is weaker at $S\sim0.1$–$0.3$ mJy. This behaviour is fully consistent with the adopted \ac{SFG} RLF: the radio-first, smoothly evolving \ac{SFG} luminosity function underpredicts the luminous tail of \ac{SFG} at $z\gtrsim1$, which suppresses the upturn when projected to flux. Since total counts depend on the RLFs and SEDs rather than on host assignment, abundance matching does not drive this feature.

\subsubsection{Residuals}
The residual panels plot $(\mathrm{model}/\mathrm{data})-1$ at the locations of the observational points. Away from the very faint and very bright limits, the residual clouds at 150\,MHz, 1.4\,GHz, and 3\,GHz are centred on zero within the quoted errors, with a typical scatter of $\pm(10\text{--}30)\,\%$ across $\sim4$ dex in $S$, indicating no frequency-dependent bias in the combination of RLFs and SEDs. Where residuals dip near the sub-mJy regime (most visibly at 1.4\,GHz, with corresponding ranges at 150\,MHz and 3\,GHz), the shortfall follows from the \ac{SFG} RLF’s high-$L$ tail rather than from halo scarcity or assignment choices. At the brightest fluxes, excursions of order $\gtrsim30\%$ reflect the rarity of luminous \ac{AGN} and the small effective survey volumes of the anchoring datasets.

\subsubsection{Implications}
The close match to multi-frequency counts implies that the global normalisations of the luminosity functions, the adopted radio SEDs, and the applied survey selections are mutually consistent: \textsc{GHOST} produces the correct number of sources as a function of flux density without frequency-dependent bias, apart from the discussed \ac{SFG} underprediction at the sub-mJy upturn. Because counts constrain the sum over populations, different \ac{SFG}/\ac{AGN} partitions can yield similar totals; discrepancies seen later in $n(z)$, the RLFs, and clustering should therefore be interpreted as redistributions across populations and redshifts rather than a failure in the overall source density. The 3\,GHz panel of Figure~\ref{fig:counts} shows the same behaviour, indicating that the spectral extrapolation from 1.4\,GHz to 3\,GHz is consistent at the precision of current counts.

\subsection{\ac{RLF}}
\label{sec:rlf}

\begin{figure*}[p]
  \centering
  \includegraphics[width=\textwidth,height=0.80\textheight,keepaspectratio]{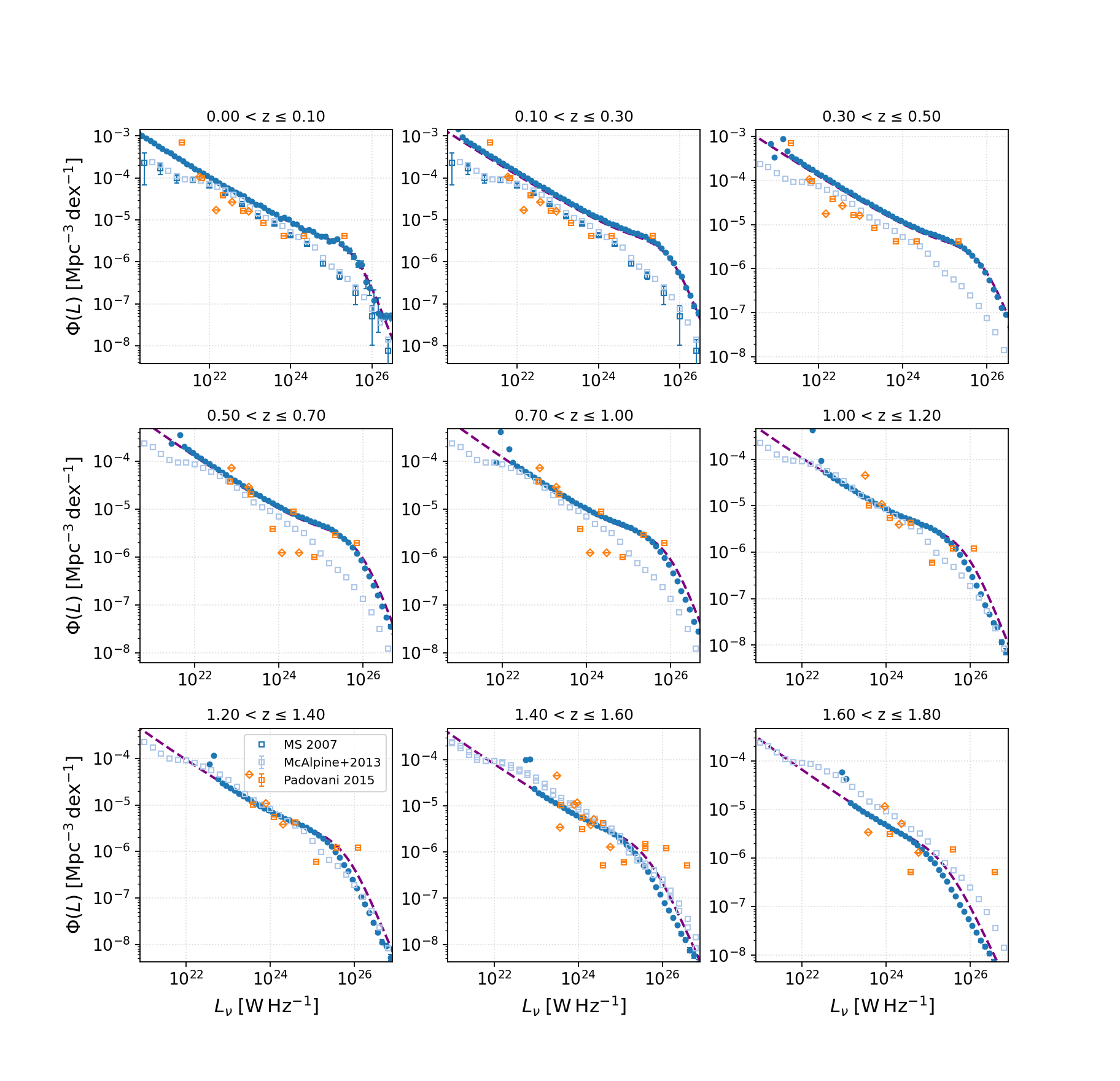}
\caption{\textbf{\ac{AGN} radio luminosity functions at 1.4\,GHz in redshift slices (raw cut).}
Each panel shows $\Phi(L_\nu)\,[\mathrm{Mpc}^{-3}\,\mathrm{dex}^{-1}]$ for the stated redshift bin.
\emph{Blue circles}: RLF measured from the simulation using the effective–volume method,
$\Phi = N/(V_{\rm eff}\,\Delta\log_{10}L)$, where $V_{\rm eff}$ accounts for the flux limit within the bin
(\S\ref{sec:rlf}; Equation~\ref{eq:phi_veff}). Error bars are $1\sigma$ Poisson uncertainties propagated through
$V_{\rm eff}$; cells with $V_{\rm eff}=0$ are omitted.
\emph{Purple dashed}: forward (projected) model $\Phi^{\rm proj}(L|z_1\!:\!z_2)$ obtained by averaging the intrinsic
\ac{AGN} RLF over the finite $\Delta z$ of the panel under the same flux selection (Equation~\ref{eq:phi_projected}).
\emph{Points in the background}: literature measurements at 1.4\,GHz (legend in the top–left panel).
Across the ten slices the simulation (blue) and the forward model (purple) show excellent agreement in both
normalisation and shape; a mild feature near $L_\nu\!\sim\!10^{26}\,\mathrm{W\,Hz^{-1}}$ reflects the combined
contribution of flat–spectrum (FSRQ/BLLac) and steep–spectrum (SSAGN) components with luminosity–dependent
evolution (see \S\ref{sec:rlf}). Axis limits in each panel are set by the extent of the observational points to aid readability: Squares indicate RL AGN; diamonds indicate RQ AGN. Filled markers are fully within the z-bin; open markers partially overlap the bin.}

  \label{fig:agn_rlf_panels}
\end{figure*}

\begin{figure*}[p]
  \centering
  \includegraphics[width=\textwidth,height=0.8\textheight,keepaspectratio]{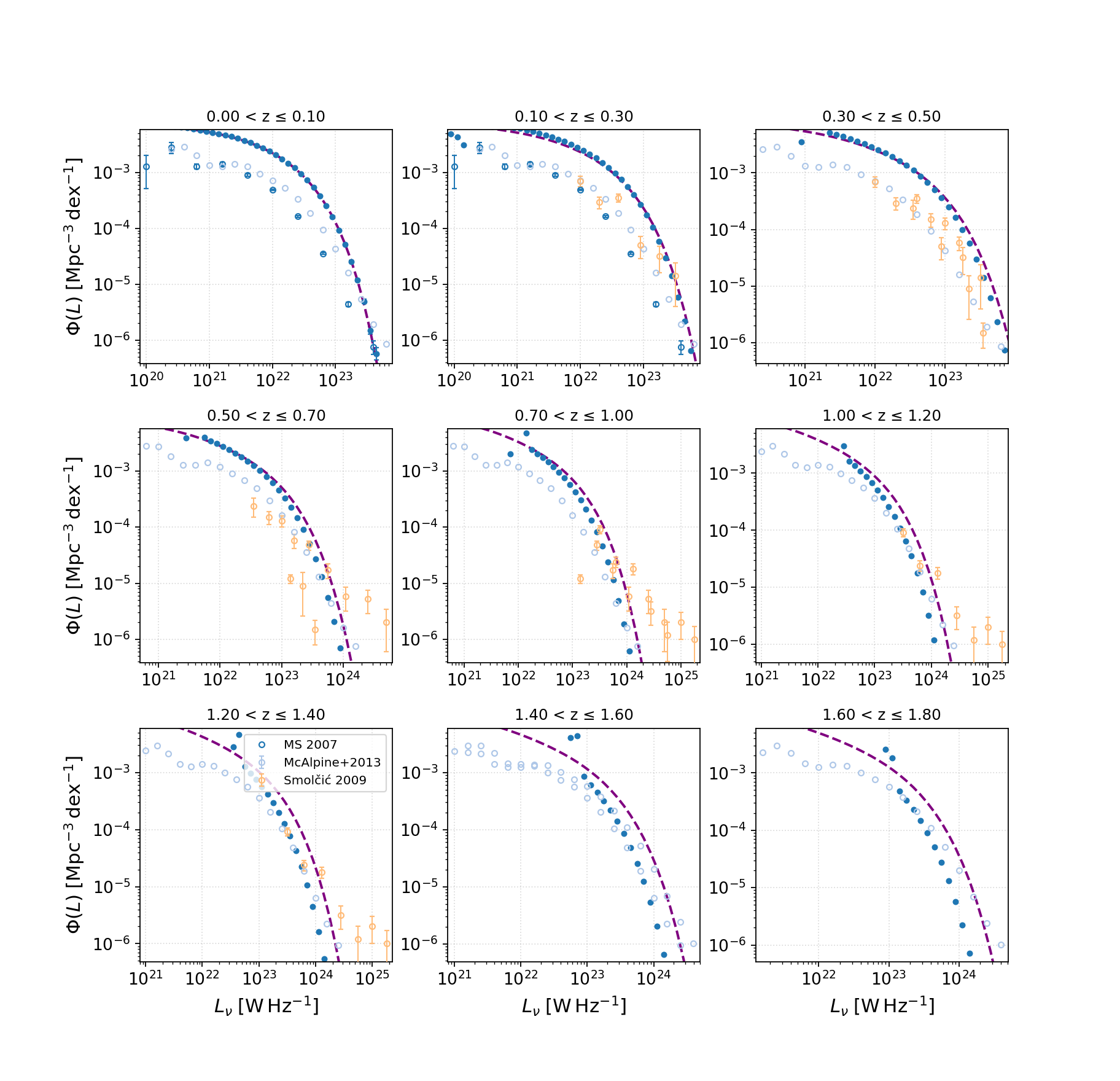}
\caption{\textbf{\ac{SFG} radio luminosity functions at 1.4\,GHz in redshift slices (raw cut).}
As in the \ac{AGN} figure, \emph{blue circles} show the simulation RLF recovered with the effective–volume estimator
$\Phi=N/(V_{\rm eff}\,\Delta\log_{10}L)$ (with $1\sigma$ Poisson errors), and \emph{orange dashed} curves are the
forward (projected) model evaluated over the same $\Delta z$ and flux selection.
\emph{Background points}: literature measurements at 1.4\,GHz (legend in the top–left panel).
At $z\!\lesssim\!1$ the model and simulation track the observed faint end well. 
At higher redshift the bright tail in the simulation/model sits below some compilations:
this follows naturally from our adopted $L$–SFR mapping (shallower high–SFR slope), the tightening luminosity
threshold imposed by the flux limit within each $\Delta z$ slice, and the resulting reduction of accessible
high–SFR systems (see \S\ref{sec:rlf}). As before, panel limits are set by the extent of the observational points; filled markers are fully within the z-bin; open markers partially overlap the bin.
}

  \label{fig:sfg_rlf_panels}
\end{figure*}

We describe the class–dependent radio luminosity function (RLF) by
$\Phi_c(L_\nu,z)$ in units of $\mathrm{Mpc}^{-3}\,\mathrm{dex}^{-1}$, where
$c\in\{\mathrm{SFG},\mathrm{SSAGN},\mathrm{FSRQ},\mathrm{BLLac}\}$.
Observed flux density at $\nu_{\rm obs}$ is related to rest–frame luminosity
through the standard $K$–correction,
\begin{equation}
  L_\nu(z,\alpha_c) = 4\pi D_L^2(z)\,S_{\nu_{\rm obs}}\,
  \Big(\frac{\nu_{\rm ref}}{\nu_{\rm obs}}\Big)^{-\alpha_c}(1+z)^{1+\alpha_c},
  \label{eq:L_of_S}
\end{equation}
so that, for a survey flux limit $S_{\rm lim}$, the luminosity boundary within
a redshift slice is
\begin{equation}
  L_{\mathrm{lim}}(z,\alpha_c) = 4\pi D_L^2(z)\,S_{{\rm lim}}\,
  \Big(\frac{\nu_{\rm ref}}{\nu_{\rm obs}}\Big)^{-\alpha_c}(1+z)^{1+\alpha_c},\label{eq:L_flux_ceiling}
\end{equation}
with $L_\nu\propto\nu^{-\alpha}$.
\paragraph{Forward (projected) model.}
When comparing to data taken over a finite redshift bin $z\in[z_1,z_2]$ we
project the intrinsic RLF under the same selection,
\begin{equation}
\begin{split}
  \Phi^{\rm proj}_c(L\,|\,z_1{:}z_2) =
  \\
  \frac{\int_{z_1}^{z_2} \Phi_c(L,z)\,
        \Theta\!\big[L-L_{\rm lim}(z,\alpha_c)\big]\,
        ({{\rm d}V}/{\rm d}z)\,{\rm d}z}
       {\int_{z_1}^{z_2} ({{\rm d}V}/{\rm d}z)\,{\rm d}z} .
  \label{eq:phi_projected}
\end{split}
\end{equation}
These orange dashed curves in Figs.~\ref{fig:agn_rlf_panels}–\ref{fig:sfg_rlf_panels}
should be regarded as the baseline “model-to-data” comparison.

\paragraph{Binned estimator from the catalogue ($V_{\rm eff}$).} The blue points are a direct estimate from the catalogue in the $[z_1,z_2]\times[L_1,L_2]$ cell using an effective volume that accounts for partial visibility of the $L$–bin across the slice. Define the bin width in dex by $\Delta\log_{10}L=\log_{10}(L_2/L_1)$. At any redshift $z$, the fraction of the luminosity bin that is observable is
\begin{equation}
\begin{split}    
f_{\rm vis}(z;L_1,L_2)=\\
\begin{cases}
1, &
L_{\rm lim}(z,\alpha_c)\le L_1,\\[2pt]
\displaystyle
\frac{\log_{10}\!\big(L_2/L_{\rm lim}(z,\alpha_c)\big)}
     {\log_{10}\!\big(L_2/L_1\big)}, &
L_1< L_{\rm lim}(z,\alpha_c) < L_2,\\[2pt]
0, &
L_{\rm lim}(z,\alpha_c)\ge L_2\, .
\end{cases}
\end{split}
\label{eq:fvis}
\end{equation}

The corresponding effective comoving volume for the cell is
\begin{equation}
  V_{\rm eff}(L_1,L_2;z_1{:}z_2) \;=\;
  \Omega \int_{z_1}^{z_2} \frac{{\rm d}V}{{\rm d}z}(z)\, f_{\rm vis}(z;L_1,L_2)\,{\rm d}z,
  \label{eq:veff}
\end{equation}
with survey solid angle $\Omega$ (sr). If $N_c$ catalogue objects of class $c$
fall in the cell, the minimum–variance, Poisson estimator is
\begin{equation}
  \widehat{\Phi}_c(L\,|\,z_1{:}z_2) \;=\;
  \frac{N_c}{V_{\rm eff}(L_1,L_2;z_1{:}z_2)\,\Delta\log_{10}L}\,,
\end{equation}
\begin{equation}
  \sigma^2\!\left[\widehat{\Phi}_c\right] \;=\;
  \frac{N_c}{\big[V_{\rm eff}\,\Delta\log_{10}L\big]^2}.
  \label{eq:phi_veff}
\end{equation} If $N<20$ for any bin, we use the estimators of \cite{Gehrels} in constructing the confidence interval.
We plot the estimate at the bin centre with vertical error bars
(from Equation~\ref{eq:phi_veff}) and omit bins where $V_{\rm eff}=0$.
Horizontal bars indicate $\Delta\log_{10}L$. For empty bins with
$V_{\rm eff}>0$ we show one–sided upper limits (Gehrels-style) when useful.

\begin{figure}
    \centering
    \includegraphics[width=0.8\linewidth]{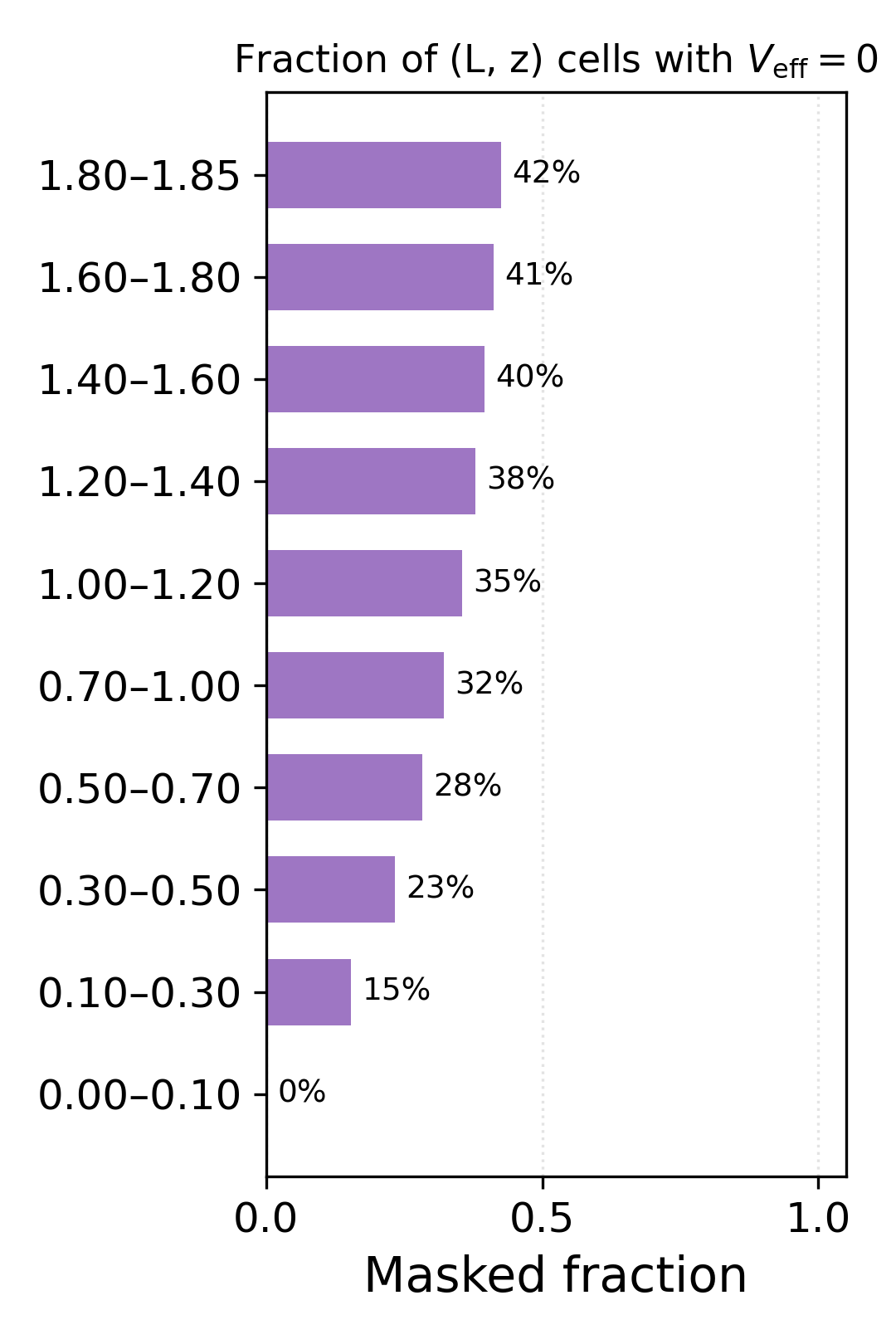}
    \caption{Fraction of $(L,z)$ cells with zero effective volume, $V_{\rm eff}=0$,
in each redshift bin for the raw selection. Bars show the average masked
fraction across AGN and SFG classes; these are indistinguishable within
$\lesssim0.5\%$ in every bin because the mask is set by the survey flux limit
(and $K$–correction) rather than population physics. The fraction rises
monotonically from $\lesssim15\%$ at $z\le0.3$ to $\simeq40\%$ by $z\sim1.8$,
reflecting the shrinking accessible luminosity range in a flux-limited survey.
Cells with $V_{\rm eff}=0$ are excluded from the panel-averaged
RLFs.}
\label{fig:veff_mask}
\end{figure}

\paragraph{Masked fraction of $(L,z)$ cells.}
Figure~\ref{fig:veff_mask} shows, for each redshift bin, the fraction of $(L,z)$ cells that are
masked because their effective survey volume is zero,
$V_{\rm eff}(z,L)=0$.
A cell is masked precisely when the visibility fraction in Equation~\ref{eq:fvis}
vanishes, $f_{\rm vis}(z;L_1,L_2)=0$; in those cells we set
$\Phi$ to NaN and exclude them from the RLF averages.
As expected for a flux–limited sample, the masked fraction is $\simeq 0$ in the
nearest slice $(0<z\le0.1)$ and then rises monotonically with redshift as the
flux boundary
$L_{\rm lim}(z,\zeta)\propto D_L^2(z)\,(1+z)^{\zeta-1}$
moves to higher luminosities.
Numerically, the fractions are $\approx 15\%$ at $0.10\text{--}0.30$,
$\approx 23\text{--}28\%$ at $0.30\text{--}0.70$,
$\approx 32\text{--}38\%$ at $0.70\text{--}1.40$,
and $\approx 40\text{--}42\%$ by $z\simeq1.8$.
The \ac{AGN} and \ac{SFG} panels are nearly identical because the $V_{\rm eff}$ mask is
set by survey geometry and the adopted spectral index used in the visibility
correction, not by population physics.
Thus the increasing masked fraction simply quantifies the shrinking accessible
$L$-range with redshift and explains the growing white space in the
high-$z$ RLF panels; it is a geometric selection effect, not a loss of
simulated sources nor a model shortcoming.

\begin{figure}
    \centering
    \includegraphics[width=\linewidth]{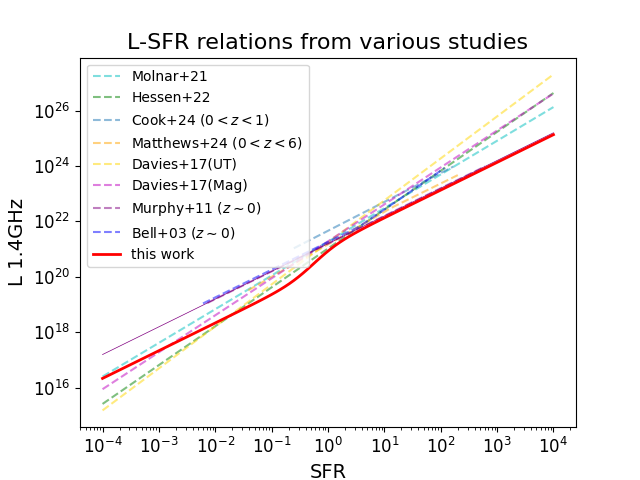}
    \caption{Comparison of 1.4\,GHz radio luminosity--SFR calibrations. Dashed curves show literature relations \citep{2021MNRAS.504..118M,2022A&A...664A..83H,2024MNRAS.531..708C,2024ApJ...966..194M,2017MNRAS.466.2312D,2011ApJ...732..126M,2003ApJS..149..289B}. The solid red line (“this work”) is our adopted mapping. At high SFR the slope of our relation is shallower than several published fits, compressing the bright tail when mapping a given $\phi_{\rm SFR}(z)$ into $L_{1.4}$ and thereby yielding fewer \ac{SFG} at $L_{1.4}\!\gtrsim\!10^{23\text{--}24}\,\mathrm{W\,Hz^{-1}}$. This behaviour contributes to the lower bright-end \ac{SFG} space densities seen at $z\gtrsim1$ in Fig.~\ref{fig:sfg_rlf_panels}. The halo mass range and abundance matching limits the upper SFR range in practice.}
    \label{fig:lsfr_comp}
\end{figure}

\paragraph{Results at $1.4$ GHz.}
Figures~\ref{fig:agn_rlf_panels} and \ref{fig:sfg_rlf_panels} show excellent
agreement between the projected model (orange dashed) and the catalogue-based,
$V_{\rm eff}$–corrected estimates (blue points):
\begin{itemize}
  \item \textbf{AGN:} The match is tight across all redshift slices and over
  the full dynamic range where the survey is complete. This validates both the
  evolutionary form of the \ac{AGN} RLF and the class mix encoded in the model.
  \item \textbf{SFG:} At $z\gtrsim 1$ the blue points fall below some literature
  bright–end measurements. This is a model outcome: our adopted $L_{1.4}$–SFR calibration is shallower at high SFR (see Figure~\ref{fig:lsfr_comp}), compressing the bright tail and yielding fewer $L_{1.4}\gtrsim 10^{23{-}24}\,\mathrm{W\,Hz^{-1}}$ systems when an evolving SFR function is mapped into radio luminosity.
  The same behaviour is consistent with the divergence of the total counts at
  the faint flux end where \ac{SFG} dominate.
\end{itemize}

A shallow “bump/inflection” occasionally visible in the orange curves near
$10^{26}\,\mathrm{W\,Hz^{-1}}$ is naturally produced by the sum of \ac{AGN} subclasses and
their evolution; the double–power–law knee of the steep–spectrum tail,
combined with the beamed (FSRQ/BLLac) component beginning to dominate
the bright end with a slightly flatter local slope, and the finite-width
slice average in Equation~\ref{eq:phi_projected}. A per-class decomposition of
$\Phi^{\rm proj}$ confirms that the total slope locally reflects the crossover of
\ac{SSAGN} and beamed contributions. The feature is therefore not required by the
data and does not affect the excellent overall agreement.

\begin{figure}
  \centering
  \includegraphics[width=\linewidth]{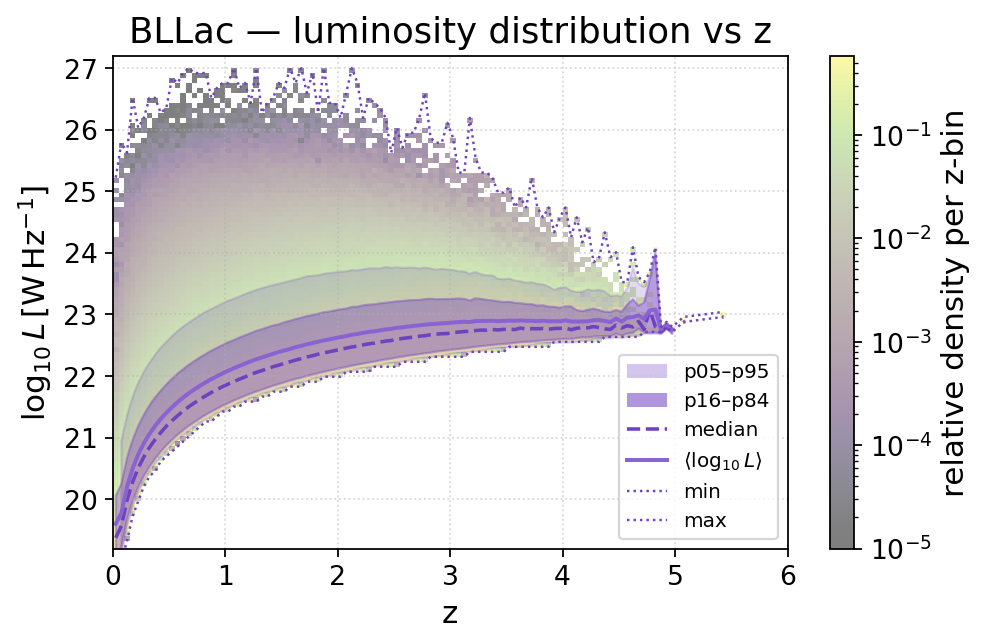}
  \caption{\textbf{BLLac — luminosity distribution vs.\ redshift.}
  Rows are normalized by their own maxima; ribbons show p16–p84 and p05–p95 of $\log_{10}L_\nu$. The narrowing and softening of the upper envelope (shown by the dashed purple line) at high $z$ reflect the flux limit and declining BL\,Lac space density. The lower minimum envelope is shown by the lower dashed line.}
  \label{fig:lz_bllac}
\end{figure}

\begin{figure}
  \centering
  \includegraphics[width=\linewidth]{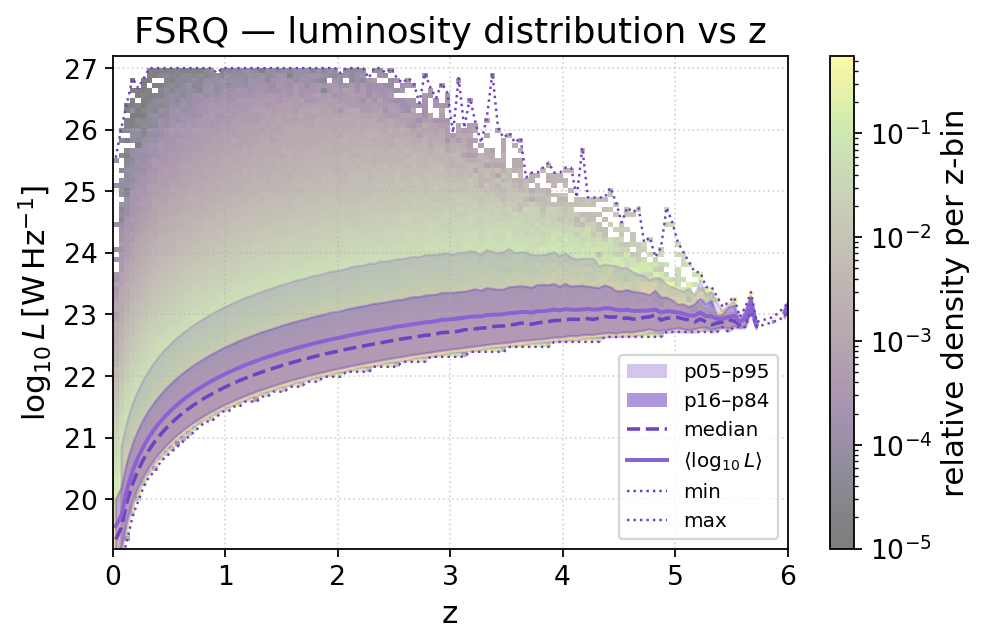}
  \caption{\textbf{FSRQ — luminosity distribution vs.\ redshift.}
  Same notation as \ref{fig:lz_bllac}. The broad luminous tail at $z\!\lesssim\!1$ and the subsequent contraction beyond $z\!\sim\!3$ are evident.}
  \label{fig:lz_fsrq}
\end{figure}

\begin{figure}
  \centering
  \includegraphics[width=\linewidth]{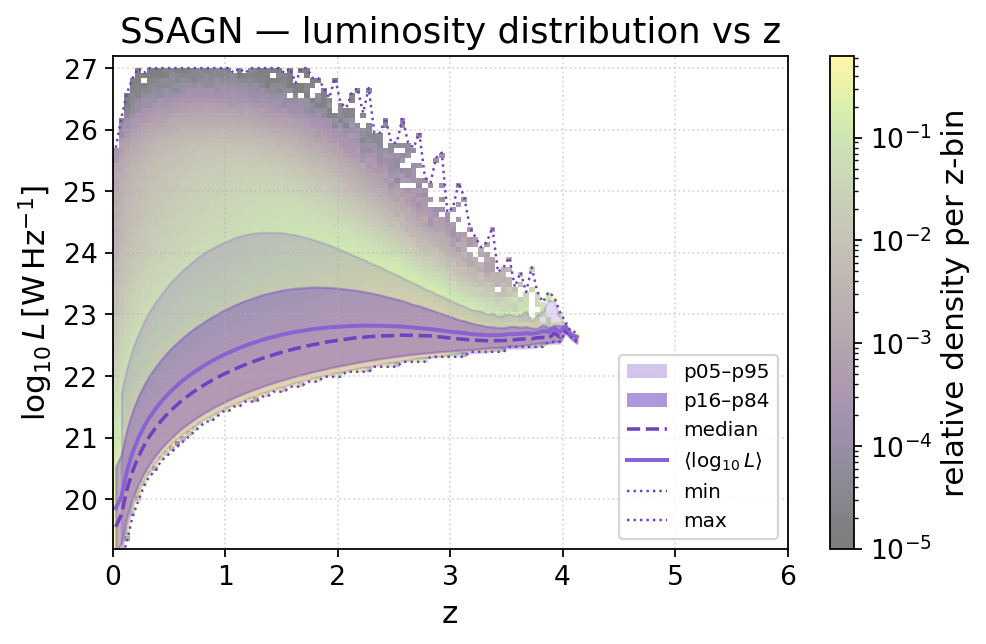}
  \caption{\textbf{\ac{SSAGN} — luminosity distribution vs.\ redshift.}
  Same notation as \ref{fig:lz_bllac}. The median rises to $z\!\sim\!1$ and then turns over, with the distribution tightening as the flux boundary dominates.}
  \label{fig:lz_ssagn}
\end{figure}

\begin{figure}
  \centering
  \includegraphics[width=\linewidth]{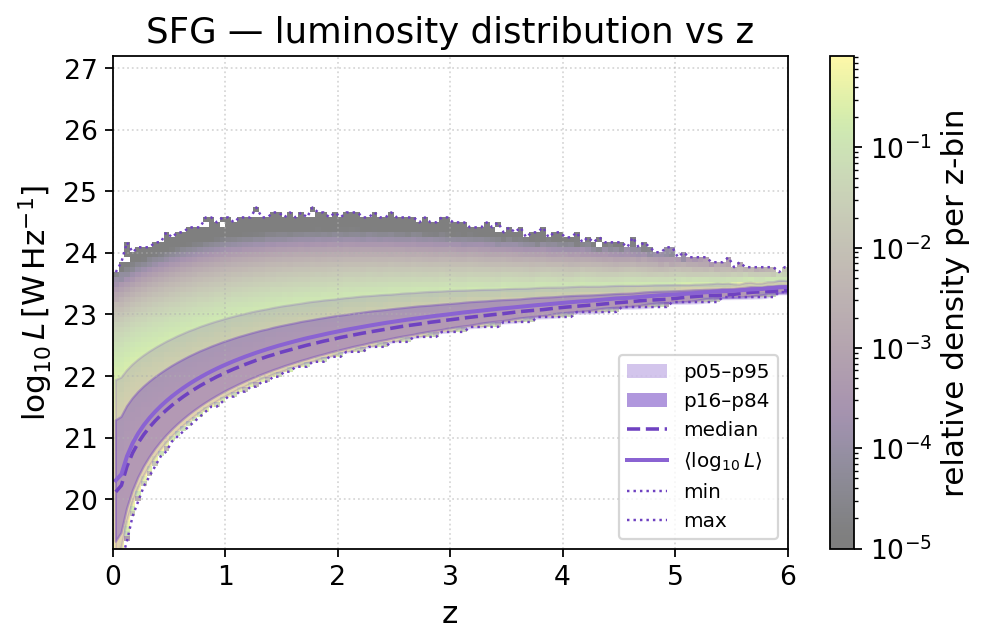}
  \caption{\textbf{\ac{SFG} — luminosity distribution vs.\ redshift.}
  Same notation as \ref{fig:lz_bllac}. The median follows the cosmic growth of typical SFRs to $z\!\sim\!2$–3 and declines gently thereafter; the accessible upper envelope drops beyond $z\!\sim\!3$ as the flux limit and host availability restrict very luminous \ac{SFG}.}
  \label{fig:lz_sfg}
\end{figure}

\section{Luminosity–redshift distributions by class}
\label{sec:lumz}

Figures~\ref{fig:lz_bllac}–\ref{fig:lz_sfg} show, for each population, the distribution of rest–frame radio luminosity versus redshift. Shading encodes the relative number density per redshift bin; over-plotted are the median and mean of $\log_{10}L$, central ($p16$–$p84$) and extended ($p05$–$p95$) percentile bands, and the minimum/maximum values realised in the catalogue. These panels are intended to summarise, at a glance, how the radio luminosity function (RLF) and the survey selection combine to populate the light–cone for each class.

\medskip\noindent
A few general features are common to all panels. The lower envelope follows the flux–limited luminosity threshold $L_{\rm lim}(z,\zeta)$ implied by the selection and $K$–corrections (Section~\ref{sec:validation-selections}), which rises rapidly with redshift and produces the Malmquist–like increase of the median at low $z$. The separation between the mean and the median indicates positive skewness due to a high–$L$ tail; the width of the percentile bands traces the intrinsic luminosity spread convolved with selection. At $z\lesssim0.5$ the accessible dynamic range in $L$ is largest and the bands are widest. As $z$ increases the faint end is progressively removed by $L_{\rm lim}$, the median rises, and the bright tail is governed by the evolving knee of the RLF and by the rarity of very luminous sources. Beyond $z\sim2$–3 the high–$L$ tail thins and the percentile bands narrow as space densities decline; small–number statistics appear as occasional spikes in the dotted “max” curves in the highest–$z$ bins.

\medskip\noindent
Star–forming galaxies exhibit the lowest median luminosities and comparatively narrow percentile bands at fixed $z$. The median increases gently with redshift, while the mean sits only slightly above it, signalling limited skewness. The density shading peaks at intermediate redshift and fades by $z\sim3$, consistent with the smoothly evolving, radio–first \ac{SFG} RLF of Section~\ref{sec:sfg-rlf}. This behaviour also explains the modest sub–mJy upturn in the counts (Section~\ref{sec:counts}): the high–$L$ \ac{SFG} tail at $z\gtrsim1$ is present but not dominant, so the projected rise in $S^{2.5}dN/dS$ is weaker than in datasets whose \ac{SFG} bright end is steeper or more numerous.

\medskip\noindent
Steep–spectrum \ac{AGN} span a wide luminosity range. Through $z\sim1$–2 the high–$L$ tail is prominent, lifting the mean well above the median and widening the $p16$–$p84$ band. The median sits roughly a dex above the \ac{SFG} median at the same redshift, reflecting the higher typical host masses of radio–loud systems (Section~\ref{sec:agn-pop}). At $z\gtrsim2.5$ the percentile bands narrow and the maximum realised luminosity drops, tracing both the \ac{SSAGN} RLF evolution and the diminishing supply of massive halos available to host them (Section~\ref{sec:abmatch}).

\medskip\noindent
Flat–spectrum subclasses show the highest luminosities at fixed redshift. FSRQ display broad percentile bands and a large mean–median gap, as expected for a beamed population with a strong bright tail; the shading peaks around $z\sim1$–2 and declines rapidly by $z\gtrsim3$ as the combination of RLF evolution and beamed fractions reduces the space density. BL\,Lac objects occupy a lower–luminosity locus than FSRQ and a somewhat narrower redshift span; their bands are tighter and the shading is more local–heavy, indicating milder evolution and smaller overall abundance. In both beamed classes the dotted “max” curve shows occasional excursions at high redshift that are consistent with small–number statistics.

\medskip\noindent
These shapes have direct consequences for selection and for the observables compared elsewhere in this paper. Because changes in survey selection functions compress or broaden the effective redshift kernel $\varphi(z)$ (Section~\ref{sec:validation-selections}), the rising $L_{\rm lim}(z)$ cuts different parts of each panel for a given flux limit. \ac{SFG} are most affected at low to intermediate redshift where their locus lies close to $L_{\rm lim}$; \ac{SSAGN} and flat–spectrum \ac{AGN} remain visible to higher $z$ but are shot–noise limited in the brightest tail. This selection geometry explains why projected clustering amplitudes and slopes shift with the survey selection functions (Section~\ref{sec:clustering}) without altering the intrinsic bias ordering inferred from the host–mass distributions: \ac{SFG} occupy lower–mass halos and populate lower $L$ at a given $z$; \ac{SSAGN} are intermediate; FSRQ/BL\,Lac trace the luminous, highly biased tail.

\medskip\noindent
For multi–tracer applications the panels make explicit that each class supplies a distinct $P(z)$ kernel and linear–bias evolution. The ordered rise of the medians and the class–to–class offsets stabilise the bias contrast required for cosmic–variance cancellation on ultra–large scales, while the thinning of the bright tails at $z\gtrsim3$ sets practical limits on how aggressively very–high–$z$ selections can be pushed before shot noise dominates. Together with the number–count agreement in Section~\ref{sec:counts} and the clustering results in Section~\ref{sec:clustering}, these luminosity–redshift maps close the loop between population prescriptions, host assignment, and the survey selections used for validation.

\section{Improvements}
\label{sec:improvements}

\noindent
\textsc{GHOST} is an all-sky radio–continuum simulation with population-dependent clustering, redshift distributions, and survey selections designed to emulate realistic wide-area datasets. Below we summarise where the current release is most useful and outline focused directions that will make it more powerful for cosmology and survey work.
\begin{itemize}
    \item Spectral assumptions and $K$-corrections. Using a single (or overly narrow) spectral-index model per population tightens the flux–luminosity mapping, especially at high $z$ where rest-frame frequencies shift. \emph{Future direction:} adopt redshift-dependent spectral-index distributions (optionally with curvature), for exact downstream $K$-corrections, and validate against multi-frequency counts.
    \item \ac{SFG} RLF bright end and the sub-mJy upturn. The radio-first \ac{SFG} RLF underpredicts the high-$L$ \ac{SFG} tail at $z\gtrsim1$, weakening the classical upturn near $S\sim0.1$–$0.3$ mJy. \emph{Future direction:} re-fit the \ac{SFG} RLF with deeper fields and include a cross-check pathway that links to SFR/IR constraints (without making them mandatory) to control the bright tail while preserving the radio-first baseline.
    \item Satellites and halo occupation. The current catalogue assigns one central per halo and no explicit satellites. This can bias small-scale $w(\theta)$ and environment trends, even if large-scale bias is stable. \emph{Future direction:} add a minimal HOD layer (central\,+\,satellites) per population, tuned to small-scale clustering while keeping the ultra–large-scale predictions unchanged.
    \item Survey realism and detection. Point-like components and simplified morphologies miss PSF/blending effects and spatially varying completeness. \emph{Future direction:} ship an “observe-at-beam’’ operator (beam, mosaics, noise/rms maps, deblending/thresholding) that produces detection catalogues, randoms, and effective-area curves consistent with pseudo-$C_\ell$ inputs.
    \item Documentation and delivered products. \emph{Future direction:} include per-source SED parameters ($\zeta$, curvature), $p(z)$ options, masks/randoms, and $n(z)$ kernels used in the figures so analyses can be reproduced exactly.
\end{itemize}

\medskip\noindent
Taken together, these items indicate where \textsc{GHOST} can grow. They also clarify how to interpret the current results: robust for wide-area cosmology (PNG, ISW, magnification) under the stated assumptions, with clear paths to refine high-$z$ populations, spectral modelling, morphology, and survey realism in future releases.

\section{Data availability}
The data and simulated catalogues will be made available on the Data Central service (https://datacentral.org.au/).

\section{Conclusions}
\label{sec:conclusion}

\noindent
\textsc{ghost} is an all-sky, clustered, survey-ready radio mock built on the \textsc{flamingo} light-cone and tailored to ultra–large-scale, multi-tracer cosmology. The catalogue ties population prescriptions to a specific halo field through rank-preserving, no-replacement abundance matching (Section~\ref{sec:abmatch}), ensuring that large-scale bias follows the halo supply. The \ac{SFG} branch is calibrated directly in radio via a smoothly evolving 1.4\,GHz RLF (Section~\ref{sec:sfg-rlf}), while the \ac{AGN} branch retains the \textsc{t-recs} functional form with coefficients refit as needed and a new host assignment on our light-cone (Section~\ref{sec:agn-pop}). Together these choices provide transparent control of bias and redshift kernels at the scales relevant for $f_{\rm NL}$.

Across 150\,MHz, 1.4\,GHz, and 3\,GHz the Euclidean-normalized differential counts are reproduced over nearly seven decades in flux (Section~\ref{sec:counts}). The model shows no frequency-dependent bias in the total counts; the subdued sub-mJy upturn is consistent with the high-luminosity tail of the radio-first \ac{SFG} RLF at $z\gtrsim1$, rather than with host-assignment choices. This anchors the overall normalisation of the luminosity functions and spectral assumptions used throughout.

Using common selections for both $n(z)$ and clustering, we emulate the survey selection functions used by representative surveys and recover their qualitative behaviour (Section~\ref{sec:nz}). Emulating LoTSS-wide, the VLA–COSMOS luminosity split of \citet{2017MNRAS.464.3271M}, and the 3\,GHz S/N threshold of \citet{Hale2018}, we find that differences between literature measurements are largely explained by how the survey selection functions compress or broaden the projection kernel $\varphi(z)$. Where a survey defines its own AGN/\ac{SFG} partition \citep{2017MNRAS.464.3271M}, we mimic it explicitly; otherwise we retain our internal classes and emulate only the detection to isolate selection from population-mix effects. This like-for-like procedure keeps $n(z)$ and clustering internally consistent and clarifies the role of classification in published comparisons.

In configuration and harmonic space, the catalogue exhibits a stable bias hierarchy with $b_{\rm \ac{SFG}}<b_{\rm \ac{SSAGN}}\lesssim b_{\rm FSRQ}\simeq b_{\rm BLLac}$ (Section~\ref{sec:cl_bias}). Power-law fits to $w(\theta)$ and template fits to $C_\ell$ agree and map cleanly onto the host–halo mass PDFs: \ac{SFG} inhabit lower-mass halos; radio-loud \ac{AGN} occupy more massive halos. Apparent slope and amplitude shifts under different selections are traced to changes in $\chi_{\rm eff}$ and $\varphi(z)$, not to a reversal of intrinsic bias ordering. This closes the loop between the population model, host assignment, and the observed large-scale clustering.

For applications, the catalogue supplies the two ingredients required by multi-tracer estimators targeting local-type primordial non-Gaussianity: a robust bias contrast between tracer families and distinct, selection-matched redshift kernels. The same full-sky, selection-aware density fields support late-time CMB cross-correlations (ISW and CMB lensing), magnification studies via add-on ray tracing, radio-dipole tests, and survey design tasks that require controlled injections of depth and masking systematics.

The main caveats and near-term improvements are well circumscribed (Section~\ref{sec:improvements}). The \ac{SFG} bright end at $z\gtrsim1$ can be revisited with deeper fields and optional SFR/IR cross-checks; spectral-index distributions can be broadened (and made redshift-dependent) to sharpen $K$-corrections; small-scale clustering and detection realism will benefit from satellites and morphology/PSF pipelines. None of these affect the principal conclusions on large scales.

In summary, \textsc{ghost} reproduces multi-frequency counts, delivers selection-matched $n(z)$, and yields a physically ordered bias hierarchy tied to the \textsc{flamingo} halo field. These properties make it a practical sandbox for end-to-end ultra–large-scale analyses. Paper~II carries the calibrated bias parameters and kernels forward to quantify $f_{\rm NL}$ sensitivity with variance-canceling multi-tracer estimators on the simulated sky.

\section*{Acknowledgments}
B.V. would like to thank the members of the \textsc{FLAMINGO} collaboration Joop Schaye, Matthieu Schaller, Roi Kugel and John Helly for the helpful discussions.

\bibliographystyle{pasa-mnras}
\bibliography{example}

\clearpage
\appendix

\section{Working equations used in \textsc{GHOST}}\label{app:trecs-eqs}
This appendix collects the definitions and working relations used throughout \textsc{GHOST} and links them to the survey selections and the observables we compare to data.

\subsection{Notation, $K$-corrections, and selection mapping}
\label{app:notation}
We define the observed flux density at frequency $\nu_{\rm obs}$ by
\begin{equation}
S_{\nu_{\rm obs}} \;=\; \frac{L_{\nu_{\rm rest}}}{4\pi D_L^2(z)}\,
\frac{1}{1+z} \,,
\qquad \nu_{\rm rest}=(1+z)\,\nu_{\rm obs},
\end{equation}
with $D_L$ the luminosity distance. For a power-law spectrum $L_\nu\propto \nu^{-\alpha}$ the usual radio $K$-correction implies
\begin{equation}
L_{\nu_{\rm rest}} \;=\; 4\pi D_L^2(z)\,S_{\nu_{\rm obs}}\,(1+z)^{1+\alpha}.
\end{equation}
Hence a survey limit in $S_{\nu_{\rm obs}}$ translates into a luminosity threshold at redshift $z$ for a class with characteristic $\alpha_c$:
\begin{equation}
L_{\rm lim}(z,\alpha_c)\;=\;4\pi D_L^2(z)\,S_{\rm lim}\,(1+z)^{1+\alpha_c}.
\end{equation}
Within any luminosity/redshift bin $[L_1,L_2]$, the fraction of sources visible at $z$ is
\begin{equation}
\begin{split}
f_{\rm vis}(z;L_1,L_2)=\\
\begin{cases}
1, &
L_{\rm lim}(z,\alpha_c)\le L_1,\\[2pt]
\displaystyle
\frac{\log_{10}\!\big(L_2/L_{\rm lim}(z,\alpha_c)\big)}
     {\log_{10}\!\big(L_2/L_1\big)}, &
L_1< L_{\rm lim}(z,\alpha_c) < L_2,\\[2pt]
0, &
L_{\rm lim}(z,\alpha_c)\ge L_2\, .
\end{cases}
\label{eq:fvis_app}
\end{split}
\end{equation}
As redshift increases, $L_{\rm lim}$ increases in $\log L$ and progressively trims the faint end of each bin; for a fixed flux limit, this effect is strongest for \ac{SFG} (larger $\zeta$, lower intrinsic $L$) and weakest for flat-spectrum \ac{AGN}. This geometric gate is the lower envelope seen in the $L$–$z$ panels and drives selection-induced changes in the slope of $w(\theta)$.

The per-steradian redshift distribution for class $X$ under a flux cut is
\begin{equation}
\begin{split}
n_X(z)\;=\;\frac{dN_X}{dz\,d\Omega}\;=\;\\\frac{dV}{dz\,d\Omega}
\int d\log_{10}L\, \Phi_X(L,z)\, \Theta[L-L_{\rm lim}(z,\alpha_X)],
\label{eq:nz_from_rlf}
\end{split}
\end{equation}
and the differential counts follow from the change of variables $L\!\leftrightarrow\!S$:
\begin{equation}
\frac{dN}{dS}(S)\;=\;\int dz\,\frac{dV}{dz\,d\Omega}\,
\left.\Phi\!\big(L,z\big)\,\frac{d\log_{10}L}{dS}\right|_{L=4\pi D_L^2 S (1+z)^{1+\alpha}},
\label{eq:counts_from_rlf}
\end{equation}
with $d\log_{10}L/dS = [\ln 10]^{-1} S^{-1}$. In practice, total counts constrain the \emph{aggregate} $\Phi$ (summing over classes), while $n_X(z)$ and clustering are sensitive to how $\Phi_X$ partitions that total across classes and redshifts.

\subsection{\ac{AGN} radio luminosity function and evolution}
\label{app:agn-rlf}
We adopt the double–power-law form of \cite{TRECS} with redshift evolution implemented via a de-evolution mapping $L(z)\!\to\!L(0)$:
\begin{equation}\label{eq:RLFAGN}
\Phi(L\,|\,z) \;=\;
\frac{n_0}
{\big(\tfrac{L(0)}{L_\star(0)}\big)^{\gamma}
 + \big(\tfrac{L(0)}{L_\star(0)}\big)^{\beta}}
\;\cdot\;
\frac{ d\log_{10} L(0) }{ d\log_{10} L(z) } .
\end{equation}
For a separable evolution $L(0)=L(z)/\mathcal{E}(z)$ the Jacobian equals unity, but we keep it explicit to allow more general mappings.

The characteristic luminosity evolves as
\begin{equation}\label{eq:charlum}
L_\star(z) \;=\; L_\star(0)\,
\mathrm{dex}\!\left[
k_{\mathrm{evo}}\, z \left(
2 z_{\mathrm{top}}
 - \frac{ z^{m_{\mathrm{evo}}}\, z_{\mathrm{top}}^{\,1 - m_{\mathrm{evo}}} }{ 1 + m_{\mathrm{evo}} }
\right)\right],
\end{equation}
while the peak redshift depends on luminosity through
\begin{equation}\label{eq:charz}
z_{\mathrm{top}}(L) \;=\; z_{\mathrm{top},0} \;+\;
\frac{ \delta z_{\mathrm{top}} }
     { 1 + \big( L(0)/L_\star(0) \big) } .
\end{equation}
Together, Equations~\ref{eq:charlum}–\ref{eq:charz} produce an early rise and late decline of $L_\star(z)$ and encode “downsizing,” with higher-luminosity \ac{AGN} peaking earlier. The faint/bright slopes $(\gamma,\beta)$ govern the curvature of $S^{2.5}\,dN/dS$ at the bright end. Given $\Phi$, expected counts in a $(\log L,z)$ bin are
\begin{equation}\label{eq:deltansij}
\Delta N_{i,j} \;=\; \Omega \,\Phi(L\,|\,z_i)\,
\left.\frac{dV}{dz}\right|_{z_i}\,
\Delta \log_{10} L_{i,j}\,\Delta z_i,
\end{equation}
the building block for both $n(z)$ and $dN/dS$ via Equations~\ref{eq:nz_from_rlf}–\ref{eq:counts_from_rlf}. The values of the constants in Equations~\ref{eq:RLFAGN}~and~\ref{eq:charz} are given in Table~\ref{tab:trecs_params}.

\begin{table}
\centering
\caption{Best-fit parameters of the \ac{AGN} evolutionary RLF adopted from \textsc{T-RECS}. 
The slopes $\gamma$ and $\beta$ correspond to $a$ and $b$ in \cite{TRECS}. 
Parameters $k_{\rm evo}$, $z_{\rm top,0}$, $\delta z_{\rm top}$, and $m_{\rm evo}$ enter Equations~\ref{eq:charlum}–\ref{eq:charz}.}
\label{tab:trecs_params}
\begin{tabular}{lccc}
\toprule
Parameter & FSRQ & BL~Lac & SS-AGN \\
\midrule
$\gamma$ (faint-end slope)        & 0.776  & 0.723  & 0.508 \\
$\beta$  (bright-end slope)       & 2.669  & 1.918  & 2.545 \\
$\log_{10} n_0\,[\mathrm{Mpc}^{-3}\,\mathrm{dex}^{-1}]$ & $-8.319$ & $-7.165$ & $-5.973$ \\
$\log_{10} L_\star(0)\,[\mathrm{W\,Hz}^{-1}]$          & 33.268 & 32.282 & 32.560 \\
$k_{\rm evo}$                     & 1.234  & 0.206  & 1.349 \\
$z_{\rm top,0}$                   & 2.062  & 1.262  & 1.116 \\
$\delta z_{\rm top}$              & 0.559  & —      & 0.705 \\
$m_{\rm evo}$                     & 0.136  & 1.000  & 0.253 \\
\bottomrule
\end{tabular}
\end{table}

\subsection{HERG/\ac{LERG} host probabilities and effective bias} \label{app:herg-lerg} Steep-spectrum \ac{AGN} are split into excitation classes using stellar-mass scalings from \cite{Janssen2012}, \begin{align} P_{\rm LERG}(M_\star|z) &\propto \left(\frac{M_\star}{M_0}\right)^{2.5} \quad M_\star\lesssim10^{11.6}M_\odot,\\ P_{\rm HERG}(M_\star|z) &\propto \left(\frac{M_\star}{M_0}\right)^{1.5}, \end{align} renormalised over candidate hosts. Mapping $M_\star$ to halo mass $M$ via the stellar–halo relation (next subsection) yields the class-effective linear bias \begin{equation} b_t(z)\;=\;\frac{\displaystyle\int dM\, n(M,z)\,b(M,z)\,W_t(M,z)} {\displaystyle\int dM\, n(M,z)\,W_t(M,z)}\,, \qquad \end{equation} \begin{equation} W_t\equiv P_t(M_\star(M,z),z), \label{eq:beff} \end{equation} with $n(M,z)$ the halo mass function and $b(M,z)$ a standard halo-bias model. Since \ac{HERG}-like weights favour higher $M_\star$, Equation~\ref{eq:beff} explains the robust bias hierarchy relative to \ac{SFG}.

\subsection{Sizes}\label{app:sizes}

Our catalogue attaches a minimalist morphology proxy to each radio \ac{AGN} solely to split the \ac{SSAGN} pool into \ac{HERG}/\ac{LERG}–like sub-classes and apply class-dependent, stellar–mass–weighted halo assignment. Sizes do not affect photometry or clustering beyond the
sub-class/host selection described here.

Following the approach of \cite{TRECS}, each \ac{AGN} is given a characteristic linear
size $D$ (kpc) via a stochastic draw from the broad, observationally motivated distribution of \cite{Dipompeo} with a weak luminosity trend. A random sky orientation is applied (uniform in $\sin\theta$). We
do not simulate shapes or images; $D$ is a single scale per source used only as
a morphology proxy.

We compute a size–ratio
\begin{equation}
  R_s \equiv \frac{d_{\rm peak}}{D_{\rm tot}},
\end{equation}
the separation of the two peak–brightness regions $d_{\rm peak}$ divided by the
end–to–end extent $D_{\rm tot}$. We adopt the usual FR cut $R_s>0.5 \Rightarrow$
FR\,II and $R_s<0.5 \Rightarrow$ FR\,I. The draw of $R_s$ is mildly
luminosity–dependent: for $L_\nu\le 10^{25.4}\,{\rm W\,Hz^{-1}}$ we take
$R_s\sim\mathcal{N}(\mu=0.18,\sigma=0.11)$, and for $L_\nu>10^{25.4}\,{\rm W\,Hz^{-1}}$
we take $R_s\sim\mathcal{N}(\mu=0.62,\sigma=0.18)$. This biases high–power sources
toward FR\,II while still allowing overlap in $R_s$ at fixed $L$.

We then assign the spectroscopic class with a \emph{deterministic} rule at fixed
$R_s$: sources with $R_s>0.5$ are tagged as \ac{HERG}, and those with $R_s\le 0.5$
as \ac{LERG}. Because $R_s$ itself is drawn with scatter, the resulting fractions
vary smoothly with luminosity; in particular
\begin{equation}
P({\rm HERG}\,|\,L)
= P(R_s>0.5\,|\,L)
= 1 - \Phi\!\left(\frac{0.5-\mu(L)}{\sigma(L)}\right),
\end{equation}
with $\Phi$ the standard normal CDF and $(\mu(L),\sigma(L))$ given above.
Hence FR\,II–\ac{LERG} and FR\,I–\ac{HERG} combinations naturally occur,
even though the label is a hard threshold in $R_s$.

\subsection{Stellar–halo mass relation (for host selection)}
\label{app:smhm}
Host selection uses a redshift-dependent double-power relation \citep{TRECS,2015ApJ...810...74A},
\begin{equation}
M_\star(M,z) \;=\; N_0(z)\,
\left[
\left(\frac{M}{M_b(z)}\right)^{\epsilon(z)}
+ \left(\frac{M}{M_b(z)}\right)^{\omega(z)}
\right]^{-1},
\end{equation}
interpolated from the fit of \citet{2015ApJ...810...74A}. This sets \ac{AGN} host probabilities (via $M_\star$) and provides the smooth proxy used to fit the \ac{SFG} $L(M)$ map. Steepening at high mass boosts \ac{HERG} bias relative to LERG; shifts in $M_b(z)$ move the characteristic host mass and reshape class-specific $n(z)$ tails.

\subsection{\ac{SFG} radio luminosities: synchrotron and free–free}
\label{app:sfg-lnu}
We model the \ac{SFG} spectrum as a sum of non-thermal synchrotron and thermal free–free components \citep{TRECS,Mancuso}. The synchrotron component is
\begin{equation}
L_{\rm sync}(\nu) \;=\;
\frac{L_{*,\rm sync}(\nu)}
     {\left(\dfrac{L_{*,\rm sync}}{\bar{L}_{\rm sync}}\right)^{\beta}
     + \left(\dfrac{L_{*,\rm sync}}{\bar{L}_{\rm sync}}\right)} ,
\end{equation}
with fiducial scaling
\begin{equation}
\begin{split}
\bar{L}_{\rm sync} \;=\; 1.9\times10^{21}
\left( \frac{\mathrm{SFR}}{M_\odot\,{\rm yr}^{-1}} \right)
\left( \frac{\nu}{\rm GHz} \right)^{-0.85}
\\
\left[ 1 + \left( \frac{\nu}{20\,{\rm GHz}} \right)^{0.5} \right]^{-1}
\;{\rm W\,Hz^{-1}}\!.
\end{split}
\end{equation}
The free–free component follows
\begin{equation}
L_{\rm ff}(\nu) \;\propto\; T^{-0.5}\, g_{\rm ff}(\nu,T)\,\nu^{-0.1},
\qquad T=10^4\,{\rm K},
\end{equation}
with Gaunt factor \citep{2014MNRAS.444..420V}
\begin{equation}
\begin{aligned}
g_{\rm ff}(\nu_{\rm GHz},T)
&= \ln\!\Bigl(
e + \exp\!\Bigl[
5.960 - \frac{\sqrt{3}}{\pi}\,
\ln\!\bigl(\nu_{\rm GHz}\,T_4^{-3/2}\bigr)
\Bigr]\Bigr),\\
T_4 &\equiv T/10^4\,{\rm K}.
\end{aligned}
\end{equation}

Below $\sim 10$–20\,GHz, synchrotron with $\zeta\!\approx\!0.8$–0.9 dominates; free–free adds a flatter $\zeta\!\approx\!0.1$ tail, making extrapolations to a few GHz slightly shallower than pure synchrotron. Together with a smoothly evolving \ac{SFG} RLF, this SED reproduces multi-frequency counts.

\subsection{Abundance matching and the working $L(M)$ map}
\label{app:abmatch}
We assign radio luminosities to halos monotonically in each redshift slice so the realised luminosity histogram reproduces the target RLF. For \ac{SFG} we fit
\begin{equation}\label{eq:LofM_app}
L(M;z) \;=\; N(z)\,
\left[
\left(\frac{M}{M_\star(z)}\right)^{\gamma}
+ \left(\frac{M}{M_\star(z)}\right)^{\beta}
\right]^{-1},
\end{equation}
apply a log-normal scatter $\sigma_{\log L}$, and perform rank-ordering without replacement. Parameters are inferred per slice with a Poisson likelihood for the RLF-implied counts. Equation~\ref{eq:LofM_app} behaves as a flexible double-slope map: $\gamma$ controls the faint-end response to host mass, $\beta$ the bright-end roll-off, $M_\star$ the transition scale, and $N$ the normalisation; the scatter compresses the effective $L$–$M$ relation and slightly lowers bias at fixed $L$. For \ac{AGN}, we first draw hosts with the mass-weighted \ac{HERG}/\ac{LERG} probabilities and then assign $L$ monotonically within the selected subset to match the class RLFs, keeping bias tied to halo mass (Equation~\ref{eq:beff}).

\subsection{From populations to observables: effective kernels and bias}
\label{app:frompop2obs}
Given the above ingredients, the selection-matched redshift kernel for class $X$ is
\begin{equation}
\begin{split}
\varphi_X(z)=\frac{n_X(z)}{\int n_X(z')\,dz'}\,,
\qquad\\ \text{where } n_X(z)\text{ is given by Eq.~\ref{eq:nz_from_rlf}}.
\end{split}
\end{equation}
For the auto–correlation of tracer $X$, the effective large–scale bias entering $C_\ell$ or $w(\theta)$ is
\begin{equation}
b_{\mathrm{eff},X}
=\left[
\frac{\displaystyle \int dz\,\varphi_X^2(z)\,b_X^2(z)\,D^2(z)}
     {\displaystyle \int dz\,\varphi_X^2(z)\,D^2(z)}
\right]^{1/2}.
\label{eq:beff2}
\end{equation}

with $D(z)$ the growth factor, that is, $\delta_m(k,z)=D(z)\delta_m(k,z=0)$, and $b_X(z)$ from Equation~\ref{eq:beff}. Changes to the survey selection alter $\varphi_X(z)$ (projection) and therefore the amplitude/slope of $w(\theta)$ without changing the intrinsic bias ordering set by $b_X(z)$.

\subsection{SFR function}
\label{app:sfrf}
We use the \ac{SFR} density function (constrained by observations) from \citep{TRECS,Mancuso}
\begin{equation}
\begin{split}
\phi_{\mathrm{SFR}}(z) \;=\; \Phi(z)\,
\left[ \frac{\mathrm{SFR}}{\mathrm{SFR}_*(z)} \right]^{1 - \zeta(z)}\\
\exp \!\left\{- \left[ \frac{\mathrm{SFR}}{\mathrm{SFR}_*(z)} \right]^{w(z)} \right\},
\end{split}
\end{equation}
with redshift evolution
\begin{align}
\log \Phi(z)\,[{\rm Mpc}^{-3}\log({\rm SFR})^{-1}]
&= -2.4 - 2.3\chi \notag\\[-2pt]
&\quad {}+ 6.2\chi^2 - 4.9\chi^3, \\
\log \mathrm{SFR}_*(z)\,[M_\odot\,{\rm yr}^{-1}]
&= 1.1 + 3.2\chi \notag\\[-2pt]
&\quad {}- 1.4\chi^2 - 2.1\chi^3, \\
\zeta(z)
&= 1.2 + 0.5\chi \notag\\[-2pt]
&\quad {}- 0.5\chi^2 + 0.2\chi^3, \\
w(z)
&= 0.7 - 0.15\chi \notag\\[-2pt]
&\quad {}+ 0.16\chi^2 + 0.01\chi^3.
\end{align}

where $\chi=\log(1+z)$. In \textsc{GHOST} the \ac{SFG} branch is radio-first; this SFRF is kept as an optional cross-check to sanity-test the bright tail of $\Phi_{\rm \ac{SFG}}$ at $z\gtrsim1$ without imposing FIR–radio priors.


\end{document}